\documentclass[twocolumn,showpacs,preprintnumbers,amsmath,amssymb]{revtex4}
\usepackage{graphicx}% Include figure files
\usepackage{dcolumn}% Align table columns on decimal point
\usepackage{bm}% bold math
\usepackage{epsf}
\usepackage{subfigure}
\usepackage{epstopdf}
\usepackage{amsmath}
\usepackage{amssymb}
\newcommand{\bra}[1]{\langle #1|}
\newcommand{\ket}[1]{|#1\rangle}
\newcommand{\braket}[2]{\langle #1|#2\rangle}

\newcommand{\mean}[1]{\langle #1 \rangle}

\renewcommand{\i}{{\rm i}}
\newcommand{\e}{{\rm e}}

\begin{document}
\title{Entropy fluctuation theorems in driven open systems: application to electron counting statistics}
\author{Massimiliano Esposito}
\altaffiliation[Also at \; ]{Center for Nonlinear Phenomena and Complex Systems,
Universite Libre de Bruxelles, Code Postal 231, Campus Plaine, B-1050 Brussels, Belgium.}
\author{Upendra Harbola}
\author{Shaul Mukamel}
\affiliation{Department of Chemistry, University of California,
Irvine, California 92697, USA.}

\date{\today}

\begin{abstract}
The total entropy production generated by the dynamics of an externally driven systems exchanging 
energy and matter with multiple reservoirs and described by a master equation is expressed as the 
sum of three contributions, each corresponding to a distinct mechanism for bringing the system out 
of equilibrium: nonequilibrium initial conditions, external driving, and breaking of detailed balance.
We derive three integral fluctuation theorems (FTs) for these contributions and show that they 
lead to the following universal inequality: an arbitrary nonequilibrium transformation always produces 
a change in the total entropy production greater or equal than the one produced if the transformation 
is done very slowly (adiabatically).
Previously derived fluctuation theorems can be recovered as special cases. 
We show how these FTs can be experimentally tested by performing the counting statistics of the electrons 
crossing a single level quantum dot coupled to two reservoirs with externally varying chemical potentials.
The entropy probability distributions are simulated for driving 
protocols ranging from the adiabatic to the sudden switching limit.
\end{abstract}

\maketitle
%%%%%%%%%%%%%%%%%%%%%%%%%%%%%%%%%%%%%%%%%%%%%%%%%%%%%%%%%%%%%%%%%%%%%%%%%%%%%%%%%%%%%%%
\section{Introduction}\label{intro}

In statistical mechanics, thermodynamic laws are recovered at the level of ensemble averages.  
The past decade has brought new insights into nonequilibrium statistical mechanics due to 
the discovery of various types of fluctuation relations valid arbitrarily far from equilibrium
\cite{Evans1,Gallavotti,Evans2,Jarzynski1,Kurchan,Lebowitz,Evans3,Crooks99,Crooks00,
Hatano99,Hatano01,Gaspard1,GaspardAndrieux,Seifert1,Seifert2,Chernyak,Broeck}.
These relations identify, at the level of the single realization of a statistical ensemble, 
the "trajectory entropy" which upon ensemble averaging reproduce the thermodynamic entropy.
They therefore quantify the statistical significance of nonthermodynamic behaviors 
which can become significant in small systems \cite{Bustamante0,GaspardAndrieux2}.
Various experimental verifications of these FTs have been reported 
\cite{Bustamante1,Bustamante2,Bustamante3,Douarche,Evans,Wrachtrup1,Wrachtrup2}.\\
In this paper, we consider an open system, described by a master 
equation (ME), exchanging matter and energy with multiple reservoirs. 
The system can be externally driven by varying its energies or the different 
temperature or chemical potentials of the reservoirs.
There are three mechanisms for bringing such a system out of equilibrium: 
preparing it in a nonequilibrium state, externally driving it, 
or putting it in contact with multiple reservoirs at different temperatures 
or chemical potentials thus breaking the detailed balance condition (DBC). 
We show that each of these mechanism makes a distinct contribution to the total 
entropy production (EP) generated by the nonequilibrium dynamics of the system.
The two first contributions are nonzero only if the system is not in its steady 
state and are therefore called nonadiabatic.
The third contribution is equal to the EP for slow transformation during which 
the system remains in a steady-state and is therefore called adiabatic.  
We derive three FTs, for the total EP and its nonadiabatic and adiabatic contribution 
and show that they lead to exact inequalities valid arbitrary far from equilibrium.
Previously derived FTs are recovered by considering 
specific types of nonequilibrium transformations. 
Steady state FTs \cite{Lebowitz,Gaspard1,GaspardAndrieux} are obtained for 
systems maintained in a nonequilibrium steady state (NESS) between reservoirs with 
different thermodynamic properties.
The Jarzynski or Crooks type FTs \cite{Jarzynski1,Crooks99,Crooks00} 
are derived for systems initially at equilibrium with a single reservoir 
which are externally driven out of equilibrium by an external force. 
The Hatano-Sasa FT \cite{Hatano99,Hatano01} is recovered for externally driven 
systems initially in a NESS with multiple reservoirs.\\
To calculate the statistical properties of the various contributions to the 
total EP and to demonstrate the FTs, we extend the generating 
function (GF) method \cite{Lebowitz} to driven open systems.  
Apart from providing clear proofs of the various FTs, this method is useful for simulations 
because it does not require to explicitly generate the stochastic trajectories.
Some additional insight is provided by using an alternative derivation 
of the FTs similar to the Crooks derivation \cite{Crooks99,Crooks00}, 
where the total EP and its nonadiabatic part can be identified in terms of 
forward-backward trajectory probabilities.
By doing so, we connect the trajectory approach previously 
used for driven closed systems \cite{Seifert1,Crooks99,Crooks00} with the 
GF approach used for steady state systems \cite{Lebowitz,Gaspard1}.\\
We propose to experimentally test these new FTs in a driven single orbital quantum dot
where the various entropy probability distributions can be measured by the 
full electron counting statistics which keeps track of the four possible types of 
electron transfer (in and out of the dot through either lead).
Such measurements of the bidirectional counting 
statistics have become feasible recently \cite{Hirayama}.
We calculate the entropy probability distributions, analyze 
their behavior as the driving is varied between the sudden and the 
adiabatic limits, and verify the validity of the FTs.\\

In section \ref{mastereq} we present our stochastic model and in section \ref{entropy} 
we describe the various contributions to the total EP generated during a 
nonequilibrium transformation and the inequalities that these contributions satisfy.
In section \ref{trajectory}, we define the various trajectory entropies which 
upon ensemble averaging give the various contributions to the EP.
We then present the GF formalism used to 
calculate the statistical properties of these trajectory entropies.
In section \ref{FT}, we derive the various FTs and the implied inequalities.
Alternative derivations of FTs in terms of forward-backward 
trajectories are given in appendix \ref{FTtrajectory}.
By considering specific nonequilibrium transformations, 
we recover most of the previously derived FTs.
Finally in section \ref{QD}, we apply our results to the full counting 
statistics of electrons in a driven quantum dot.
Conclusions are drawn in section \ref{conclusions}.

%%%%%%%%%%%%%%%%%%%%%%%%%%%%%%%%%%%%%%%%%%%%%%%%%%%%%%%%%%%%%%%%%%%%%%
\section{The master equation}\label{mastereq}

We consider an externally driven open system exchanging 
particles and energy with multiple reservoirs.
Each state $m$ of the system has a given energy 
$\epsilon_{m}$ and $N_{m}$ particles. 
The total number of states $m$ is finite and equal to $M$.
The probability to find the system in a state $m$ 
at time $t$ is denoted by $p_m(t)$.
The evolution of this probability is described by the ME
\begin{eqnarray}
\dot{p}_m(t) = \sum_{m'} W_{m,m'}(\lambda_t) p_{m'}(t) \;,
\label{aaaaa}
\end{eqnarray}
where the rate matrix satisfies 
\begin{eqnarray}
\sum_{m} W_{m,m'}(\lambda_t)=0 \label{aaaaarate} \;.
\end{eqnarray}
We assume that if a transition from $m$ to $m'$ can occur,
the reversed transition from $m'$ to $m$ can also occur.  
Various parameters, such as the energies $\epsilon_{m}$ of the 
system or the chemical potential $\mu_{\nu}$ and the 
temperature $\beta_{\nu}^{-1}$ of the $\nu$ reservoir can be 
varied in time externally according to a known protocol.
This is described by the dependence of 
the rate matrix on several time-dependent parameters $\lambda_t$.
If the transition rates are kept constant, the system will eventually 
reach the unique steady state solution $p_{m}^{\rm st}(\lambda)$
which satisfies $\dot{p}_{m}^{\rm st}(\lambda)=0$ \cite{note1}.\\

The transition rates will be expressed as sums of contributions 
from different reservoirs $\nu$ 
\begin{eqnarray}
W_{m,m'}(\lambda) = \sum_{\nu} W_{m,m'}^{(\nu)}(\lambda) \;,
\label{aaaab}
\end{eqnarray}
each satisfying 
\begin{eqnarray}
\frac{W_{m',m}^{(\nu)}(\lambda)}{W_{m,m'}^{(\nu)}(\lambda)}
&=& \exp{ \bigg\{ \beta_{\nu}(\lambda) \big[ 
\big(\epsilon_m(\lambda)-\epsilon_{m'}(\lambda)\big) \big. \bigg.} \nonumber\\
&&\hspace{1.5cm}\bigg. \big. -\mu_{\nu}(\lambda) (N_{m}-N_{m'}) \big] \bigg\}  \;.
\label{aaaad}
\end{eqnarray}

If all reservoirs have the same thermodynamic properties (temperature 
$\beta^{-1}$ and chemical potential $\mu$), the steady state distribution  
coincide with the equilibrium distribution $p_{m}^{\rm st}(\lambda)=p_{m}^{eq}(\lambda)$ 
which satisfies the detailed balanced condition (DBC) 
\begin{eqnarray}
W_{m,m'}^{(\nu)}(\lambda) p_{m'}^{\rm eq}(\lambda) 
= W_{m',m}^{(\nu)}(\lambda) p_{m}^{\rm eq}(\lambda) \;.
\label{aaaae}
\end{eqnarray}
As a consequence of (\ref{aaaad}) and (\ref{aaaae}), the equilibrium 
distribution then assumes the grand canonical form
\begin{eqnarray}
p_{m}^{eq}(\lambda) 
= \frac{\exp{[-\beta(\lambda) \big( \epsilon_{m}(\lambda) - 
\mu(\lambda) N_{m} \big)]}}{\Xi(\lambda)} \label{aaaac}\;,
\end{eqnarray}
where $\Xi(\lambda)$ is the grand canonical partition function.
However, in the general case where the reservoirs have different 
$\beta^{(\nu)}$ and $\mu^{(\nu)}$, the DBC does not hold and 
$p_{m}^{\rm st}(\lambda)$ is a NESS.

\iffalse
We introduce the probability flux due to matter or energy 
exchange between the system and the $\nu$ reservoir during 
transitions between state $m$ and $m'$ 
\begin{eqnarray}
J_{m',m}^{(\nu)}(t) \equiv W_{m,m'}^{(\nu)}(\lambda_t) p_{m'}(t) 
- W_{m',m}^{(\nu)}(\lambda_t) p_{m}(t) \;.
\label{aaaaf}
\end{eqnarray}
We can rewrite Eq. (\ref{aaaaa}) as
\begin{eqnarray}
\dot{p}_m(t) 
= \sum_{\nu, m'} J_{m',m}^{(\nu)}(t) \;.\label{aaaag}
\end{eqnarray}
At steady state as well as at equilibrium, we have 
that $\sum_{\nu, m'} J_{m',m}^{(\nu)}(t)=0$.
However, for a NESS state
$J_{m',m}^{(\nu)}(t) \neq 0$, whereas at equilibrium 
$J_{m',m}^{(\nu)}(t) = 0$.
\fi

%%%%%%%%%%%%%%%%%%%%%%%%%%%%%%%%%%%%%%%%%%%%%%%%%%%%%%%%%%%%%%%%%%%%%%
\section{The entropies}\label{entropy}

The {\it Gibbs entropy} of the system is a state function defined as
\begin{eqnarray}
S(t) \equiv - \sum_{m} p_{m}(t) \ln p_{m}(t) \;.
\label{baaaa}
\end{eqnarray}
Using (\ref{aaaaa}) and (\ref{aaaaarate}), the {\it system EP} reads
\begin{eqnarray}
\dot{S}(t) &=& - \sum_{m} \dot{p}_{m}(t) \ln p_{m}(t) \nonumber\\
&=& - \sum_{m,m'} W_{m,m'}(\lambda_t) p_{m'}(t) 
\ln \frac{p_{m}(t)}{p_{m'}(t)} \;.
\label{baaaab}
\end{eqnarray}
This can be partitioned as 
\cite{Prigogine,GrootMazur,Schnakenberg,Gaspard1,Seifert1}
\begin{eqnarray}
\dot{S}(t)=\dot{S}_{tot}(t)-\dot{S}_r(t)  \;,
\label{baaab}
\end{eqnarray}
with the {\it total EP}
\begin{eqnarray}
\dot{S}_{tot}(t) &\equiv& - \sum_{m,m',\nu} 
W_{m,m'}^{(\nu)}(\lambda_t) p_{m'}(t)\label{baaac} \\
&& \hspace{3cm}
\ln \frac{W_{m',m}^{(\nu)}(\lambda_t) p_{m}(t)} 
{W_{m,m'}^{(\nu)}(\lambda_t) p_{m'}(t)} \geq 0 \nonumber
\end{eqnarray}
and the {\it reservoir EP} (also called medium entropy or entropy flow)
\begin{eqnarray}
\dot{S}_r(t) \equiv - \sum_{m,m',\nu} 
W_{m,m'}^{(\nu)}(\lambda_t) p_{m'}(t) \ln 
\frac{W_{m',m}^{(\nu)}(\lambda_t)}{W_{m,m'}^{(\nu)}(\lambda_t)} \;.
\label{baaad}
\end{eqnarray}
We note that $\dot{S}_{tot}(t) \geq 0$
follows from $W_{m,m'}^{(\nu)}(\lambda_t) p_{m'}(t)>0$ if $m'\neq m$ 
and $\ln x \leq x-1$ for $x > 0$ (if $m'=m$ the log in zero), 
by using the fact that 
$\sum_{m,\nu} W_{m',m}^{(\nu)}(\lambda_t) p_{m}^{st}(\lambda_t)=0$
and $\sum_{m,\nu} W_{m,m'}^{(\nu)}(\lambda_t)=0$.
$\dot{S}(t)$ is the contribution to $\dot{S}_{tot}(t)$ coming from the  
changes in the system probability distribution and $\dot{S}_r(t)$ is 
the contribution coming from matter and energy exchange processes 
between the system and its reservoirs.\\

We further separate the reservoir EP into two components \cite{Hatano01,Seifert2}
\begin{eqnarray}
\dot{S}_r(t) \equiv \dot{S}_{ex}(t) + \dot{S}_{a}(t)\;,
\label{baaae}
\end{eqnarray}
with the {\it excess EP}
\begin{eqnarray}
\dot{S}_{ex}(t) &\equiv& - \sum_{m,m'} 
W_{m,m'}(\lambda_t) p_{m'}(t) \ln 
\frac{p_{m'}^{st}(\lambda_t)}{p_{m}^{st}(\lambda_t)}
\label{baaag} \\
&=&\sum_{m} \dot{p}_{m}(t) \ln p_{m}^{st}(\lambda_t) \nonumber
\end{eqnarray}
and the {\it adiabatic EP} (also called housekeeping entropy \cite{Hatano01,Seifert2})
\begin{eqnarray}
\dot{S}_{a}(t) &\equiv& - \sum_{m,m',\nu} 
W_{m,m'}^{(\nu)}(\lambda_t) p_{m'}(t) \label{baaaf}\\
&& \hspace{2.5cm} \ln 
\frac{W_{m',m}^{(\nu)}(\lambda_t) p_{m}^{st}(\lambda_t)}
{W_{m,m'}^{(\nu)}(\lambda_t) p_{m'}^{st}(\lambda_t)} \geq 0 \;.
\nonumber
\end{eqnarray}
The positivity of (\ref{baaaf}) follows from the same reason as (\ref{baaac}).
If a transformation is done very slowly, the system remains at all 
times in the steady state distribution $p_{m}(t)=p_{m}^{st}(\lambda_t)$. 
Such a transformation is called adiabatic.
We then have $\dot{S}_{tot}(t)=\dot{S}_{a}(t)$ and $\dot{S}_{ex}(t)=-\dot{S}(t)$.  
%The excess EP $\dot{S}_{ex}(t)$ represents the contribution to the reservoir EP 
%due to nonadiabatic effects ($\dot{S}_{ex}(t)=0$ during an adiabatic transformation).
We also notice that $\dot{S}_{a}(t)=0$ when the DBC is satisfied.\\

We next define the state function quantity
\begin{eqnarray}
S_{b}(t) \equiv  - \sum_{m} p_{m}(t) 
\ln \frac{p_{m}(t)}{p_{m}^{st}(\lambda_t)} \label{baaaha}
\end{eqnarray}
which is obviously zero when the system is at steady state.
When considering a transformation between steady states, 
$\Delta S_{b}(T,0) = \int_{0}^{T} d\tau \dot{S}_{b}(\tau)=S_{b}(T)-S_{b}(0)=0$.
We call $\dot{S}_{b}(t)$ the {\it boundary EP} (the terminology 
will be explain shortly) and separate it in two parts 
\begin{eqnarray}
\dot{S}_{b}(t) = \dot{S}_{na}(t) -  \dot{S}_d(t)   \label{baaahb} \;,
\end{eqnarray}
where the {\it nonadiabatic EP} is 
\begin{eqnarray}
\dot{S}_{na}(t) 
&\equiv& - \sum_{m} \dot{p}_{m}(t) \ln 
\frac{p_{m}(t)}{p_{m}^{st}(\lambda_t)} \label{baaah} \\
&&\hspace{-0.5cm}= 
-\sum_{m,m'} W_{m,m'}(\lambda_t) p_{m'}(t) \ln 
\frac{p_{m}(t) p_{m'}^{st}(\lambda_t)}
{p_{m}^{st}(\lambda_t) p_{m'}(t)} \geq 0 \;, \nonumber
\end{eqnarray}
and the {\it driving EP} is
\begin{eqnarray}
\dot{S}_d(t) \equiv \sum_{m} p_{m}(t) \dot{\phi}_m(\lambda_t) \;,
\label{baaajb}
\end{eqnarray}
with
\begin{eqnarray}
\phi_m(\lambda_t) \equiv - \ln p_{m}^{\rm st}(\lambda_t) \;.
\label{caaanb}
\end{eqnarray}
The positivity of (\ref{baaah}) is again shown in the same way as for (\ref{baaac}) and (\ref{baaaf}).
If no external driving acts on the system, $\lambda$ is time 
independent and from (\ref{baaajb}), $\dot{S}_d(t)=0$.
For an adiabatic transformations, since $p_{m}(t)=p_{m}^{st}(\lambda_t)$, from (\ref{baaaha}) 
and (\ref{baaah}), we see that $\dot{S}_{na}(t)=0$ as well as $\dot{S}_{b}(t)=0$. 
From (\ref{baaahb}), this also means that $\dot{S}_d(t)=0$.
Therefore, $\dot{S}_d(t) \neq 0$ only for nonadiabatic driving.
Using (\ref{baaah}) with (\ref{baaaab}) and (\ref{baaag}), we find 
\begin{eqnarray}
\dot{S}_{na}(t) = \dot{S}_{ex}(t) + \dot{S}(t)
= \dot{S}_{tot}(t) - \dot{S}_{a}(t) \label{baaakk}\;.
\end{eqnarray}
It is clear from the last equality why we call $\dot{S}_{na}(t)$ the nonadiabatic EP.
The inequality $\dot{S}_{ex}(t) \geq - \dot{S}(t)$ which follows from the 
first line is a generalization of the "second law of steady state 
thermodynamics" \cite{Oono,HTasaki,Hatano01} derived for 
transitions between steady states.

We next summarize our results 
\begin{eqnarray}
\dot{S}_{tot}(t) &=& \dot{S}_{na}(t) + \dot{S}_{a}(t) \geq 0 \label{baaai}\\
\dot{S}_{na}(t) &=& \dot{S}_{d}(t) + \dot{S}_{b}(t) \geq 0 \label{baaak}\\
\dot{S}_{a}(t) &\geq& 0 \label{baaaj} \;.
\end{eqnarray}
The total EP is always positive and can be separated into 
two positive contributions, adiabatic (which are nonzero 
only when the DBC is violated) and nonadiabatic effects. 
The latter can be due to a nonadiabatic external driving acting on the system 
or to the fact that one considers transformation during which the system is 
initially or finally not in a steady state.
We therefore have a minimum EP principle stating that the total EP 
for arbitrary nonequilibrium transformations takes its 
minimal value if the transformation is done adiabatically (very slowly).
The equality sign in (\ref{baaai}) is satisfied for adiabatic 
transformations which occur at equilibrium.
The equality sign in (\ref{baaak}) holds for adiabatic transformations.
The equality sign in (\ref{baaaj}) only occurs when the DBC is satisfied.

%%%%%%%%%%%%%%%%%%%%%%%%%%%%%%%%%%%%%%%%%%%%%%%%%%%%%%%%%%%%%%%%%%%%%%
\section{Trajectory entropies}
\label{trajectory}

The evolution described by the ME can be represented by 
an ensemble of stochastic trajectories involving instantaneous jumps 
between states. This will allow us to define fluctuating (trajectory) entropies.\\

\subsection{Definitions}

\begin{figure}[h]
\centering
\rotatebox{0}{\scalebox{0.5}{\includegraphics{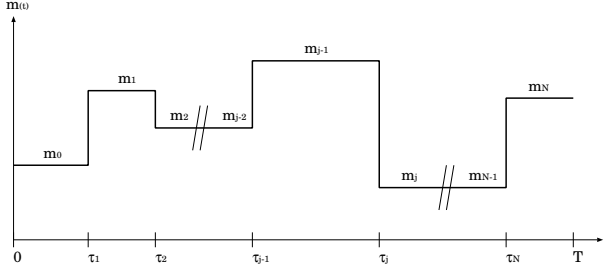}}}
\caption{Representation of a trajectory $m_{(\tau)}$.}
\label{figTraj}
\end{figure}

We denote a trajectory taken by the system between $t=0$ and $t=T$ by
$m_{(\tau)}=\{0-m_0 \stackrel{\tau_1}{\rightarrow} m_1 \stackrel{\tau_2}{\rightarrow} 
\cdots m_{j-1} \stackrel{\tau_j}{\rightarrow} m_{j} \stackrel{\tau_{j+1}}{\rightarrow} 
\cdots m_{N-1} \stackrel{\tau_N}{\rightarrow} m_{N}-T\}$.
At $t=0$ the system is in $m_0$, and stays there until it jumps at $\tau_1$ to $m_1$, 
{\it etc.}, jumps at $\tau_N$ from $m_{N-1}$ to $m_N$ and stays in $m_N$ until $t=T$ 
[see Fig. \ref{figTraj}]. $N$ is the total number of jumps during this trajectory.\\
%%%%%%%%%%%%%%%%%%%%%%%%%
%%%%%%%%%%%%%%%%%%%%%%%%%
We next introduce various type of 'trajectory entropy production' (TEP). 
We will see at the end of this section that
when ensemble averaged, these correspond to the various 
EP defined in section \ref{entropy}.\\
The {\it trajectory Gibbs entropy} is defined as 
\begin{eqnarray}
s[m_{(\tau)},t] \equiv - \ln p_{m_{(\tau)}}(t) \;, \label{trajentr_aaaaa}
\end{eqnarray}
where $p_{m_{(\tau)}}(t)$ represents the value of $p_{m}(t)$ 
along the trajectory $m_{(\tau)}$. 
The {\it system TEP} is given by
\begin{eqnarray}
\dot{s}[m_{(\tau)},t] = 
-\frac{\dot{p}_{m}(t)}{p_{m}(t)} \bigg\vert_{m_{(\tau)}} 
+ \sum_{j=1}^{N} \delta(t-\tau_j) \ln 
\frac{p_{m_{j-1}}(\tau_{j})}{p_{m_{j}}(\tau_{j})} \;.
\label{trajentr_aaaab}
\end{eqnarray}
The first term represents the smooth changes of $s[m_{(\tau)},t]$ along the horizontal 
segments of the trajectory on Fig. \ref{figTraj} during which the system is in a well 
defined state. These changes are only due to the time dependence of the probability 
to be on a given state. 
The notation $\vert_{m_{(\tau)}}$ means that the $m$ in the expression 
changes depending on which horizontal segment along the trajectory one considers.
The second term represents the discrete changes of $s[m_{(\tau)},t]$ 
along the vertical segments of the trajectory.
These changes are singular and only due to the change in the system state.\\
Separating the trajectory system TEP similarly as the system 
EP in section \ref{entropy}, we get
\begin{eqnarray}
\dot{s}_{tot}[m_{(\tau)},t] =\dot{s}[m_{(\tau)},t] + \dot{s}_r[m_{(\tau)},t] 
\label{trajentr_aaaac}\;,
\end{eqnarray}
where the {\it total TEP} is
\begin{eqnarray}
\dot{s}_{tot}[m_{(\tau)},t] &\equiv&
-\frac{\dot{p}_{m}(t)}{p_{m}(t)} \bigg\vert_{m_{(\tau)}} 
\label{trajentr_aaaaf}\\
&&\hspace{-0.5cm}+\sum_{j=1}^{N} \delta(t-\tau_j) \ln 
\frac{W_{m_{j},m_{j-1}}^{(\nu_j)}(\lambda_{\tau_j}) p_{m_{j-1}}(\tau_{j})}
{W_{m_{j-1},m_{j}}^{(\nu_j)}(\lambda_{\tau_j}) p_{m_{j}}(\tau_{j})} \nonumber\;,
\end{eqnarray}
and the {\it reservoir TEP}
\begin{eqnarray}
\dot{s}_r[m_{(\tau)},t] \equiv
\sum_{j=1}^{N} \delta(t-\tau_j) \ln 
\frac{W_{m_{j},m_{j-1}}^{(\nu_j)}(\lambda_{\tau_j})}
{W_{m_{j-1},m_{j}}^{(\nu_j)}(\lambda_{\tau_j})} \;.
\label{trajentr_aaaag}
\end{eqnarray}

We further separate the reservoir TEP into
\begin{eqnarray}
\dot{s}_{r}[m_{(\tau)},t] =
\dot{s}_{a}[m_{(\tau)},t] + \dot{s}_{ex}[m_{(\tau)},t] 
\label{trajentr_aaaad}\;,
\end{eqnarray}
with the {\it adiabatic TEP} 
\begin{eqnarray}
\dot{s}_{a}[m_{(\tau)},t] \equiv 
\sum_{j=1}^{N} \delta(t-\tau_j) \ln 
\frac{W_{m_{j},m_{j-1}}^{(\nu_j)}(\lambda_{\tau_j}) 
p_{m_{j-1}}^{\rm st}(\lambda_{\tau_{j}})}
{W_{m_{j-1},m_{j}}^{(\nu_j)}(\lambda_{\tau_j}) 
p_{m_{j}}^{\rm st}(\lambda_{\tau_{j}})} \;,
\label{trajentr_aaaah}
\end{eqnarray}
and the {\it excess TEP}
\begin{eqnarray}
\dot{s}_{ex}[m_{(\tau)},t] \equiv 
\sum_{j=1}^{N} \delta(t-\tau_j) \ln 
\frac{p_{m_{j}}^{\rm st}(\lambda_{\tau_{j}})}
{p_{m_{j-1}}^{\rm st}(\lambda_{\tau_{j}})} \;.
\label{trajentr_aaaai}
\end{eqnarray}
The {\it nonadiabatic TEP} 
\begin{eqnarray}
\dot{s}_{na}[m_{(\tau)},t] &\equiv& 
-\frac{\dot{p}_{m}(t)}{p_{m}(t)} \bigg\vert_{m_{(\tau)}} 
\label{trajentr_aaaaj}\\
&&+ \sum_{j=1}^{N} \delta(t-\tau_j) \ln 
\frac{p_{m_{j-1}}(\tau_{j}) p_{m_{j}}^{\rm st}(\lambda_{\tau_{j}})}
{p_{m_{j-1}}^{\rm st}(\lambda_{\tau_{j}}) p_{m_{j}}(\tau_{j})} 
\nonumber
\end{eqnarray} 
is made of the sum of two terms
\begin{eqnarray}
\dot{s}_{na}[m_{(\tau)},t]
\equiv \dot{s}_d [m_{(\tau)},t] + \dot{s}_{b}[m_{(\tau)},t] 
\label{trajentr_aaaaja} 
\end{eqnarray}
the {\it boundary TEP}
\begin{eqnarray}
\dot{s}_{b}[m_{(\tau)},t] &\equiv& 
-\frac{\dot{p}_{m}(t)}{p_{m}(t)} \bigg\vert_{m_{(\tau)}} 
- \dot{\lambda}_{t} \frac{\partial \phi_{m_{(\tau)}}(\lambda_{t})}{\partial \lambda_{t}} 
\label{trajentr_aaaajb}\\
&&+ \sum_{j=1}^{N} \delta(t-\tau_j) \ln 
\frac{p_{m_{j-1}}(\tau_{j}) p_{m_{j}}^{\rm st}(\lambda_{\tau_{j}})}
{p_{m_{j-1}}^{\rm st}(\lambda_{\tau_{j}}) p_{m_{j}}(\tau_{j})} \;,
\nonumber 
\end{eqnarray} 
and the {\it driving TEP}
\begin{eqnarray}
\dot{s}_d[m_{(\tau)},t] &\equiv& 
\dot{\lambda}_{t} \frac{\partial \phi_{m_{(\tau)}}(\lambda_{t})}{\partial \lambda_{t}} 
\nonumber \;.
\end{eqnarray} 
As in section \ref{entropy}, since
\begin{eqnarray}
\dot{s}_{na}[m_{(\tau)},t] = \dot{s} [m_{(\tau)},t] + \dot{s}_{ex}[m_{(\tau)},t] 
\label{trajentr_aaaae}\;,
\end{eqnarray}
we get
\begin{eqnarray}
\dot{s}_{tot}[m_{(\tau)},t] = \dot{s}_{na}[m_{(\tau)},t] + \dot{s}_{a}[m_{(\tau)},t] 
\label{trajentr_aaaacb} \;.
\end{eqnarray}

We generically denote these TEP by $a[m_{(\tau)},t]$.
The change of $a[m_{(\tau)},t]$ along a 
trajectory $m_{(\tau)}$ of length $T$ is given by  
\begin{eqnarray}
\Delta a[m_{(\tau)},T] &=& \int_{0}^{T} dt \dot{a}[m_{(\tau)},t]
\label{TrajaveAaaaag} \;.
\end{eqnarray}
Notice that $\Delta s[m_{(\tau)},T]$ and $\Delta s_{b}[m_{(\tau)},T]$ 
are state function TEP
\begin{eqnarray}
\Delta s [m_{(\tau)},T] = \ln \frac{p_{m_0}(0)}{p_{m_N}(T)} 
= s_{m_N}(T) - s_{m_0}(0) \;, \label{caaaj}
\end{eqnarray}
where $s_{m}(t) \equiv - \ln p_{m}(t)$, and 
\begin{eqnarray}
\Delta s_{b} [m_{(\tau)},T]
&=& \big(s_{m_N}(T)-\phi_{m_N}(\lambda_{T})\big) \label{caaajb} \\
&&\hspace{1.5cm}- \big(s_{m_0}(0)-\phi_{m_0}(\lambda_{0})\big)  \;.\nonumber
\end{eqnarray}
The other TEP are path functions.\\ 

\subsection{Deriving trajectory entropies from measured currents}\label{measure}

Using (\ref{aaaad}), the reservoir TEP can be expressed as
\begin{eqnarray}
&&\dot{s}_r[m_{(\tau)},t]= \label{trajentr_aaaagb}\\
&&\hspace{0.5cm}-\sum_{\nu} \beta_{\nu}(\lambda_{t}) 
\bigg( I_{heat}^{(\nu)}[m_{(\tau)},t] - \mu_{\nu}(\lambda_{t}) 
I_{mat}^{(\nu)}[m_{(\tau)},t] \bigg)\;,\nonumber
\end{eqnarray}
where the heat current between the $\nu$ reservoir and the system is 
\begin{eqnarray}
I_{heat}^{(\nu)}[m_{(\tau)},t] = \sum_{j=1}^{N} \delta_{\nu,\nu_j}(t-\tau_j) 
\big( \epsilon_{m_{j}}(\lambda_{\tau_j})-\epsilon_{m_{j-1}}(\lambda_{\tau_j}) \big) 
\label{heatcurr}
\end{eqnarray}
and the matter current between the $\nu$ reservoir and the system 
\begin{eqnarray}
I_{mat}^{(\nu)}[m_{(\tau)},t] = \sum_{j=1}^{N} \delta_{\nu,\nu_j}(t-\tau_j) 
\big(N_{m_{j}}-N_{m_{j-1}}\big) \;.
\label{mattercurr}
\end{eqnarray}
The currents are positive if the system energy (matter) increases.
$\delta_{\nu,\nu_j}(t-\tau_j)$ is a Dirac distribution centered 
at time $\tau_j$ only if the transition is due to the reservoir $\nu=\nu_i$. 
Otherwise it is zero. 
We thus confirm that the reservoir EP is the entropy associated to 
system-reservoir exchange processes.\\

We assume that the parametric time dependence of the 
energies, temperatures and chemical potentials is known.
Except in degenerate cases for which two different transitions between states 
have the same energy difference and number of particle difference, 
the trajectory of the system can be uniquely determined by measuring the 
heat and matter currents between the system and the reservoirs.
The system steady state probability distribution can be calculated by 
recording the steady state currents for sufficiently long times for 
different values of the energies, temperatures or chemical potentials. 
The driven system probability distribution can in principle be calculated 
by reproducing the measurement of the currents multiple times. 
All trajectory entropies containing the logarithm of the transition rates can 
be expressed in term of a combination of the reservoir EP (directly measurable 
via current) and other trajectory entropies which can be expressed in term of 
the system probability distribution (actual or steady state). 
Therefore, provided the current measurements can be repeated often enough to 
get a good statistics, all the trajectory entropies are in principle measurable. 

\subsection{Statistical properties using generating functions}\label{statprop}

The GF formalism allows to compute the probability distributions and all statistical 
properties of the TEP without having to generate the trajectories themselves.  
It further provides a direct means for proving the FTs.\\   
The GF associated with the changes of $a[m_{(\tau)},t]$ along a trajectory is given by
\begin{eqnarray}
G(\gamma,t) \equiv \mean{\exp{\{\gamma \Delta a[m_{(\tau)},t]\}}} \label{TrajaveAaaaah} \;,
\end{eqnarray}
where $\mean{\cdot}$ denotes an average over all possible trajectories.
The probability that the system follows a trajectory with the constraint 
$\Delta A=\Delta a[m_{(\tau)},t]$ at time $t$, can be obtained from the GF using
\begin{eqnarray}
P(\Delta A,t) &\equiv& \mean{ \delta(\Delta A-\Delta a[m_{(\tau)},t]} \label{TrajaveAaaaai}\\ 
&=& \frac{1}{2 \pi} \int_{-\infty}^{\infty} 
d\gamma \e^{-\i \gamma \Delta A} G(\i \gamma,t) \nonumber \;.
\end{eqnarray}
By inverting (\ref{TrajaveAaaaai}), we get
\begin{eqnarray}
G(\i \gamma,t) = \int_{-\infty}^{\infty} d\Delta A 
\e^{\i \gamma \Delta A} P(\Delta A,t) \;. \label{TrajaveAaaaaj} 
\end{eqnarray}
The moments of the distribution are given by derivatives of the GF 
\begin{eqnarray}
\mean{\Delta a^k[m_{(\tau)},t]} = \left. \frac{\partial^k G(\gamma,t)}
{\partial \gamma^k} \right|_{\gamma=0} \label{TrajaveAaaaak}  \;, \; \ \ k=1,2,\cdots \;.
\end{eqnarray}

In order to compute the GF, we recast it in the form
\begin{eqnarray}
G(\gamma,t) = \sum_{m} g_{m}(\gamma,t) 
\label{TrajaveAaaaal} \;,
\end{eqnarray}
where 
\begin{eqnarray}
g_{m}(\gamma,t) = p_{m}(t) \mean{\exp{\{\gamma \Delta a[m_{(\tau)},t]\}}}_m 
\label{TrajaveAaaaalb} 
\end{eqnarray}
is the product of the probability to find the system in state $m$ at time $t$
multiplied by the expectation value of $\exp{\{\gamma \Delta a[m_{(\tau)},t]\}}$
conditional on the system being in state $m$ at time $t$.
Since $\Delta a[m_{(\tau)},t]=0$ for a trajectory of length $t=0$, we have 
$g_{m}(\gamma,0)=p_{m}(0)$. We also have $G(0,t)=1$ and $g_{m}(0,t)=p_{m}(t)$.\\
The time derivative of (\ref{TrajaveAaaaal}) gives
\begin{eqnarray}
\dot{G}(\gamma,t)= \sum_{m} \dot{g}_{m}(\gamma,t) \;,
\label{TrajaveAaaaao}
\end{eqnarray}
where $\dot{g}_{m}(\gamma,t)$ depends on the TEP of interest. 
Below, we will derive equations of motion for the $g_{m}(\gamma,t)$'s 
associated to the various TEP.

\subsubsection{State function trajectory entropy production}

The generating function associated with a state function TEP
$\Delta a[m_{(\tau)},t]=a_{m}(t)-a_{n}(0)$ like $\Delta s[m_{(\tau)},t]$ or 
$\Delta s_{b}[m_{(\tau)},t]$ may be straightforwardly obtained using 
(\ref{TrajaveAaaaal}) with (\ref{TrajaveAaaaalb}). We get
\begin{eqnarray}
G(\gamma,t) = \sum_{m,n} \exp{\{\gamma \big(a_m(t)-a_n(0)\big) \}} p_{m}(t) p_{n}(0)
\label{TrajaveAaaaalc} \;.
\end{eqnarray}

\subsubsection{Excess trajectory entropy production}

$s_{ex}[m_{(\tau)},t]$ acquires an amount 
$s_{ex}(m,m')=\ln \big(p_{m}^{\rm st}(\lambda_t) / 
p_{m'}^{\rm st}(\lambda_t) \big)$ each time a transition 
from a state $m'$ to $m$ occurs and it remains constant 
along a given state $m$ of the system. This means that 
\begin{eqnarray}
\dot{g}_m^{(ex)}(\gamma,t) &=& \sum_{m'} 
W_{m,m'}(\lambda_t) p_{m'}(t)  \label{Genaaaae}\\
&&\hspace{-0.5cm}+ \exp{\{\gamma s_{ex}(m,m') \}} 
\mean{\exp{\{\gamma \Delta s_{ex}[m_{(\tau)},t]\}}}_{m'} 
\nonumber \;,
\end{eqnarray} 
which using (\ref{TrajaveAaaaalb}) can be rewritten 
\begin{eqnarray}
\dot{g}_m^{(ex)}(\gamma,t) = \sum_{m'} 
\bigg( \frac{p_{m}^{\rm st}(\lambda_t)}{p_{m'}^{\rm st}(\lambda_t)} 
\bigg)^{\gamma} W_{m,m'}(\lambda_t) g_{m'}^{(ex)}(\gamma,t) \label{Genaaaaeb}\;.
\end{eqnarray} 
%Notice that using the change of variable 
%$\tilde{g}_{m}(\gamma,t) \equiv 
%g_{m}(\gamma,t)/(p_{m}^{\rm st}(t))^{\gamma}$, 
%we can rewrite (\ref{Genaaaae}) as
%\begin{eqnarray}
%\dot{\tilde{g}}_m(\gamma,t) &=& 
%\gamma \dot{\phi}_{m}(\lambda_t) \tilde{g}_{m}(\gamma,t) 
%\label{Genaaaad}\\ &&\hspace{1cm}
%+\sum_{m'} W_{m,m'}(\lambda_t) \tilde{g}_{m'}(\gamma,t) \;.
%\nonumber
%\end{eqnarray}
%cccccccccccccccccccccccccccccccccccccccccccccccccccccccccccccc
%Notice also that if $\gamma=-1$, (\ref{Genaaaae}) is
%\begin{eqnarray}
%\dot{g}_m(-1,t) = \sum_{m'} 
%W_{m,m'}(\lambda_t) \frac{p_{m'}^{\rm st}(\lambda_t)}
%{p_{m}^{\rm st}(\lambda_t)} g_{m'}(-1,t) \label{Genaaaaf}
%\end{eqnarray}

\subsubsection{Reservoir trajectory entropy production and currents}

Each time a transition along the system trajectory occurs, $s_{r}[m_{(\tau)},t]$ 
acquires an amount
$s_{r}(m,m')=\ln \big(W_{m,m'}^{(\nu)}(\lambda_t) / 
W_{m',m}^{(\nu)}(\lambda_t) \big)$.
Similarly to the excess entropy, we get
\begin{eqnarray}
\dot{g}_m^{(r)}(\gamma,t) = \sum_{\nu,m'} 
\bigg( \frac{W_{m,m'}^{(\nu)}(\lambda_t)}{W_{m',m}^{(\nu)}(\lambda_t)} 
\bigg)^{\gamma} W_{m,m'}^{(\nu)}(\lambda_t) g_{m'}^{(r)}(\gamma,t) \;.
\label{Genaaaag}
\end{eqnarray}
This equation has been used in the study of steady 
state FTs \cite{Lebowitz,Gaspard1,GaspardAndrieux1} .\\

In (\ref{trajentr_aaaagb}), we expressed the reservoir TEP in terms of currents. 
The time integrated individual currents give the heat and matter transfer 
between the $\nu$ reservoir and the system
$q_{heat}^{(\nu)}[m_{(\tau)},t]=\int_{0}^{t}dt' I_{heat}^{(\nu)}[m_{(\tau)},t']$
and
$q_{mat}^{(\nu)}[m_{(\tau)},t]=\int_{0}^{t}dt' I_{mat}^{(\nu)}[m_{(\tau)},t']$.
Their statistics can be calculated using 
\begin{eqnarray}
\dot{g}_m(\vec{\gamma},t) &=& \sum_{\nu,m'} 
\exp{\{ \gamma^{(\nu)}_{heat} 
\big( \epsilon_{m}(\lambda_{t})-\epsilon_{m'}(\lambda_{t}) \big) \}} \nonumber\\
&&\hspace{-1.2cm}\exp{\{ \gamma^{(\nu)}_{mat} \big( N_{m}-N_{m'} \big) \}} 
W_{m,m'}^{(\nu)}(\lambda_t) g_{m'}(\vec{\gamma},t) \;,
\label{Genaaaagcurr}
\end{eqnarray}
where $\vec{\gamma}$ is a vector who's elements are the different 
$\gamma^{(\nu)}_{heat}$'s and $\gamma^{(\nu)}_{mat}$'s.
The GF calculated from (\ref{Genaaaagcurr}) is therefore associated with the join 
probability distribution for having a certain heat and matter transfer with each reservoir.
%Lets denote this generating function $G_{cur}(\vec{\gamma},t)$.
%By defining $\vec{\tilde{\gamma}}$ where 
%$\tilde{\gamma}^{(\nu)}_{heat}=-\gamma^{(\nu)}_{heat}/\beta^{(\nu)}$
%and $\tilde{\gamma}^{(\nu)}_{mat}=\gamma^{(\nu)}_{heat}/(\beta^{(\nu)} \mu^{(\nu)})$
%we get that $G_{cur}(\vec{\tilde{\gamma}},t)=G_{r}(\vec{\gamma},t)$.

\subsubsection{Adiabatic trajectory entropy production}

Each time a transition along the system trajectory occurs, $s_{a}[m_{(\tau)},t]$
acquires an amount
$s_{a}(m,m')=\ln \big(W_{m,m'}^{(\nu)}(\lambda_t) p_{m'}^{\rm st}(\lambda_t)/ 
W_{m',m}^{(\nu)}(\lambda_t) p_m^{\rm st}(\lambda_t) \big)$.
We therefore get
\begin{eqnarray}
\dot{g}_m^{(a)}(\gamma,t) &=&  \label{Genaaaai}\\
&&\hspace{-1cm}\sum_{\nu,m'} 
\bigg( \frac{W_{m,m'}^{(\nu)}(\lambda_t) p_{m'}^{\rm st}(\lambda_t)}
{W_{m',m}^{(\nu)}(\lambda_t) p_m^{\rm st}(\lambda_t)} \bigg)^{\gamma} 
W_{m,m'}^{(\nu)}(\lambda_t) g_{m'}^{(a)}(\gamma,t)\nonumber \;.
\end{eqnarray}
%Using the change of variable $\tilde{g}_{m}(\gamma,t) \equiv g_{m}(\gamma,t) 
%(p_{m}^{\rm st}(\gamma,t))^{\gamma}$, (\ref{Genaaaai}) can be rewritten as 
%\begin{eqnarray}
%\dot{\tilde{g}}_m(\gamma,t) &=& - \gamma \dot{\phi}_{m}(\lambda_t) 
%\tilde{g}_{m}(\gamma,t) \label{Genaaaaib}\\ &&\hspace{0cm}+\sum_{\nu,m'} 
%\bigg( \frac{W_{m,m'}^{(\nu)}(\lambda_t)}{W_{m',m}^{(\nu)}(\lambda_t)} 
%\bigg)^{\gamma} W_{m,m'}^{(\nu)}(\lambda_t) \tilde{g}_{m'}(\gamma,t) 
%\nonumber
%\end{eqnarray}
%ccccccccccccccccccccccccccccccccccccccccccccccccccccccccccccccccc
%Notice also that if $\gamma=-1$ the (\ref{Genaaaai}) becomes
%\begin{eqnarray}
%\dot{g}_m(-1,t) = \sum_{m'} 
%W_{m',m}(\lambda_t) \frac{p_{m}^{\rm st}(\lambda_t)}
%{p_{m'}^{\rm st}(\lambda_t)} g_{m'}(-1,t) \label{Genaaaaj}
%\end{eqnarray}

\subsubsection{Total trajectory entropy production}

Each time a transition along the system trajectory occurs, 
$s_{tot}[m_{(\tau)},t]$ acquires an amount
$\ln \big(W_{m,m'}^{(\nu)}(\lambda_t) p_{m'}(t)/ 
W_{m',m}^{(\nu)}(\lambda_t) p_m(t) \big)$. 
In addition it also changes by an amount $-d \big(\ln p_{m}(t)\big) / dt$ 
during an infinitesimally small time on a given state $m$ of the system.
Combining the two, we have
\begin{eqnarray}
\dot{g}_m^{(tot)}(\gamma,t) &=& -
\gamma \bigg(\frac{\dot{p}_{m}(t)}{p_{m}(t)} \bigg) 
g_{m}^{(tot)}(\gamma,t) \label{Genaaaak}\\
&&\hspace{-1.3cm} +\sum_{\nu,m'} 
\bigg( \frac{W_{m,m'}^{(\nu)}(\lambda_t) p_{m'}(t)}
{W_{m',m}^{(\nu)}(\lambda_t) p_m(t)} \bigg)^{\gamma} 
W_{m,m'}^{(\nu)}(\lambda_t) g_{m'}^{(tot)}(\gamma,t)\nonumber\;.
\end{eqnarray}
%Using (\ref{Genaaaak}), we notice that 
%$\tilde{g}_{m}(\gamma,t) \equiv g_{m}(\gamma,t) p_{m}^{\gamma}(t)$
%obeys the same evolution equation as (\ref{Genaaaag}) but with 
%an initial condition $\tilde{g}_m(\gamma,0)=p_{m}^{\gamma+1}(0)$.
%cccccccccccccccccccccccccccccccccccccccccccccccccccccccccccccccccc
%\begin{eqnarray}
%\dot{\tilde{g}}_m(\gamma,t) = \sum_{m'} 
%\bigg( \frac{W_{m,m'}^{(\nu)}(\lambda_t)}{W_{m',m}^{(\nu)}(\lambda_t)} 
%\bigg)^{\gamma} W_{m,m'}(\lambda_t) \tilde{g}_{m'}(\gamma,t) 
%\label{Genaaaakb}
%\end{eqnarray} 

\subsubsection{Non-adiabatic trajectory entropy production}

Like the total TEP, $s_{na}[m_{(\tau)},t]$ acquires an amount
$\ln \big(p_{m}^{\rm st}(\lambda_t) p_{m'}(t)/ 
p_{m'}^{\rm st}(\lambda_t) p_{m}(t) \big)$ each time a transition 
from a states $m'$ to $m$ occurs, and also changes by an 
amount $-d \big(\ln p_{m}(t)\big) / dt$ during an 
infinitesimally small time on a given state $m$.
This gives
\begin{eqnarray}
\dot{g}_m^{(na)}(\gamma,t) &=& 
-\gamma \bigg(\frac{\dot{p}_{m}(t)}{p_{m}(t)} \bigg) 
g_{m}^{(na)}(\gamma,t)  \label{Genaaaan}\\ &&\hspace{-0.9cm}+ \sum_{m'} 
\bigg( \frac{p_{m}^{\rm st}(\lambda_t) p_{m'}(t)}
{p_{m'}^{\rm st}(\lambda_t) p_{m}(t)} \bigg)^{\gamma} 
W_{m,m'}(\lambda_t) g_{m'}^{(na)}(\gamma,t)\nonumber\;.
\end{eqnarray}
%Using (\ref{Genaaaan}), we notice that 
%$\tilde{g}_{m}(\gamma,t) \equiv g_{m}(\gamma,t) p_{m}^{\gamma}(t)$
%obeys the same evolution equation as (\ref{Genaaaae}) but with 
%the initial condition $\tilde{g}_m(\gamma,0)=p_{m}^{\gamma+1}(0)$.\\
%cccccccccccccccccccccccccccccccccccccccccccccccccccccccccccccccccccccc
%\begin{eqnarray}
%\dot{\tilde{g}}_m(\gamma,t) = \sum_{m'} 
%\bigg( \frac{p_{m}^{\rm st}(\lambda_t)}{p_{m'}^{\rm st}(\lambda_t)} 
%\bigg)^{\gamma} W_{m,m'}(\lambda_t) \tilde{g}_{m'}(\gamma,t) 
%\label{Genaaaao}
%\end{eqnarray} 

\subsubsection{Driving trajectory entropy production}

Since $\Delta s_{d}[m_{(\tau)},t]$ exclusively accumulates along the 
segments of the system trajectory, we get
\begin{eqnarray}
\dot{g}_m^{(d)}(\gamma,t) &=& 
\gamma \dot{\phi}_{m}(\lambda_t) g_{m}^{(d)}(\gamma,t) 
\label{Genaaaad}\\ &&\hspace{1cm}
+\sum_{m'} W_{m,m'}(\lambda_t) g_{m'}^{(d)}(\gamma,t) \;.
\nonumber
\end{eqnarray}

It follows from (\ref{TrajaveAaaaak}) that the average change of a TEP 
is obtained from its GF by differentiation with respect to $\gamma$ at $\gamma=0$.
By differentiating the GF evolution equations of this section one recover the 
evolution equation for the EPs of section \ref{entropy}.
The EPs are therefore the ensemble average of the TEPs introduced in this section 
$\dot{A}(t)=\mean{\dot{a}[m_{(\tau)},t]}$ and $\Delta A(T,0) =\mean{\Delta a[m_{(\tau)},T]}$.

%%%%%%%%%%%%%%%%%%%%%%%%%%%%%%%%%%%%%%%%%%%%%%%%%%%%%%%%%%%%%%%%%%%%%%%
%%%%%%%%%%%%%%%%%%%%%%%%%%%%%%%%%%%%%%%%%%%%%%%%%%%%%%%%%%%%%%%%%%%%%%
\section{Fluctuation theorems}\label{FT}

%%%%%%%%%%%%%%%%%%%%%%%%%%%%%%%%%%%%%%%%%%%%%%%%%%%%%%%%%%%%%%%%%%%%%%
\subsection{General integral fluctuation theorems}\label{FTgen}

We can easily verify that $g_m(\gamma=-1,t)=p_m(t)$ is the solution 
of the evolution equations (\ref{Genaaaak}) and (\ref{Genaaaan}). 
It immediately follows from probability conservation and 
(\ref{TrajaveAaaaao}) that $\dot{G}_{tot}(-1,t)=\dot{G}_{na}(-1,t)=0$.
Summing both side of (\ref{Genaaaai}) over $m$, we also verify that $\dot{G}_{a}(-1,t)=0$.
Because $g_m^{(z)}(-1,0)=p_m(0)$, $G_z(-1,0)=1$ where $z=tot,na,a$.
Therefore, we find that $G_{z}(-1,t)=1$.
Using (\ref{TrajaveAaaaah}), this results in the three FTs 
\begin{eqnarray}
&&\mean{\exp{\{- \Delta s_{tot}[m_{(\tau)},t] \}}}=1 \;, \label{FTi}\\
&&\mean{\exp{\{- \Delta s_{na}[m_{(\tau)},t] \}}}=1 \;, \label{FTx}\\
&&\mean{\exp{\{- \Delta s_{a}[m_{(\tau)},t] \}}}=1 \;. \label{FThk}
\end{eqnarray}
These FTs hold irrespective of the initial condition and the type of driving. 
Using Jensen's inequality $\mean{\e^{x}} \geq \e^{\mean{x}}$,
they imply the inequalities (\ref{baaai})-(\ref{baaaj}).\\

Eq. (\ref{FTi}) is the generalization to open systems of 
the integral FT for the TEP obtained earlier 
for closed systems \cite{Seifert1}. 
The TEP (\ref{trajentr_aaaaf}) needs to specify which 
reservoir is responsible for the transitions occurring along the 
trajectory (by labeling the rates with reservoir index).  
This point, made earlier for open system at steady state \cite{Gaspard1}, 
is generalized here for driven systems with an arbitrary initial condition.
Eq. (\ref{FTx}) will be shown in next section to reduce to the integral 
Hatano-Sasa FT \cite{Hatano01} for systems initially in a steady state.
Eq. (\ref{FThk}) generalizes the integral FT for the adiabatic entropy 
\cite{Seifert2} previously derived for closed system initially in a steady state.\\

Some additional insights can be gained by an alternative proof of the FTs 
(\ref{FTi}) and (\ref{FTx}) which use a forward-backward trajectory 
picture of the dynamics.
This is given for completeness in appendix \ref{FTtrajectory}.

%%%%%%%%%%%%%%%%%%%%%%%%%%%%%%%%%%%%%%%%%%%%%%%%%%%%%%%%%%%%%%%%%%%%%%%
\subsection{Transitions between steady states}\label{HSFT}

We consider a system initially (t=0) at steady state and subjected 
to an external driving force between $t=t_{di}$ and $t=t_{df}$.
For $t_{di}>t>0$, the system remains in the steady state corresponding to $\lambda=\lambda_{t_{di}}$.
The time protocol of $\lambda_t$ during the driving $t_{df} > t > t_{di}$ is arbitrary.
If $t_{tr}$ is the characteristic transient time needed for the system 
to reach a steady state from an arbitrary distribution, for $t>t_{df}+t_{tr}$, 
the system is in the new steady state corresponding to $\lambda_{t_{df}}$.
The system is measured between $t=0$ and $t=T$.

\subsubsection{Fluctuation theorem for the reservoir entropy production}\label{FTe}

We restrict our analysis to cases where the system is at steady state 
at $t=0$ and the driving starts at least a time $t_{tr}$ after the 
measurement started: $t_{di}>t_{tr}$.\\
We use the braket notation where $\ket{p(t)}$ is the probability vector 
with components $p_m(t)$ and $\hat{{\cal W}}_t$ denotes the rate matrix. 
$\ket{I}$ denotes a vector with all components equal to one.
The ME (\ref{aaaaa}) now reads 
\begin{eqnarray}
\ket{\dot{p}(t)} = \hat{{\cal W}}_t \ket{p(t)} \label{VecME} \;.
\end{eqnarray}
The generating function for the reservoir EP (\ref{Genaaaag}) for
$\gamma=-1$ evolves according to the adjoint equation of (\ref{VecME})
\begin{eqnarray}
\ket{\dot{g}^{(r)}(-1,t)} = \hat{{\cal W}}^{\dagger}_t \ket{g^{(r)}(-1,t)}\;. \label{VecMEad}
\end{eqnarray}
The initial condition of (\ref{VecME}) and (\ref{VecMEad}) is
$\ket{p(0)}=\ket{g^{(r)}(-1,0)}=\ket{p^{\rm st}(\lambda_{t_{di}})}$. 
The formal solution of (\ref{VecMEad}) for $t<t_{di}$ 
before the driving starts reads 
\begin{eqnarray}
G_r(-1,t) = \bra{I} \exp{\{\hat{{\cal W}}^{\dagger} t \}} \ket{p^{\rm st}_m} 
\end{eqnarray}
We now insert a closure relation in term of right and left eigenvectors
of the adjoint rate matrix between the evolution operator and the initial condition.
Because the rate matrix and its adjoint have the same eigenvalues (all negative and one zero), 
for $t_{tr}<t<t_{di}$, only the right and left eigenvector associated with the zero eigenvalue survive. 
Since the right [left] eigenvector of $\hat{{\cal W}}^{\dagger}$ is $\ket{I}$ 
[$\bra{p^{\rm st}_m(\lambda_{t_{di}})}$], we get for $t_{tr}<t<t_{di}$
\begin{eqnarray}
G_r(-1,t) = \bra{I} \exp{\{\hat{{\cal W}}^{\dagger} t \}} \ket{I} 
\braket{p^{\rm st}}{p^{\rm st}} \;.
\end{eqnarray}
For longer times, even when the system starts to be driven, 
$\ket{I}$ remains invariant under the time evolution operator
as can be seen using (\ref{aaaaarate}) in (\ref{VecMEad}).
We get 
\begin{eqnarray}
G_r(-1,T) = M \braket{p^{\rm st}}{p^{\rm st}} \; \   \ \textrm{for} \  \ T \geq t_{tr} \;,
\end{eqnarray}
where $M=\braket{I}{I}$ is the total number of states. 
This implies the following integral FT for the reservoir TEP
\begin{eqnarray}
M \geq \mean{\exp{\{-\Delta s_r[m_{(\tau)},T] \}}} =
M \sum_{m=1}^{M} \big( p^{\rm st}_m (\lambda_{t_{di}}) \big)^2 \geq 1 \;.
\label{SSaaaai}
\end{eqnarray}
The equality on the l.h.s (r.h.s) is satisfied if 
$p^{\rm st}_m (\lambda_{t_{di}})=\delta_{n,m}$ ($p^{\rm st}_m (\lambda_{t_{di}})=1/M$).
Jensen's inequality implies $\Delta S_r(T,0) \geq 0$.
Note that since $\Delta s[m_{(\tau)},T]$ is a state function, it is easily verified that
\begin{eqnarray}
M \sum_{m=1}^{M} p_m^2(0) = \mean{\exp{\{\Delta s[m_{(\tau)},T] \}}} \label{SSaaaaj} \;.
\end{eqnarray} 

\subsubsection{The Hatano-Sasa fluctuation theorem}\label{HSsub}

We assume that the driving starts at the same time or later as the measurement ($t_{di} \geq 0$). 
We define $T'>t_{df}+t_{tr}$.\\
We have pointed out at the end of section \ref{entropy} that for transitions 
between steady states $\Delta S_{b}(T',0)=0$, so that 
\begin{eqnarray}
\Delta S_{na}(T',0) = \Delta S_{d}(T',0) = \Delta S_{d}(t_{df},t_{di}) \geq 0
\label{caaanaa}
\end{eqnarray}
We used the fact that $\Delta S_d(T',0)$ starts (stops) evolving at $t_{di}$ ($t_{df}$).
The same is true at the trajectory level, since from (\ref{caaajb}) 
we have $\Delta s_{b} [m_{(\tau)},T']=0$ and therefore
\begin{eqnarray}
\Delta s_{na}[m_{(\tau)},T'] = \Delta s_d[m_{(\tau)},T'] \;,
\label{caaana}
\end{eqnarray}
where
\begin{eqnarray}
\Delta s_d[m_{(\tau)},T'] &\equiv& \sum_{j=0}^{N}  \ln 
\frac{p_{m_{j}}^{\rm st}(\lambda_{\tau_j})}
{p_{m_{j}}^{\rm st}(\lambda_{\tau_{j+1}})}\label{caaav}\\
&=& \sum_{j=0}^{N} \big( \phi_{m_{j}}(\lambda_{\tau_{j+1}}) 
- \phi_{m_{j}}(\lambda_{\tau_{j}}) \big) \nonumber \\
&=& \int_{0}^{T'} dt \dot{\lambda}_{t} 
\frac{\partial \phi_{m_{(\tau)}}(\lambda_{t})}{\partial \lambda_{t}} 
\nonumber \\
&=& \int_{t_{di}}^{t_{df}} dt \dot{\lambda}_{t} 
\frac{\partial \phi_{m_{(\tau)}}(\lambda_{t})}{\partial \lambda_{t}} 
\nonumber \;.
\end{eqnarray} 
The integrand in the third line contributes only during 
the time intervals between jumps provided the driving is changing.
Therefore if without loss of generality we choose the measurement time such that 
$T \geq t_{df}$ (if $T<t_{df}$ one can redefine $t_{df}$ as equal to $T$),
$\Delta s_d[m_{(\tau)},T]=\Delta s_d[m_{(\tau)},T']$.
This means that for a transition between steady states, the FT (\ref{FTx}) 
reduces to the Hatano-Sasa FT \cite{Hatano01}
\begin{eqnarray}
\mean{\exp{\{-s_{d}[m_{(\tau)},T] \}}}=1 \label{FTy} \;. 
\end{eqnarray}
Alternatively, (\ref{FTy}) can be proved from (\ref{Genaaaad}) 
by showing that when $\gamma=-1$, $g_{m}^{(d)}(-1,t)=\exp{\{-\phi_{m}(\lambda_t)\}}
=p_{m}^{\rm st}(\lambda_t)$ is solution of (\ref{Genaaaad}).

The FT (\ref{FTy}) holds for an arbitrary driving protocol.
Let us consider the two extremes. For an {\it adiabatic} (infinitely 
slow) driving, the inequality in (\ref{caaanaa}) becomes an equality.
In the other extreme of a {\it sudden} driving, where 
$\lambda_{t}= \lambda_{t_{di}} + \Theta(t-t_{di}) (\lambda_{t_{df}}-\lambda_{t_{di}})$ 
and $\dot{\lambda}_{t}=\delta(t_{di})(\lambda_{t_{df}}-\lambda_{t_{di}})$, 
\begin{eqnarray}
\Delta s_d[m_{(\tau)},T] = \phi_{m_0}(\lambda_{t_{df}}) - \phi_{m_0}(\lambda_{t_{di}}) 
\label{caabj} 
\end{eqnarray}
becomes a state function and its average takes the simple form
\begin{eqnarray}
\Delta S_d(T,0) = \sum_{m} p_{m}^{\rm st}(\lambda_{t_{di}}) 
\big( \phi_{m}(\lambda_{t_{df}}) - \phi_{m}(\lambda_{t_{di}}) \big)\;.
\label{caabl}
\end{eqnarray}
Using (\ref{trajentr_aaaae}) with (\ref{caaana}), and since
\begin{eqnarray}
\Delta s[m_{(\tau)},T]
= \phi_{m_T}(\lambda_{t_{df}}) - \phi_{m_0}(\lambda_{t_{di}}) \;,
\label{caabi}
\end{eqnarray}
we find that
\begin{eqnarray}
\Delta s_{ex}[m_{(\tau)},T]
= \phi_{m_0}(\lambda_{t_{df}})- \phi_{m_T}(\lambda_{t_{df}}) 
\label{caabk}
\end{eqnarray}
also becomes a state function and its average becomes
\begin{eqnarray}
\Delta S_{ex}(T,0) &=&
\sum_{m_0,m_T} p_{m_0}^{\rm st}(\lambda_{t_{di}}) p_{m_T}^{\rm st}(\lambda_{t_{df}})
\nonumber\\&&\hspace{1cm}
\big( \phi_{m_0}(\lambda_{t_{df}}) - \phi_{m_T}(\lambda_{t_{df}}) \big) \;.
\label{caabm}
\end{eqnarray}

%%%%%%%%%%%%%%%%%%%%%%%%%%%%%%%%%%%%%%%%%%%%%%%%%%%%%%%%%%%%%%%%%%%%%
\subsection{Transitions between equilibrium states} \label{singleres}

For a system coupled to a single reservoir (or multiple reservoirs 
with identical thermodynamical properties), the DBC (\ref{aaaae}) is satisfied. 
A non-driven system in an arbitrary state will reach after some transient 
time $t_{\rm tr}$ the equilibrium grand canonical distribution (\ref{aaaac}).
We again choose $T'>t_{df}+t_{tr}$ and $T \geq t_{df}$.
From the TEP of section \ref{trajectory}, we find in this case
\begin{eqnarray}
\Delta s_{a}[m_{(\tau)},T'] &=& 0 \label{daaaab} \\
\Delta s_{r}[m_{(\tau)},T'] &=& \Delta s_{ex}[m_{(\tau)},T'] \nonumber\\
\Delta s_{tot}[m_{(\tau)},T'] &=& \Delta s_{na}[m_{(\tau)},T'] \;.\nonumber
\end{eqnarray}
The two FT (\ref{FTi}) and (\ref{FTx}) become identical 
and the FT (\ref{FThk}) becomes trivial.
Using (\ref{aaaac}), we also find that Eq. (\ref{caaanb}) becomes
\begin{eqnarray}
\phi_{m}(\lambda) = \beta(\lambda) \{ \epsilon_{m}(\lambda) 
- \mu(\lambda) N_{m} - \Omega(\lambda) \} \;, \label{daaaa}
\end{eqnarray}
where $\Omega(\lambda)=-\ln \Xi(\lambda)$ is 
the thermodynamic grand canonical potential.\\

We next consider transitions between equilibrium states,
so that the procedure is the same as in \ref{HSsub} but with
the DBC (\ref{aaaae}) now satisfied.
We therefore have $\Delta s_{na}[m_{(\tau)},T']=\Delta s_d[m_{(\tau)},T']$.
The driving implies externally modulating the system energies, the 
chemical potential or the temperature of the reservoir.\\

When driving the system energy, using (\ref{caaav}) and (\ref{daaaa}), we find
\begin{eqnarray}
\Delta s_d[m_{(\tau)},T] = \Delta s_d[m_{(\tau)},T'] 
= \beta w[m_{(\tau)}] - \beta \Delta \Omega
\label{daaab}
\end{eqnarray}
where the work is given by $w[m_{(\tau)}] = \int_{t_{di}}^{t_{df}} 
dt \dot{\epsilon}_{m_{(\tau)}}(\lambda_t)$ and 
$\Delta \Omega=\Omega[\epsilon(t_{df})]-\Omega[\epsilon(t_{di})]$.
Both FT, (\ref{FTi}) and (\ref{FTx}), lead to the same Jarzynski 
relation \cite{Jarzynski1} 
\begin{eqnarray}
\mean{\exp{\{-\beta w[m_{(\tau)}]}\}} = \exp{\{-\beta \Delta \Omega \}} \;.
\label{daaad}
\end{eqnarray}

When driving the reservoir chemical potential,
Eq. (\ref{caaav}) and (\ref{daaaa}) give
\begin{eqnarray}
\Delta s_d[m_{(\tau)},T] = \Delta s_d[m_{(\tau)},T'] 
= \beta \tilde{w}[m_{(\tau)}] - \beta \Delta \Omega \label{daaaf} 
\end{eqnarray}
where $\tilde{w}[m_{(\tau)}] = -\int_{t_{di}}^{t_{df}} dt \dot{\mu} N_{m_{(\tau)}}$
and $\Delta \Omega=\Omega[\mu(t_{df})]-\Omega[\mu(t_{di})]$.
Both FT, (\ref{FTi}) and (\ref{FTx}), now lead to
\begin{eqnarray}
\mean{\exp{\{- \beta \tilde{w}[m_{(\tau)}] \}}} = \exp{\{-\beta \Delta \Omega \}} \;.
\label{daaah}
\end{eqnarray}

The case where reservoir temperature is driven can be calculated similarly.

\iffalse
Driving the reservoir temperature
Using (\ref{aaaac}), Eq. (\ref{caaanb}) becomes
\begin{eqnarray}
\phi_{m}(\lambda) = \beta(\lambda) \big( \epsilon_{m} 
- \mu N_{m} - \Omega(\lambda) \big) \;.
\label{daaai}
\end{eqnarray}
Eq. (\ref{caaav}) gives
\begin{eqnarray}
\Delta y[m_{(\tau)}] &=& \int_{0}^{T} d\tau \dot{\lambda}_{\tau}
\frac{\partial \beta(\lambda_{\tau})}{\partial \lambda_{\tau}} 
\big( \epsilon_{m_{(\tau)}} - \mu N_{m_{(\tau)}} \big) \label{daaaj}\\
&&- \big( \beta(\lambda_{T}) \Omega (\lambda_{T}) 
- \beta(\lambda_{0}) \Omega (\lambda_{0}) \big) \;. \nonumber
\end{eqnarray}
Using (\ref{caaba}), we get
\begin{eqnarray}
\mean{\exp{\{ \int_{0}^{T} d\tau \dot{\lambda}_{\tau}
\frac{\partial \beta(\lambda_{\tau})}{\partial \lambda_{\tau}} 
\big( \epsilon_{m_{(\tau)}} - \mu N_{m_{(\tau)}} \big) \}}} 
&=& \label{daaak}\\ &&\hspace{-5cm}
\exp{\{ \beta(\lambda_{T}) \Omega (\lambda_{T}) 
- \beta(\lambda_{0}) \Omega (\lambda_{0}) \}} \;.\nonumber
\end{eqnarray}
\fi

%%%%%%%%%%%%%%%%%%%%%%%%%%%%%%%%%%%%%%%%%%%%%%%%%%%%%%%%%%%%%%%%%%%%%
\subsection{No driving: steady state fluctuation theorem}\label{SS}

In a NESS, the relations of section \ref{trajectory} give
\begin{eqnarray}
\Delta s_{na}[m_{(\tau)},t]&=&\Delta s_d[m_{(\tau)},t]=0 \label{SSaaaaa}\\
\Delta s [m_{(\tau)},t]&=& -\Delta s_{ex}[m_{(\tau)},t] \nonumber\\
\Delta s_{tot} [m_{(\tau)},t]&=& \Delta s_{a}[m_{(\tau)},t] \nonumber \;.
\end{eqnarray}
Furthermore, since $\Delta S(t,0)=0$, we get
\begin{eqnarray}
\Delta S_{tot}(t,0) = \Delta S_{r}(t,0) \geq 0  \label{SSaaaab}\;.
\end{eqnarray} 
We shall rewrite the GF evolution equation for the reservoir TEP (\ref{Genaaaag}) 
in the bracket notation 
\begin{eqnarray}
\ket{\dot{g}^{(r)}(\gamma,t)} = \hat{{\cal V}}(\gamma) \ket{g^{(r)}(\gamma,t)} \;,
\label{SSaaaac}
\end{eqnarray}
so that
\begin{eqnarray}
G_r(\gamma,t) = \bra{I} \exp{\{\hat{{\cal V}}(\gamma) t\}} 
\ket{p^{\rm st}} \;, \label{SSaaaad}
\end{eqnarray}
where $\ket{I}$ is a vector with all elements equal to one.
Since from (\ref{Genaaaag}) the generator has the property 
$\hat{{\cal V}}(\gamma)=\hat{{\cal V}}^{\dagger}(-\gamma-1)$, 
its eigenvalues have the symmetry $s_{\xi}(\gamma)=s_{\xi}(-\gamma-1)$.
Furthermore, since $\exp{\{\hat{{\cal V}}(\gamma) t\}}$ is a positive matrix,
the Frobenious-Perron theorem \cite{Stirzaker,Norris,Kampen} ensures that 
all eigenvalues are negative or zero and that the left and the right eigenvectors, 
$\ket{\xi_0(\gamma)}$ and $\ket{\tilde{\xi}_0(\gamma)}$, associated with the 
largest eigenvalue $s_{\xi_0}(\gamma)$ exist. 
Adopting the normalization $\braket{\tilde{\xi}_0(\gamma)}{\xi_0(\gamma)}=1$,
we find for long times
\begin{eqnarray}
G_r(\gamma,t) \stackrel{t \to \infty}{=}
\exp{\{s_{\xi_0}(\gamma) t\}} \braket{I}{\tilde{\xi_0}(\gamma)} 
\braket{\xi_0(\gamma)}{p^{\rm st}} \;. \label{SSaaaaea} 
\end{eqnarray}
and that
\begin{eqnarray}
G_r(-\gamma-1,t) &\stackrel{t \to \infty}{=}& 
\exp{\{s_{\xi_0}(\gamma) t\}} \label{SSaaaaeb} \\ 
&&\braket{I}{\tilde{\xi_0}(-\gamma-1)} 
\braket{\xi_0(-\gamma-1)}{p^{\rm st}} \;. \nonumber
\end{eqnarray}
This means that the cumulant generating function
\begin{eqnarray}
F_r(\gamma) \equiv \lim_{t \to \infty} \frac{1}{t} \ln G_r(\gamma,t)
\label{SSaaaaf}
\end{eqnarray}
satisfies the symmetry
\begin{eqnarray}
F_r(\gamma) = F_r(-\gamma-1) \;.
\label{SSaaaag}
\end{eqnarray}
Using the theory of large fluctuations this symmetry implies the 
detailed steady state FT \cite{Lebowitz,Gaspard1}
\begin{eqnarray}
\frac{{\rm P}(\Delta S_r)}
{{\rm P}(-\Delta S_r)} \stackrel{t \to \infty}{=} \e^{\Delta S_r} \;,
\label{SSaaaah}
\end{eqnarray}
where ${\rm P}(\Delta S_r)$ is the probability for a trajectory
of the system to produce a reservoir TEP equal to $\Delta S_r$.\\
$\Delta s_r[m_{(\tau)},t]$ grows in average with time because it 
depends on the number of jumps along the trajectory. 
However, $\Delta s[m_{(\tau)},t]$ is bounded. 
The FT (\ref{SSaaaah}) can therefore be viewed as a consequence 
of the detailed FT for $\Delta S_{tot}$ derived in (\ref{caaap}). 
The long time limit is needed in order to neglect the contribution
from $\Delta s[m_{(\tau)},t]$ to $\Delta s_{tot}[m_{(\tau)},t]$.\\
The FT (\ref{SSaaaai}) remains valid at steady state.
FTs for currents can also be derived 
\cite{GaspardAndrieux2,GaspardAndrieux1,EspositoHarbola2}.

%The relation (\ref{SSaaaai}) therefore indicates that the fluctuations 
%of $e^{-\Delta s_{tot}[m_{(\tau)}]}$ and $e^{\Delta s[m_{(\tau)}]}$ become 
%uncorrelated for $t \gg t_{tr}$
%\begin{eqnarray}
%\mean{e^{-\Delta s_{tot}[m_{(\tau)}] + \Delta s[m_{(\tau)}]}}
%\stackrel{t \gg t_{tr}}{=} 
%\mean{e^{-\Delta s_{tot}[m_{(\tau)}] }} 
%\mean{e^{\Delta s[m_{(\tau)}] }} \;.
%\label{SSaaaak}
%\end{eqnarray}

%At long times, since $\Delta s_{r} [m_{(\tau)}]$ 
%depends on the number of transitions occurring in the system, 
%and since $\Delta s [m_{(\tau)}]$ only depends on the initial 
%and final distribution of the system and is therefore bounded, 
%we expect
%\begin{eqnarray}
%\lim_{T \to \infty} \Delta s_{tot} [m_{(\tau)}] = \Delta s_{r} [m_{(\tau)}]
%\label{daaal}
%\end{eqnarray}

%%%%%%%%%%%%%%%%%%%%%%%%%%%%%%%%%%%%%%%%%%%%%%%%%%%%%%%%%%%%%%%%%%%%%%%%%%%%%%%%%%%%%%
\section{Entropy fluctuations for electron transport trough a single level quantum dot}\label{QD}

We have seen in section \ref{measure} that the various entropies can be calculated 
by measuring the different currents between the system and the reservoirs. 
The counting statistics of electrons through quantum dots has recently raised  
considerable theoretical 
\cite{Levitov,Rammer,Nazarov,Utsumi,EspositoHarbola2} as 
well as experimental \cite{Lu,Fujisawa,Bylander,Gustavsson,Hirayama} interest.
The single electrons entering and exiting a quantum dot connected to two leads can be measured.
One can therefore calculate all currents, deduce the system trajectories 
and calculate the various trajectory entropies presented earlier.\\

We will analyze the probability distribution for the various trajectory 
entropies in a single level quantum dot of energy $\epsilon$ connected 
to two leads with different chemical potentials $\mu_{\nu}(t)$, where $\nu=l,r$.  
We neglect spin so that the dot can either be empty $0$ or filled $1$. 
The ME is of the form (\ref{aaaaa})
\cite{Nazarov,Beenakker,Bonet,Elste,Datta1,Datta2,EspositoHarbola1}
\begin{eqnarray}
\left(
\begin{array}{c}
\dot{p}_{1}(t) \\ 
\dot{p}_{0}(t) 
\end{array}
\right) =
\left(
\begin{array}{cc}
- v_t &  w_t \\
v_t   & - w_t
\end{array}
\right)
\left(
\begin{array}{c}
p_{1}(t) \\ 
p_{0}(t) 
\end{array}
\right) \;, \label{yaaaa}
\end{eqnarray}
where
\begin{eqnarray}
v_t &=& \sum_{\nu} v^{(\nu)}_t
= \sum_{\nu} a_{\nu} \big( 1-f_{\nu}(t) \big) \nonumber\\
w_t &=& \sum_{\nu} w^{(\nu)}_t
= \sum_{\nu} a_{\nu} f_{\nu}(t) \;. \label{yaaab}
\end{eqnarray}
The coefficients $a_{\nu}$ characterize the coupling 
between the dot and the lead $\nu$ with Fermi distribution 
$f_{\nu}(t) \equiv 1/(\exp{\{\beta \{\epsilon-\mu_{\nu}(t)\} }\}+1)$.
If $\hbar=1$, $[a]=[{\rm energy}]=[{\rm time}^{-1}]$. 
By renormalizing energies by $\epsilon$, all parameters of our model become dimensionless.
The steady state distribution of the system is
\begin{eqnarray}
p_1^{({\rm st})} = \frac{w_t}{v_t+w_t} \; \ \ \;, \  \ \; 
p_0^{({\rm st})} = \frac{v_t}{v_t+w_t} \;, \label{yaaad}
\end{eqnarray}
and the steady state currents are given by 
\begin{eqnarray}
\mean{I}_{1,3}^{({\rm st})}=
\frac{v^{(l,r)}_t w_t}{v_t+w_t} \; \ \ \;, \  \ \; 
 \mean{I}_{2,4}^{({\rm st})}=
\frac{w^{(l,r)}_t v_t}{v_t+w_t} \;.
\label{yaaae}
\end{eqnarray}
We switch the chemical potential of the left lead 
$\mu_{l}(t)= \mu_{0} + V(t)$ using the protocol
\begin{eqnarray}
V(t) = \frac{V_f-V_i}{2} \tanh \big( c (t-t_m)+1 \big)
\label{potential_profile}
\end{eqnarray}
while holding the right lead chemical potential fixed $\mu_{r}(t)= \mu_{0}$.
We can therefore calculate all the trajectory entropies' probability 
distributions using the GF method described in section \ref{statprop}.
We solve numerically the evolution equations for the $g_m(\i \gamma,t)$'s 
associated with the different entropies for different values of $\gamma$
with the initial condition $g_m(\i \gamma,0)=p_m(0)$.
After calculating the $G(\i \gamma,t)$'s using (\ref{TrajaveAaaaal}), the probability 
distribution is obtained by a numerical inverse Fourier transform (\ref{TrajaveAaaaai}).
In all calculations we used $\beta=5$, $\epsilon=1$, $a_{l}=0.2$ and $a_{r}=0.1$.\\

We start by analyzing the different contributions to the EP as defined in section \ref{entropy}
for the $\mu_l(t)$ protocol shown in Fig. (\ref{aver})a. \\
The system is initially in a nonequilibrium distribution different from the steady state. 
The solution of the ME (\ref{yaaaa}) as well as its steady state solution are displayed in Fig. \ref{aver}b.
Between $t=0$ and $t=20$, $\mu_{l}(t)$ is essentially constant and the system undergoes an 
exponential relaxation to the steady state.
Between $t=20$ and $t=50$, $\mu_{l}(t)$ changes from $\mu_{0}+V_i$ to $\mu_{0}+V_f$ fast enough for the system 
distribution to start differing again from the instantaneous steady state distribution (adiabatic solution). 
After $t=50$, $\mu_{l}(t)$ remains constant and the system again undergoes a transient 
relaxation to the new steady state corresponding to $\mu_{0}+V_f$.
Fig. \ref{aver}c shows the time dependent EP $\dot{S}_{tot}$ and its adiabatic 
$\dot{S}_{a}$ and nonadiabatic contribution $\dot{S}_{na}$.
As predicted, these three quantities are always positive [see (\ref{baaai})-(\ref{baaaj})].  
We also demonstrate that $\dot{S}_{na}$ only contributes when nonadiabatic effects are significant 
i.e. when the actual probability distribution is different from the steady state one 
[$p_m(t) \neq p_m^{(st)}(\mu(t))$].
$\dot{S}_{a}=0$ only once at $t \approx 33$, when $\mu_{l}(t)=\mu_{r}(t)=0.5$
and the DBC is satisfied. Otherwise the DBC is broken and $\dot{S}_{a}>0$.
In Fig. \ref{aver}d, we present the two contributions to the nonadiabatic EP $\dot{S}_{na}$, 
the driving EP $\dot{S}_d$ and the boundary EP $\dot{S}_{b}$ [see (\ref{baaak})]. 
The driving EP $\dot{S}_d$ only contributes when $\mu_{l}(t)$ changes in time.
One can also guess that $\Delta S_{b}=\int_{20}^{60}dt \dot{S}_{b}=0$ due to the fact that the 
change of boundary EP during an interval between two steady state is zero. 
$\Delta S_{b}=\int_{0}^{20}dt \dot{S}_{b} \neq 0$ because the system is initially not in a steady state.
Fig. \ref{aver}e shows the alternative partitioning of the nonadiabatic EP into 
the system EP $\dot{S}$ and the excess EP $\dot{S}_{ex}$ [see (\ref{baaakk})].
Finally the splitting of the total EP in the reservoir EP and the system EP [see (\ref{baaab})] 
is shown in Fig. \ref{aver}f. 
We see that at steady state $\dot{S}=0$ so that $\dot{S}_{tot}=\dot{S}_r$.\\

We next study the statistical properties of the different TEP for transitions between steady states.
The probability distributions are obtained using the GF method presented in section \ref{statprop}.
The five driving protocols used to change $\mu_l(t)$ from $\mu_{0}+V_i$ to $\mu_{0}+V_f$ 
are presented in Fig. \ref{profilefig}.
They range from sudden switch in (i) to slow (adiabatic) switch in (v). 
The system is always initially in the steady state corresponding to $\mu_{0}+V_i$.
We will consider measurements which end when the system reaches its new steady state at $\mu_{0}+V_f$.
Different measurement times are represented by a,b,c,d.\\

In Fig. \ref{PS}, we display $P(\Delta S)$. 
Since $\Delta s$ is a state function, $P(\Delta S)$ is the same for the various protocol.
Because we consider a two level system, $\Delta s$ can only take four possible values which 
correspond to the four possible change in the system state between its initial and final condition.
The transitions $0 \to 0$ and $0 \to 1$ are much more probable because the probability to 
initially find the system in the empty state $0$ is much higher ($0.96$) 
than finding it in the filled state $1$ ($0.04$). 
The transition $0 \to 0$ is more probable than $0 \to 1$ because the system 
has a final probability $0.64$ to be in its empty state and $0.36$ to be in its filled state.\\

In the left column of Fig. \ref{PSexPSd}, we depict $P(\Delta S_{d})$.
Here $\Delta s_{b}=0$ so that $\Delta s_{d}=\Delta s_{na}$ [see (\ref{baaahb})].  
All curves (i)-(v) satisfy the FT (\ref{FTx}).
For the sudden switch (i), $\Delta s_{d}$ becomes a state function 
which only depends on the initial state of the system [see (\ref{caabj})]. 
$\Delta s_{d}$ can therefore take two possible values corresponding to the 
empty $0$ or filled $1$ orbital with a respective probability $0.96$ or $0.04$.
When the driving speed slows down in (ii) the peaks are broadened.
In the adiabatic limit (V), $P(\Delta S_{d})$ becomes a broad distribution with zero average.
In the right column of Fig. \ref{PSexPSd}, we depict $P(\Delta S_{ex})$. 
For sudden switch (i), $\Delta s_{ex}$ turn to a state function which only 
depends on the final steady state distribution [see (\ref{caabk})]. 
It is clear from (\ref{caabk}) that the transitions $0 \to 0$ and $1 \to 1$ leads to 
$\Delta s_{ex}=0$ and $1 \to 0$ and $0 \to 1$ to the same $\Delta s_{ex}$ with opposite sign.
The probabilities to observe these transitions follow from the fact that 
the system is initially more likely to be in $0$ (prob $0.96$) than in $1$. 
The probability for the final state $0$ ($1$) is $0.64$ ($0.36$). 
Therefore the most likely transition is $0 \to 0$ followed from $0 \to 1$.
As the driving speed slows down like in (ii), the peaks get broadened. 
Since $\Delta s_{ex}=\Delta s_{d}-\Delta s$ [see (\ref{baaakk})] and since 
in the adiabatic switch limit (v) $P(\Delta S_{d})$ is centered around zero,
$P(\Delta S_{ex})$ (v) has the same peak structure as $P(-\Delta S)$ 
[see Fig. \ref{PS}], but broadened by $\Delta s_{d}$.\\

In Fig. \ref{PSe}, we display $P(\Delta S_r)$ for different measurement times and protocols.
Plots with same driving but different measurement times [(i)a and (ii)d or (ii)a and (ii)d 
or (iii)b and (iii)d] show the evolution of $P(\Delta S_r)$ in the final steady state.
The plots (ii)-(v) satisfy the FT (\ref{SSaaaai}) which for our parameters imply
$\mean{\exp{\{-\Delta s_r\}}}=1.846$. 
The FT is not satisfied for (i) because the driving starts 
at the same time as the measurement [see section \ref{FTe}]. 
To understand the structure of $P(\Delta S_r)$, we time integrate (\ref{trajentr_aaaagb}) and 
use the fact that in our model the heat current is proportional 
to the matter current between the $\nu$ reservoir and the system 
$I^{(\nu)}=I_{mat}^{(\nu)}=I_{heat}^{(\nu)}/\epsilon$ where
\begin{eqnarray}
I^{(\nu)}(t) =\sum_{j=1}^{N} \delta_{\nu,\nu_j}(t-\tau_j) 
\big(N_{m_{j}}-N_{m_{j-1}}\big) \;.
\label{mattercurrbis}
\end{eqnarray}
We get
\begin{eqnarray}
\Delta s_r(t)= - \beta \sum_{\nu} \int_{0}^{t} d\tau
\bigg( \epsilon - \mu_{\nu}(\tau) \bigg) I^{(\nu)}(\tau) \;.
\end{eqnarray}
In the sudden switch limit (i), we get
\begin{eqnarray}
\Delta s_r(t)= - \beta \sum_{\nu}
\bigg( \epsilon - \mu_{\nu}(T) \bigg) {\cal N}^{(\nu)}(t) \;.
\end{eqnarray}
where ${\cal N}^{(\nu)}= \int_{0}^{t} d\tau I^{(\nu)}(\tau)$ is the net number 
of electron transferred from the reservoir $\nu$ to the system between $0$ and $t$. 
This explains why $\Delta s_r$ in (i) only take discrete value which are multiples of each other. 
The distance between the peaks of $2.5$ observed in (i) is due to 
the right lead only $\beta (\epsilon - \mu_{r})=2.5$ because our 
parameters are such that $\beta (\epsilon - \mu_{l}(T))=0$].
The new peaks which appear in (ii) with a spacing $0.125$ are due 
to the fact that the driving starts some time after the measurement 
so that $\beta (\epsilon - \mu_{l}(0))=3.75$ also contributes.
As the driving speed slows down (iii)-(v), the discrete structure 
broadens and $\Delta s_r$ can take continuous values.\\

In Fig. \ref{PSa}, we display $P(\Delta S_a)$. 
All curves satisfy the FT (\ref{FThk}). 
The verification (not shown) is best done on the GF ($G_a(-1,t)=1$) because the 
numerical accuracy of the tail of the distribution is not sufficient.  
The peak structure of $P(\Delta S_{a})$ can be understood from $P(\Delta S_{r})$ 
and $P(\Delta S_{ex})$ because $\Delta s_{r}=\Delta s_{a}+\Delta s_{ex}$.
This is particularly clear for the sudden switch (i) where 
the possible values of the entropies are strongly restricted.
Indeed, in (i) each peak of $P(\Delta S_{r})$ is split in three smaller 
peaks which have the same structure as $P(\Delta S_{ex})$.
As the speed of the driving decreases (ii)-(v), the peak structure disappears.\\

In Fig. \ref{PSi}, we show $P(\Delta S_{tot})$. 
All curves satisfy the FT (\ref{FTi}) [verification was done on the GF (not shown)].  
The structure of $P(\Delta S_{tot})$ can be understood using $P(\Delta S_{r})$ 
and $P(\Delta S)$ because $\Delta s_{tot}=\Delta s_{r}+\Delta s$.
This is clear for the sudden switch (i) where the peaks of $P(\Delta S_{r})$ are 
split in smaller peaks which have the structure of $P(\Delta S)$.\\

%%%%%%%%%%%%%%%%%%%%%%%%%%%%%%%%%%%%%%%%%%%%%%%%%%%%%%%%%%%%%%%%%%%%%%%%%%%%%%%%%%%%%%%
\section{Conclusions}\label{conclusions}

For a driven open system in contact with multiple reservoirs and described 
by a master equation, we have proposed a partitioning of the trajectory 
entropy production into two parts. 
One contributes when the system is not in its steady state 
and contains two contributions due to the external driving and the deviation 
from steady state in the initial and final probability distribution of the system.
The second part comes from breaking of the detailed balance condition 
by the multiple reservoirs and becomes equal to the total entropy production 
when the system remains in its steady state all throughout the nonequilibrium process.
Both parts as well as the total entropy production satisfy an general integral 
fluctuation theorem which imposes positivity on their ensemble average. 
This partitioning also provides a simple way to identify which part of the 
entropy production contributes during a specific type of nonequilibrium 
process [see Fig. \ref{shema}].
Previously derived integral fluctuation theorems can be recovered from our 
three general fluctuation theorems and in addition we derived a new integral 
fluctuation theorem for the part of the entropy production due to exchange 
processes between the system and its reservoirs (reservoir entropy production).
Our results strictly apply to systems described by a master equation (\ref{aaaaa}).
However, as has often be the case for previous fluctuation relations, 
one could expect similar results to hold for other types of dynamics.
For electron transport trough a single level quantum dot between two 
reservoir with time dependent chemical potentials, we have simulated and 
analyzed in detail the probability distributions of the various trajectory 
entropies and showed how they can be measured in electron counting statistics experiments.

%%%%%%%%%%%%%%%%%%%%%%%%%%%%%%%%%%%%%%%%%%%%%%%%%%%%%%%%%%%%%%%%%%%%%%%%%%%%%%%%%%%%%%%
\section*{Acknowledgments}

The support of the National Science Foundation (Grant No. CHE-0446555) 
and NIRT (Grant No. EEC 0303389) is gratefully acknowledged.
M. E. is partially supported by the FNRS Belgium
(collaborateur scientifique).\\

%%%%%%%%%%%%%%%%%%%%%%%%%%%%%%%%%%%%%%%%%%%%%%%%%%%%%%%%%%%%%%%%%%%%%%%%%%%%%%%%%%%%%%%%
\appendix

%%%%%%%%%%%%%%%%%%%%%%%%%%%%%%%%%%%%%%%%%%%%%%%%%%%%%%%%%%%%%%%%%%%%%%
\section{Fluctuation theorems in terms of forward backward 
trajectory probabilities} \label{FTtrajectory}

We show that the FT (\ref{FTi}) and (\ref{FTx}) have an interesting interpretation in 
term of the ratio of the probability of a forward dynamics generating a given trajectory 
and the probability of the time-reversed trajectory during some backward dynamics.
This is an alternative to the GF approach which connects the detailed 
form to the integral form of the FTs.\\

The forward dynamics is described by the ME (\ref{aaaaa}).
We introduce the probability (in trajectory space)
${\cal P}[m_{(\tau)}]$ that the system follows a trajectory $m_{(\tau)}$ 
\begin{eqnarray}
{\cal P}[m_{(\tau)}] &=& p_{m_0}(0) \bigg[ \prod_{j=1}^{N} 
\label{caaaa}\\ &&\hspace{-2cm} 
\exp{\bigg(\int_{\tau_{j-1}}^{\tau_{j}} d\tau' 
W_{m_{j-1},m_{j-1}}(\lambda_{\tau'}) \bigg)} 
W_{m_{j},m_{j-1}}^{(\nu_{j})}(\lambda_{\tau_j}) \bigg] \nonumber 
\\&&\hspace{0cm} \exp{\big(\int_{\tau_{N}}^{T} d\tau'
W_{m_{N},m_{N}}(\lambda_{\tau'}) \big)} \;.\nonumber
\end{eqnarray}
The $W_{m_{j},m_{j-1}}^{(\nu_{j})}(\lambda_{\tau_j})$ factors in 
this expression represent the probability that the system undergoes 
a given transition whereas the exponentials describe the probability 
for the system to remain in a given state between two successive jumps.
Summation over all possible trajectories will be denoted
by $\sum_{m_{(\tau)}}$. It consists of time-ordered integrations
over the $N$ time variables $\tau_{j}$ from $0$ 
to $T$ (this gives the probability of having a path
with $N$ transitions) and then summing over all possible 
$N$ from $0$ to $\infty$.  
Normalization in trajectory-space implies that
$\sum_{m_{(\tau)}} {\cal P}[m_{(\tau)}] = 1$.\\

The backward dynamics is described on the 
time interval $t=[0,T]$ by the ME
\begin{eqnarray}
\dot{\tilde{p}}_m(t) = \sum_{m'} 
\tilde{W}_{m,m'}(\lambda_{T-t}) \tilde{p}_{m'}(t)
\label{caaab}
\end{eqnarray}
where the new rate matrix $\tilde{W}_{m,m'}(\lambda_{t})$
satisfies $\sum_{m} \tilde{W}_{m,m'}(\lambda_{t})=0$.
We require that the parametric time dependence (via the 
driving protocol $\lambda_{t}$) of the rate matrix in Eq. 
(\ref{caaab}) is time-reversed compared that of Eq. (\ref{aaaaa}) 
and that the diagonal part of the rate matrix in Eq. (\ref{aaaaa}) 
and (\ref{caaab}) is the same 
\begin{eqnarray}
\tilde{W}_{m,m}(\lambda_{t})=W_{m,m}(\lambda_{t}) \;.
\label{caaaba}
\end{eqnarray}
This still leaves room for different choices 
of $\tilde{W}_{m,m'}(\lambda_{t})$.
We will later specify two choice of $\tilde{W}_{m,m'}(\lambda_{T-t})$ 
[(\ref{caaaoa}) and (\ref{caaax})] that will result in two FTs.\\
We define the time-reversed trajectory of $m_{(\tau)}$ by $\bar{m}_{(\tau)}=$ 
$\{0-m_N \stackrel{T-\tau_N}{\rightarrow} m_{N-1} \stackrel{T-\tau_{N-1}}{\rightarrow} 
\cdots m_{j} \stackrel{T-\tau_j}{\rightarrow} m_{j-1} \stackrel{T-\tau_{j-1}}{\rightarrow} 
\cdots m_{1} \stackrel{T-\tau_1}{\rightarrow} m_{0}-T\}$.
The probability $\tilde{{\cal P}}[\bar{m}_{(\tau)}]$ 
that the system described by (\ref{caaab}) follows the 
time-reversed trajectory $\bar{m}_{(\tau)}$ is given by
\begin{eqnarray}
\tilde{{\cal P}}[\bar{m}_{(\tau)}] &=& 
\tilde{p}_{m_N}(0) \bigg[ \prod_{j=1}^{N} 
\label{caaac}  \\&&\hspace{-1.5cm} 
\exp{\bigg(\int_{T-\tau_{N-j+2}}^{T-\tau_{N-j+1}} d\tau' 
\tilde{W}_{m_{N-j+1},m_{N-j+1}}(\lambda_{T-\tau'}) \bigg)} 
\nonumber \\&&\hspace{2cm}
\tilde{W}_{m_{N-j},m_{N-j+1}}^{(\nu_{N-j+1})}
(\lambda_{T-\tau_{N-j+1}}) \bigg] 
\nonumber \\&&\hspace{1cm} \exp{\big(\int_{T-\tau_{1}}^{T} d\tau' 
\tilde{W}_{m_{0},m_{0}}(\lambda_{T-\tau'}) \big)} \;,\nonumber
\end{eqnarray}
where $\tau_{N+1}=T$.
Normalization in the reverse path ensemble implies 
$\sum_{\bar{m}_{(\tau)}} \tilde{{\cal P}}[\bar{m}_{(\tau)}] = 1$.\\

We consider the ratio of the two probabilities 
(\ref{caaaa}) and (\ref{caaac}), 
\begin{eqnarray}
r[m_{(\tau)}] &\equiv& 
\ln \frac{{\cal P}[m_{(\tau)}]}{\tilde{{\cal P}}[\bar{m}_{(\tau)}]} 
\label{caaad} \;.
\end{eqnarray}
Due to (\ref{caaaba}), the contributions from the exponentials 
(which represent the probabilities to remain on a given state) 
in (\ref{caaad}) cancel, so that 
\begin{eqnarray}
r[m_{(\tau)}]= \ln \frac{p_{m_0}(0)}{\tilde{p}_{m_N}(0)} +
\sum_{j=1}^{N} \ln \frac{W_{m_{j},m_{j-1}}^{(\nu_{j})}(\lambda_{\tau_j})}
{\tilde{W}_{m_{j-1},m_{j}}^{(\nu_{j})}(\lambda_{\tau_{j}})} 
\;.\label{caaae}
\end{eqnarray}
We can partition (\ref{caaae}) in the form
\begin{eqnarray}
r[m_{(\tau)}]
&=& \ln \frac{p_{m_0}(0)}{\tilde{p}_{m_N}(0)} +
\sum_{j=1}^{N} \ln \frac{p_{m_{j}}^{\rm st}(\lambda_{\tau_j})}
{p_{m_{j-1}}^{\rm st}(\lambda_{\tau_j})} 
\label{caaaf}\\
&&+\bigg(\sum_{j=1}^{N} 
\ln \frac{p_{m_{j-1}}^{\rm st}(\lambda_{\tau_j}) 
W_{m_{j},m_{j-1}}^{(\nu_{j})}(\lambda_{\tau_j})}
{p_{m_{j}}^{\rm st}(\lambda_{\tau_j}) 
\tilde{W}_{m_{j-1},m_{j}}^{(\nu_{j})}(\lambda_{\tau_{j}})}\bigg) \;.
\nonumber
\end{eqnarray} 
We assume for the moment that $r[m_{(\tau)}]$ can be expressed 
exclusively in terms of quantities of the dynamics (\ref{aaaaa})
i.e. a recipe has to be provided to express the tilde 
quantities in (\ref{caaae}) [$\tilde{p}_{m_N}(0)$ and 
$\tilde{W}_{m_{j-1},m_{j}}^{(\nu_{j})}(\lambda_{\tau_{j}})$] 
in terms of non-tilde quantities.\\
In analogy with (\ref{caaad}), we define 
\begin{eqnarray}
\tilde{r}[\bar{m}_{(\tau)}] \equiv
\ln \frac{\tilde{{\cal P}}[\bar{m}_{(\tau)}]}
{{\cal P}[m_{(\tau)}]}
\label{caaag}
\end{eqnarray} 
for the tilde dynamics.
The previous recipe also implies that $\tilde{r}[\bar{m}_{(\tau)}]$ 
can be exclusively expressed in terms of quantities of the tilde 
dynamics (\ref{caaab}).
Eq. (\ref{caaad}) together with (\ref{caaag}) implies that 
$r[m_{(\tau)}]=-\tilde{r}[\bar{m}_{(\tau)}]$.\\

The probability ${\rm P}(R)$ to observe a trajectory such that 
$r[m_{(\tau)}]=R$ during the forward dynamics is related to the 
probability $\tilde{{\rm P}}(-R)$ to observe a trajectory such 
that $\tilde{r}[\bar{m}_{(\tau)}]=-R$ during the backward dynamics
\begin{eqnarray}
{\rm P}(R) &\equiv& 
\sum_{m_{(\tau)}} {\cal P}[m_{(\tau)}] \delta(R-r[m_{(\tau)}]) 
\label{caaah}\\
&=& \sum_{m_{(\tau)}} \tilde{{\cal P}}[\bar{m}_{(\tau)}] 
\e^{r[m_{(\tau)}]} \delta(R-r[m_{(\tau)}]) \nonumber \\
&=&\e^{R} \sum_{m_{(\tau)}} \tilde{{\cal P}}[\bar{m}_{(\tau)}] 
\delta(R-r[m_{(\tau)}])\nonumber \\
&=&\e^{R} \sum_{\bar{m}_{(\tau)}} \tilde{{\cal P}}[\bar{m}_{(\tau)}] 
\delta(R+\tilde{r}[\bar{m}_{(\tau)}]) \nonumber \\
%&=&\e^{R} \sum_{m_{(\tau)}} \tilde{{\cal P}}[m_{(\tau)}] 
%\delta(R+\tilde{r}[m_{(\tau)}]) \nonumber \\
&=&\e^{R} \tilde{{\rm P}}(-R) \nonumber \;.
\end{eqnarray}
By integrating $\e^{-R}{\rm P}(R)=\tilde{{\rm P}}(-R)$ 
over $R$, we get 
\begin{eqnarray}
\mean{\e^{-r[m_{(\tau)}] }}=1 \label{caaai} \;.
\end{eqnarray}
It follows from Jensen's inequality 
$\mean{\e^{x}} \geq \e^{\mean{x}}$, 
that $\mean{r[m_{(\tau)}] } \geq 0$.\\

We now make a first choice of $\tilde{W}_{m,m'}(\lambda_t)$ 
in the backward dynamics (\ref{caaab})
\begin{eqnarray}
\tilde{W}_{m,m'}^{(\nu)}(\lambda_{t})=W_{m,m'}^{(\nu)}(\lambda_{t}) \;.
\label{caaaoa}
\end{eqnarray}
In this case the backward dynamics is identical to the original 
one, except that the driving protocol is time reversed.
If we also choose the initial conditions of the backward 
dynamics to be the final conditions of the forward dynamics 
$\tilde{p}_{m}(0) = p_{m}(T)$, using (\ref{caaae}) 
and (\ref{trajentr_aaaaf}), we find 
\begin{eqnarray}
r[m_{(\tau)}] = \Delta s_{tot} [m_{(\tau)},T] \;.
\label{caaao}
\end{eqnarray}
The FT (\ref{FTi}) previously derived using GFs now 
follows from Eq. (\ref{caaai}).
Using (\ref{caaah}), we also get the detailed form of the FT 
\begin{eqnarray}
\frac{{\rm P}(\Delta S_{tot})}
{\tilde{{\rm P}}(-\Delta S_{tot})} = \e^{\Delta S_{tot}} \;.
\label{caaap}
\end{eqnarray}

We now make a second choices of $\tilde{W}_{m,m'}(\lambda_t)$ 
in (\ref{caaab})
\begin{eqnarray}
\tilde{W}_{m,m'}^{(\nu)}(\lambda_{t}) &=&
W_{m',m}^{(\nu)}(\lambda_{t}) \frac{p_{m}^{\rm st}(\lambda_{t})}
{p_{m'}^{\rm st}(\lambda_{t})} \label{caaax} \;.
\end{eqnarray}
In the theory of MEs, (\ref{caaab}) with (\ref{caaax}) 
is called the time reversal ME of (\ref{aaaaa}) 
\cite{Norris,Stirzaker}. 
We again choose the initial condition of the tilde dynamics 
to be the final conditions of the original dynamics 
$\tilde{p}_{m}(0) = p_{m}(T)$. 
Using (\ref{caaaf}) with (\ref{caaax}) and (\ref{trajentr_aaaaj}), 
we get
\begin{eqnarray}
r[m_{(\tau)}] = \Delta s_{na}[m_{(\tau)},T] \;.
\label{caaay}
\end{eqnarray}
The previously derived FT (\ref{FTx}) follows now from (\ref{caaai}).
From Eq. (\ref{caaah}) we find the detailed form of the FT
\begin{eqnarray}
\frac{{\rm P}(\Delta S_{na})}
{\tilde{{\rm P}}(-\Delta S_{na})} = \e^{\Delta S_{na}} \;.
\label{caaaz}
\end{eqnarray}
We can interpret the change in the total TEP during the $0$ to $T$ 
time interval as the logarithm of the (forward) probability 
that the driven system follows a given trajectory divided by the backward 
probability that the system, initially in the final probability distribution 
of the forward evolution, and driven in a time reversed way compared to the 
forward evolution, follows the time-reversed trajectory.\\  
The nonadiabatic TEP is interpreted as the logarithm of the (forward) probability 
that the driven system follows a given trajectory divided by the backward probability 
that the system, initially in the final probability distribution of the forward 
evolution, and described by the time-reversed ME, follows the time-reversed trajectory.\\
It should be noted that the backward ME (\ref{caaab}) with (\ref{caaax}) is 
different from the backward ME (\ref{caaab}) with (\ref{caaaoa}) only for systems 
interacting with multiple reservoirs which break the DBC. 
Only in this case (\ref{FTi}) is different from (\ref{FTx}).

%%%%%%%%%%%%%%%%%%%%%%%%%%%%%%%%%%%%%%%%%%%%%%%%%%%%%%%%%%%%%%%%%%%%%%%%%%%%%%%%%%%%%%%

%%%%%%%%%%%%%%%%%%%%%%%%%%%%%%%%%%%%%%%%%%%%%%%%%%%%%%%%%%%%%%%%%%%%%%%%%%%%%%%%%%%%%%
%%%%%%%%%%%%%%%%%%%%%%%%%
\begin{figure}[p]
\centering
\begin{tabular}{c@{\hspace{0.5cm}}c@{\hspace{0.5cm}}c}
\rotatebox{0}{\scalebox{0.25}{\includegraphics{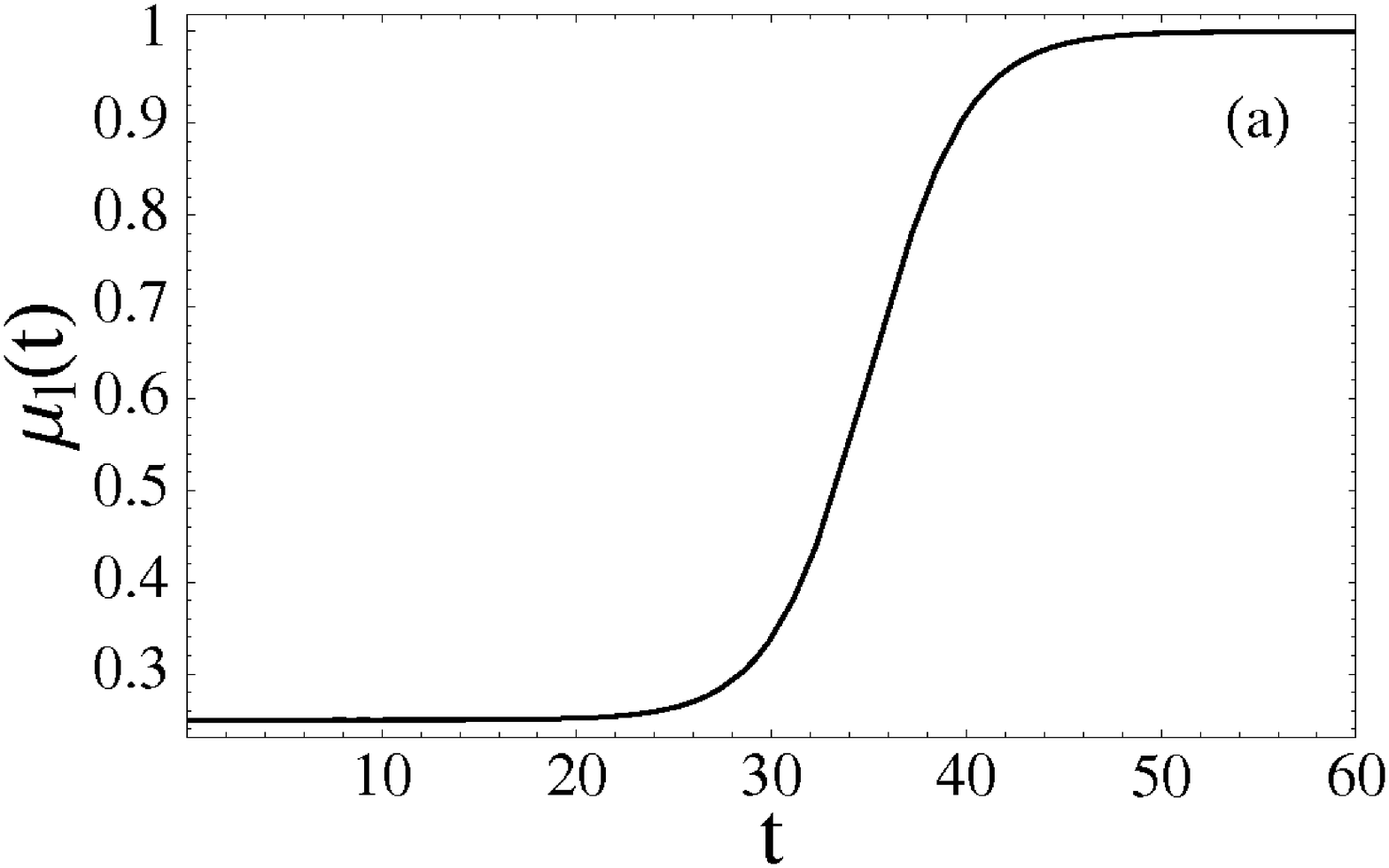}}} &
\rotatebox{0}{\scalebox{0.25}{\includegraphics{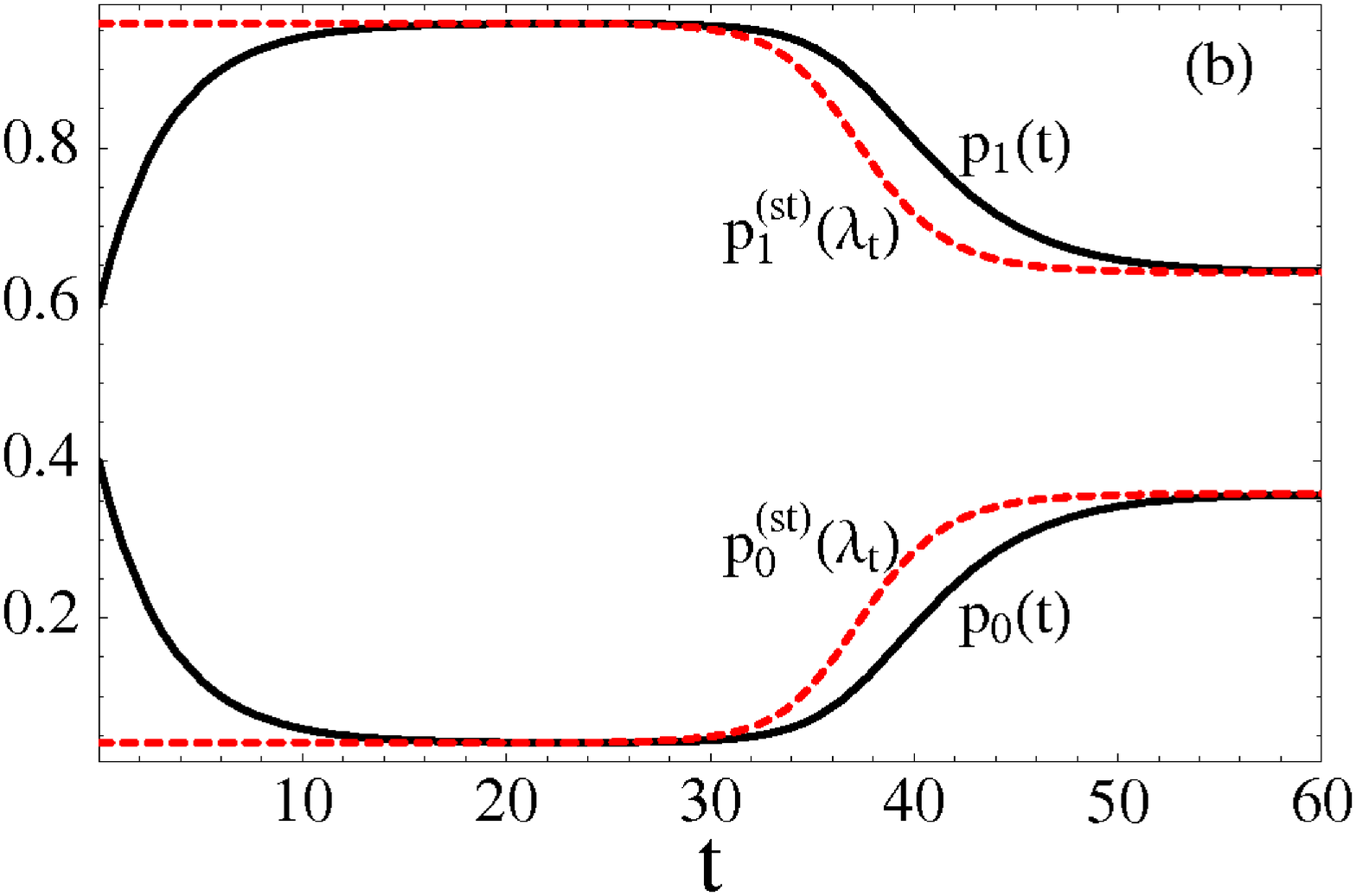}}} \\
\rotatebox{0}{\scalebox{0.25}{\includegraphics{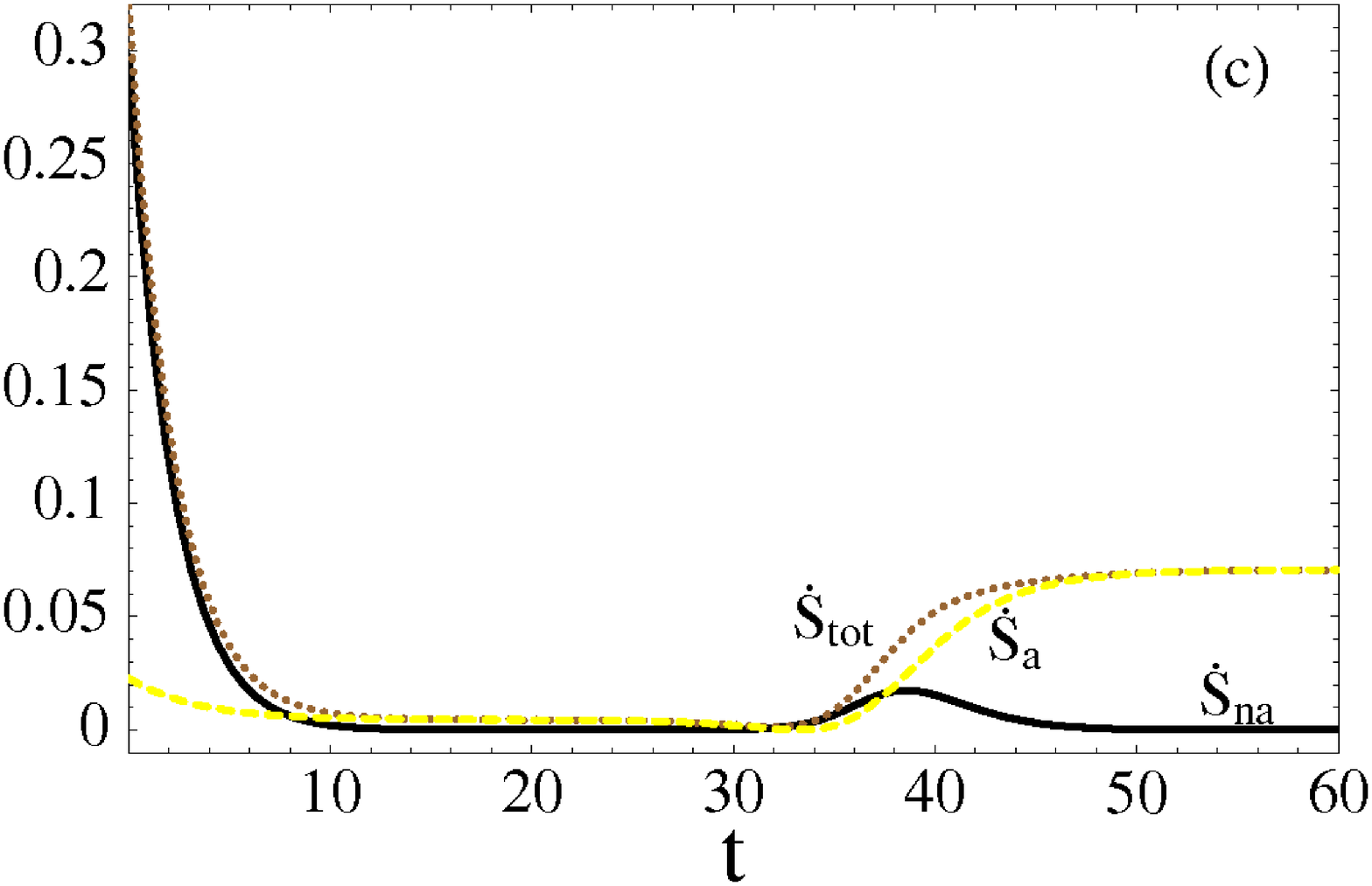}}} &
\rotatebox{0}{\scalebox{0.25}{\includegraphics{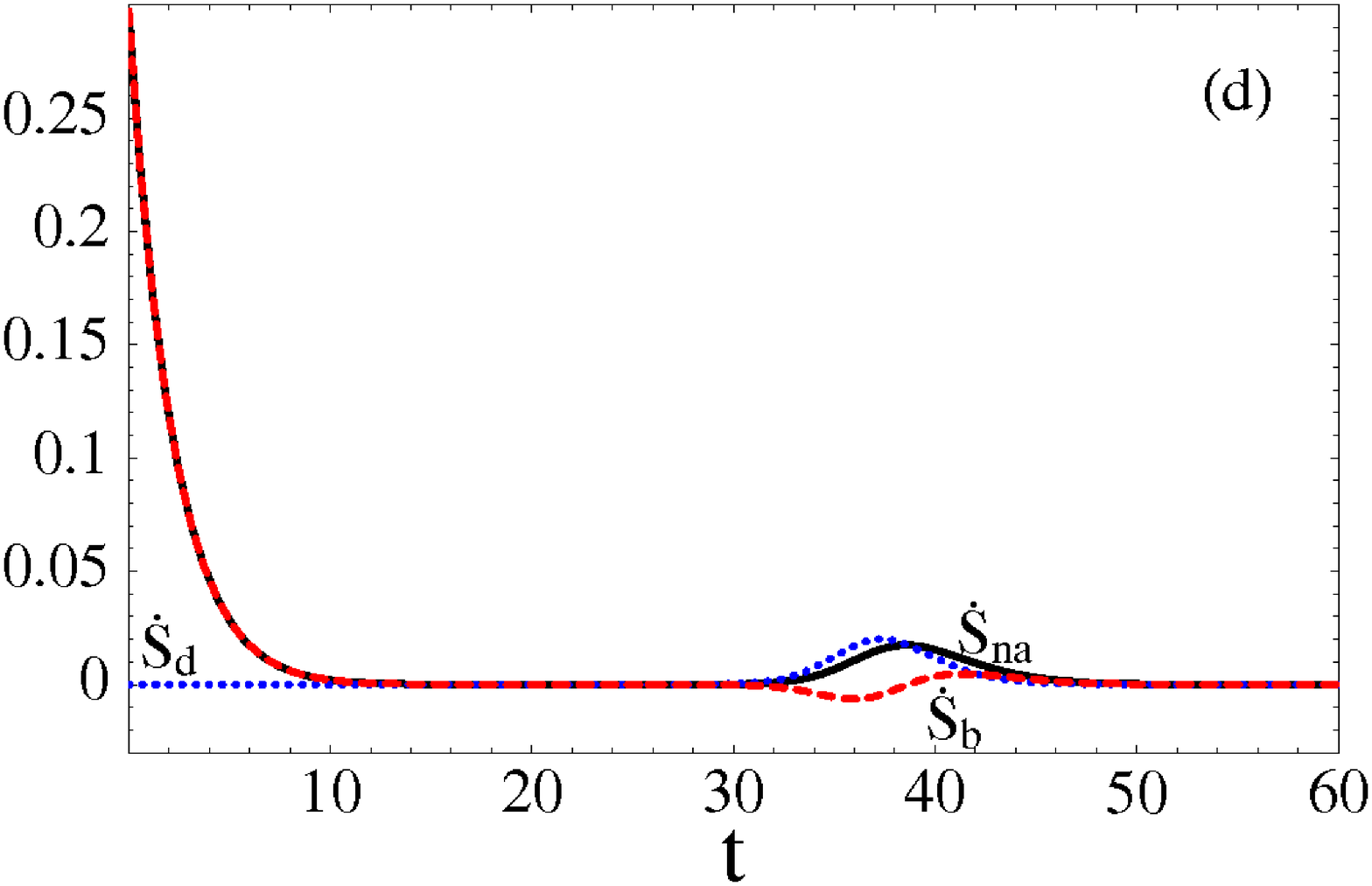}}} \\
\rotatebox{0}{\scalebox{0.25}{\includegraphics{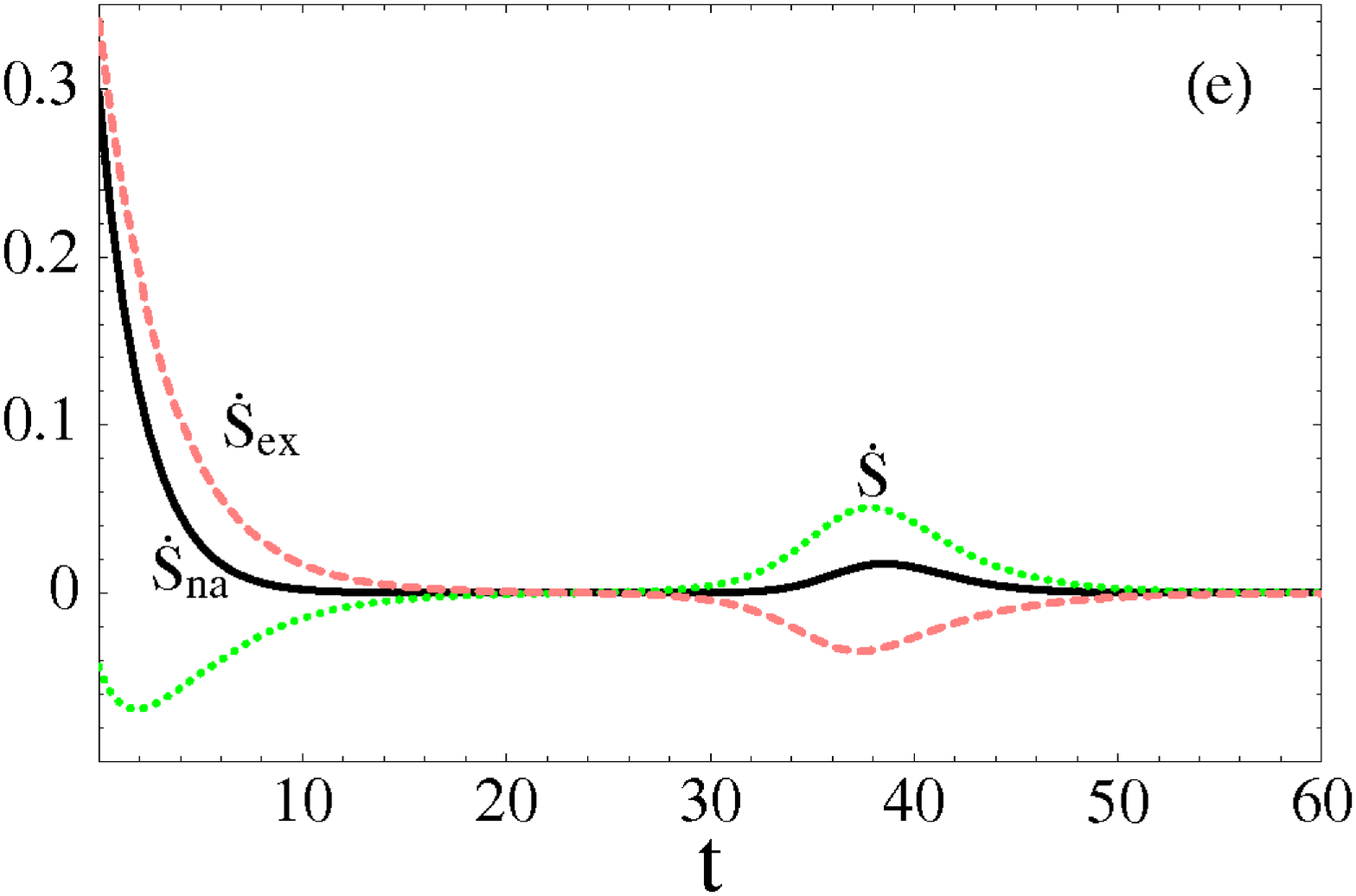}}} &
\rotatebox{0}{\scalebox{0.25}{\includegraphics{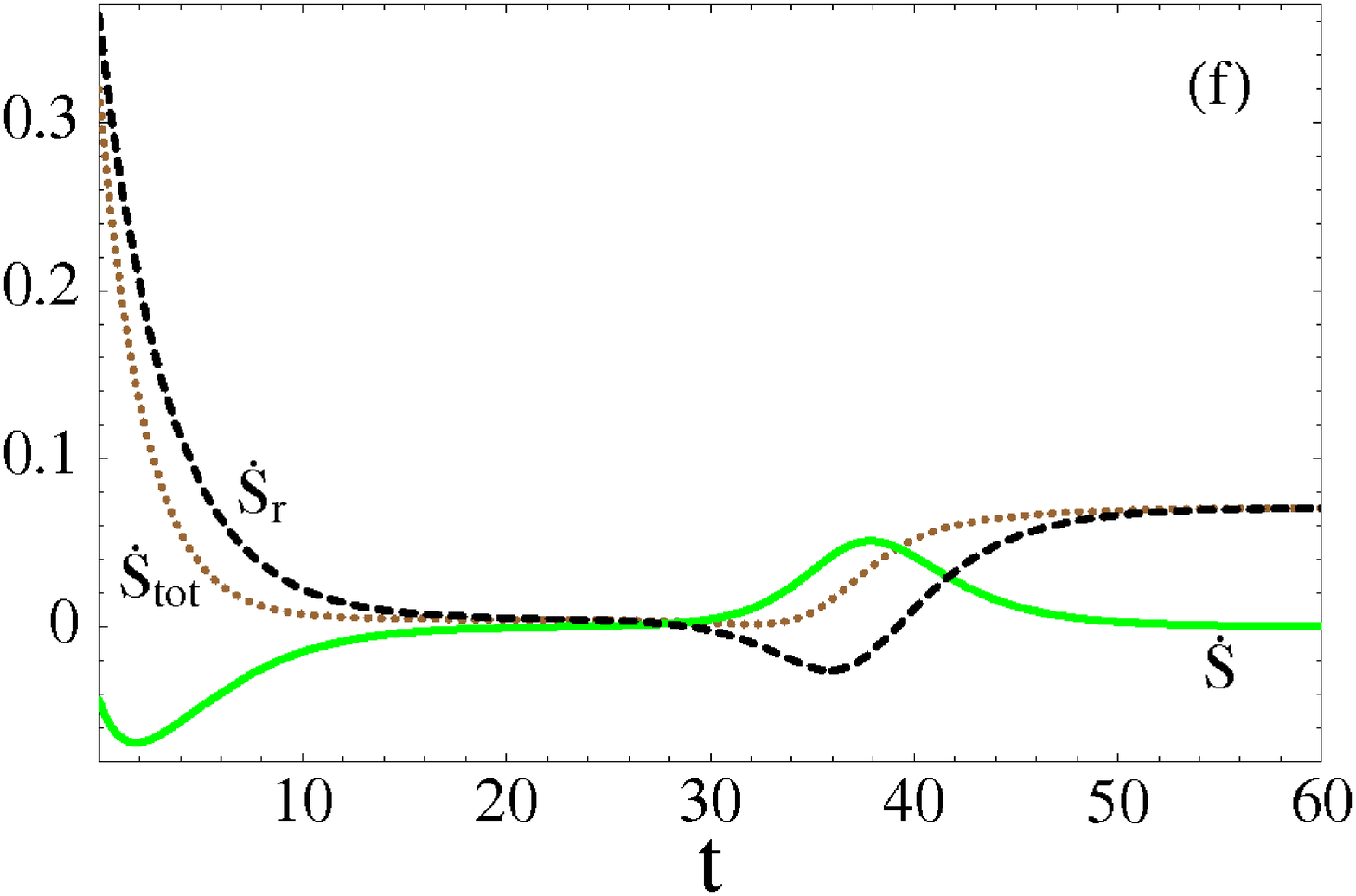}}} 
\end{tabular}
\caption{(Color online)
(a) Driving protocol of the left lead chemical potential $\mu_{l}(t)= \mu_{0} + V(t)$
where $V(t)$ follows (\ref{potential_profile}) with 
$V_i=-0.25$, $V_f=0.5$, $c=0.2$, $t_m=0.2$ and $\mu_{0} = 0.5$.
The right lead chemical potential remains constant at $\mu_{r}= \mu_{0}$. 
(b) Solid: The probability distribution of the dot obtained by
solving the ME (\ref{yaaaa}) with initial condition $p_{0}(0)=0.4$ and $p_{1}(0)=0.6$.
Dotted: The adiabatic probability distribution.
(c) Decomposition of $\dot{S}_{tot}(t)$ according to (\ref{baaai}).
(d) Decomposition of $\dot{S}_{na}(t)$ according to  (\ref{baaak}).
(e) Decomposition of $\dot{S}_{na}(t)$ according to (\ref{baaakk}).
(f) Decomposition of $\dot{S}_{tot}(t)$ according to (\ref{baaab}).}
\label{aver}
\end{figure}
%%%%%%%%%%%%%%%%%%%%%%%%%
%%%%%%%%%%%%%%%%%%%%%%%%%
\begin{figure}[p]
\centering
\rotatebox{0}{\scalebox{0.3}{\includegraphics{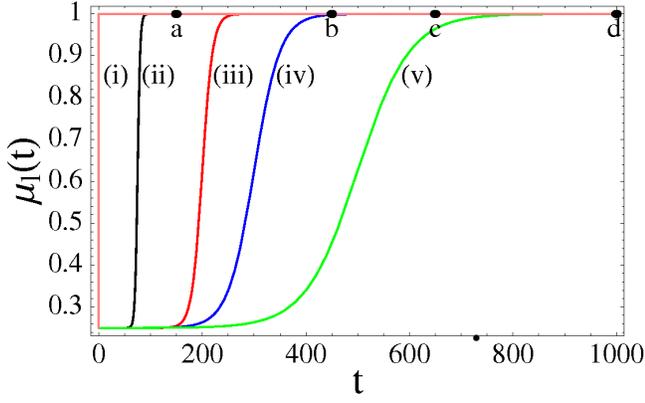}}}
\caption{(Color online) 
Five driving protocols for the left lead chemical potential $\mu_{l}(t)= \mu_{0} + V(t)$
where $V(t)$ follows (\ref{potential_profile}) with $V_i=-0.25$, $V_f=0.5$ and
(ii) $c=0.2$, $t_m=75$, (iii) $c=0.05$, $t_m=200$, (iv) $c=0.02$, $t_m=300$ 
(v) $c=0.01$, $t_m=500$. $\mu_{r}(t)= \mu_{0}=0.5$. 
(i) is the sudden switch limit $\mu_{l}(t)= \mu_{0} + \Theta(t) V_f$ 
[limit $c \to \infty$ with $t_m=0$ in (\ref{potential_profile})].
The system is initially at steady state where $p_0^{st}=0.96$ and $p_1^{st}=0.04$.
In the final steady state $p_0^{st}=0.64$ and $p_1^{st}=0.36$.
$a,b,c,d$ correspond to different measurement times.
In all calculations $\beta=5$, $\epsilon=1$, $a_{l}=0.2$ and $a_{r}=0.1$.}
\label{profilefig}
\end{figure}
%%%%%%%%%%%%%%%%%%%%%%%%%
%%%%%%%%%%%%%%%%%%%%%%%%%
\begin{figure}[p]
\centering
\rotatebox{0}{\scalebox{0.3}{\includegraphics{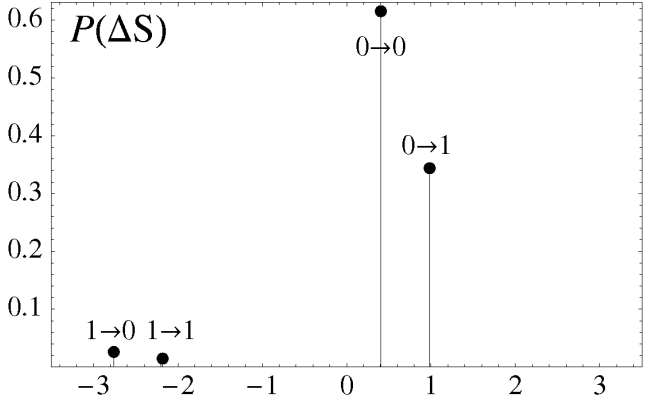}}}
\caption{
Probability distribution of the change in the system TEP 
for the protocols shown in Fig. \ref{profilefig}.} 
\label{PS}
\end{figure}
%%%%%%%%%%%%%%%%%%%%%%%%%
%%%%%%%%%%%%%%%%%%%%%%%%%
\begin{figure}[p]
\centering
\begin{tabular}{c@{\hspace{0.5cm}}c@{\hspace{0.5cm}}c}
\rotatebox{0}{\scalebox{0.25}{\includegraphics{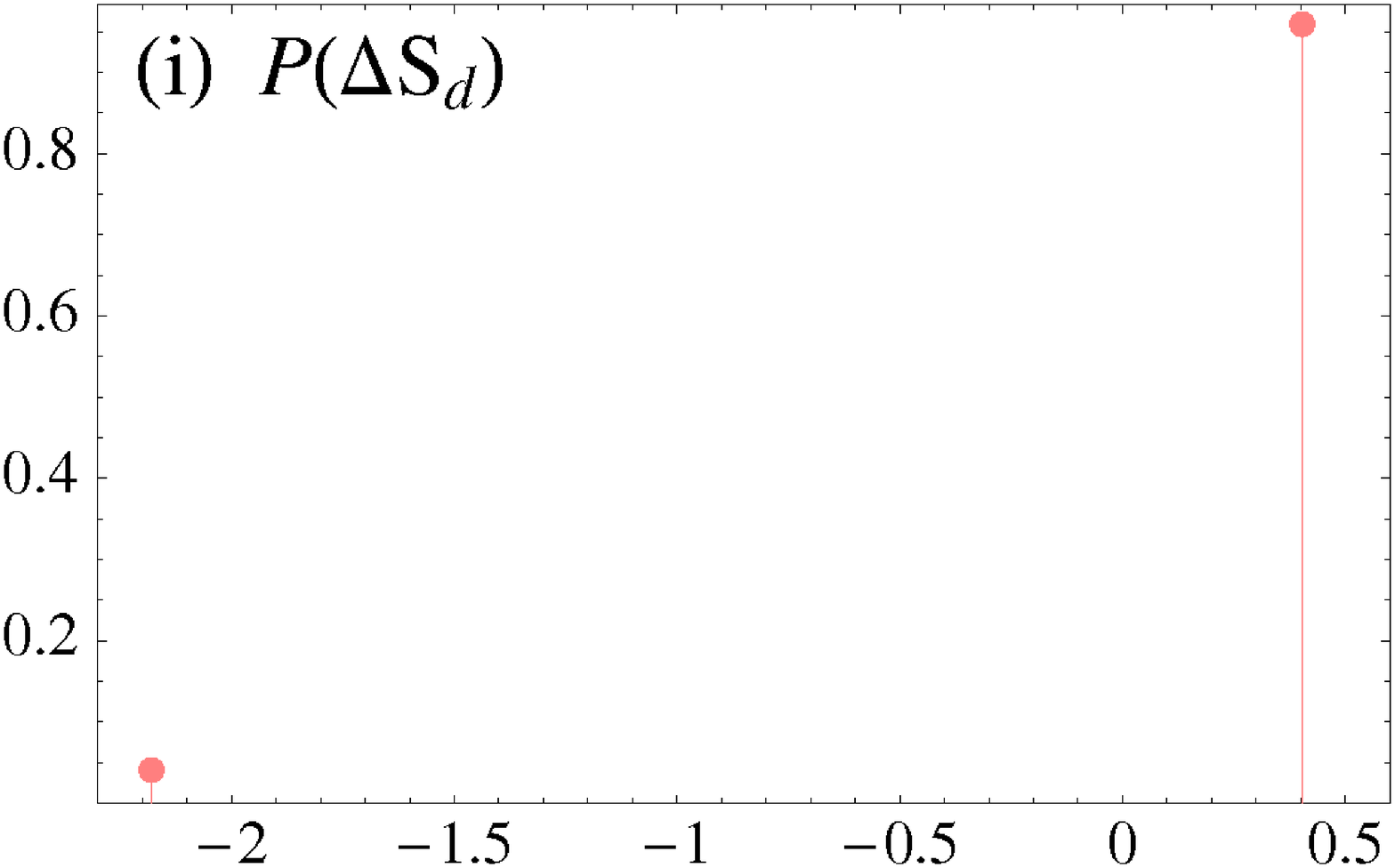}}} &
\rotatebox{0}{\scalebox{0.25}{\includegraphics{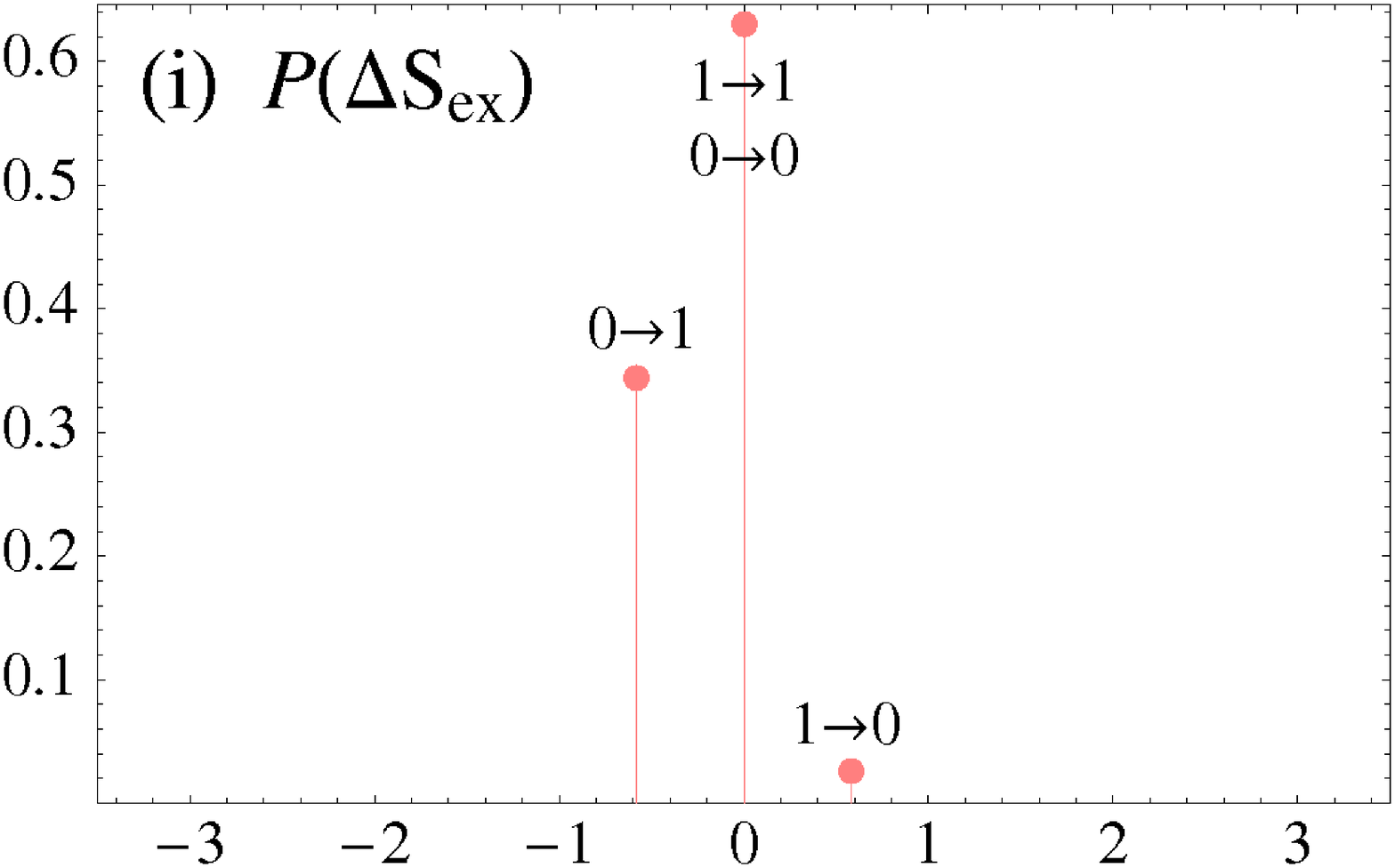}}} \\
\rotatebox{0}{\scalebox{0.25}{\includegraphics{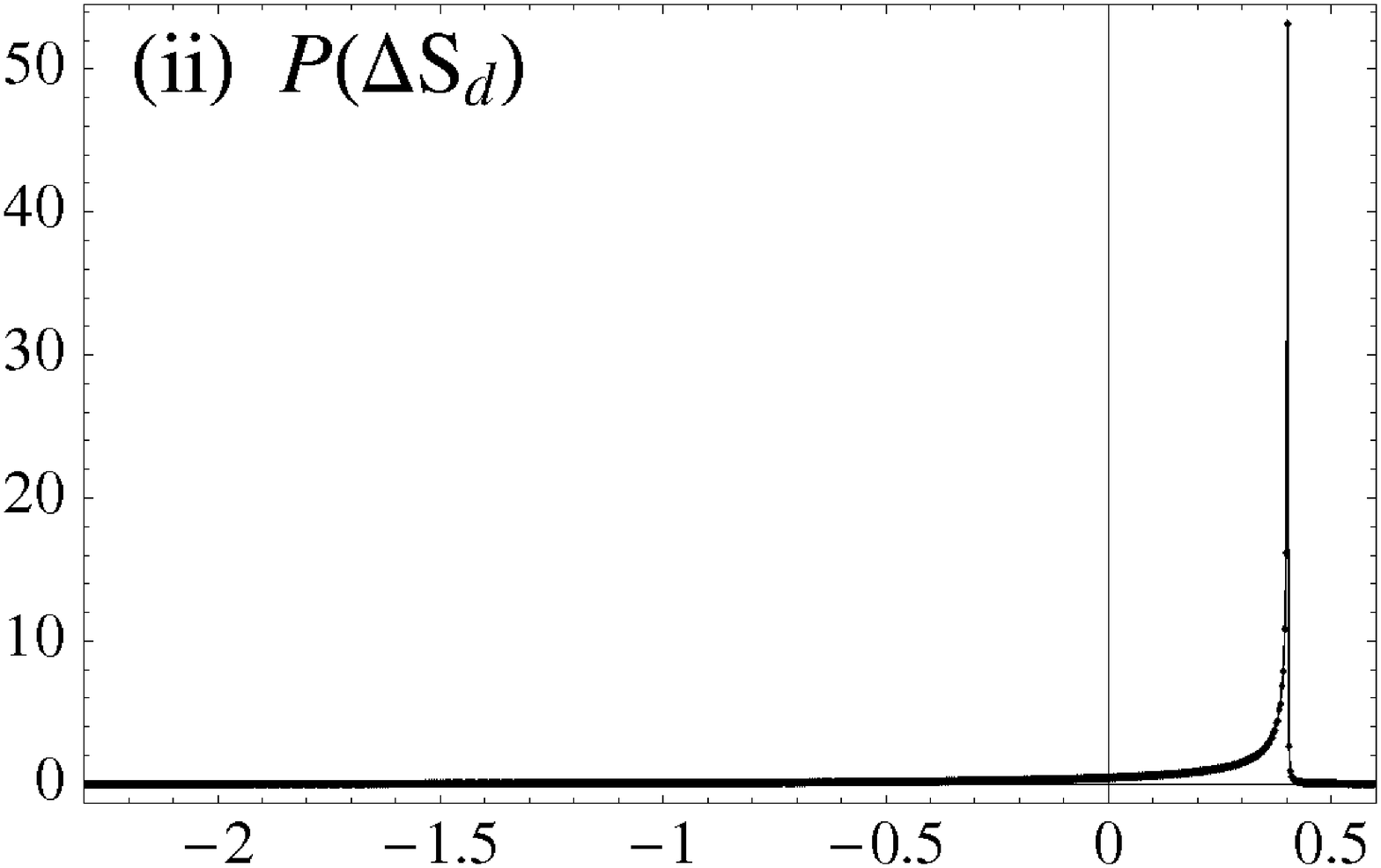}}} &
\rotatebox{0}{\scalebox{0.25}{\includegraphics{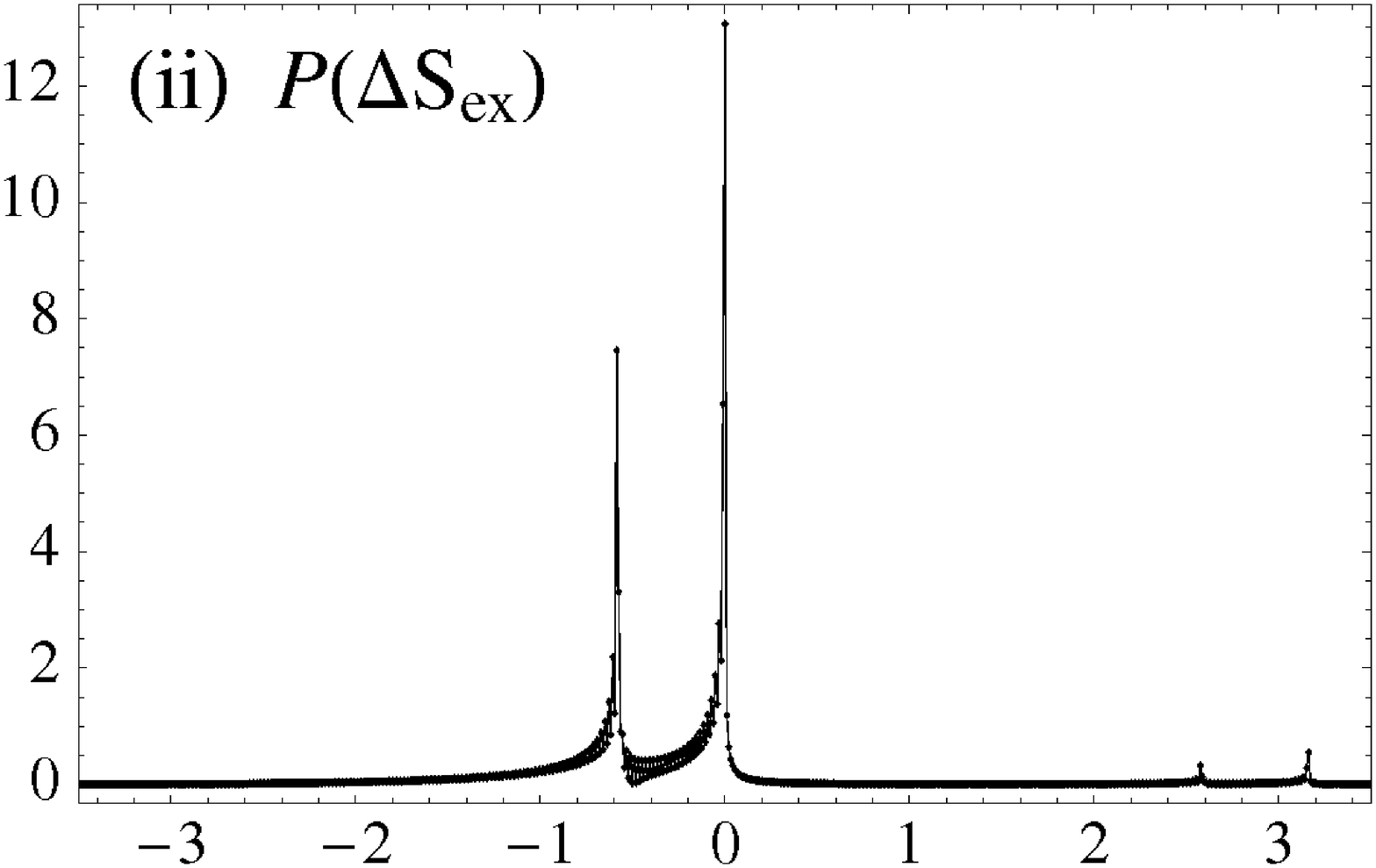}}} \\
\rotatebox{0}{\scalebox{0.25}{\includegraphics{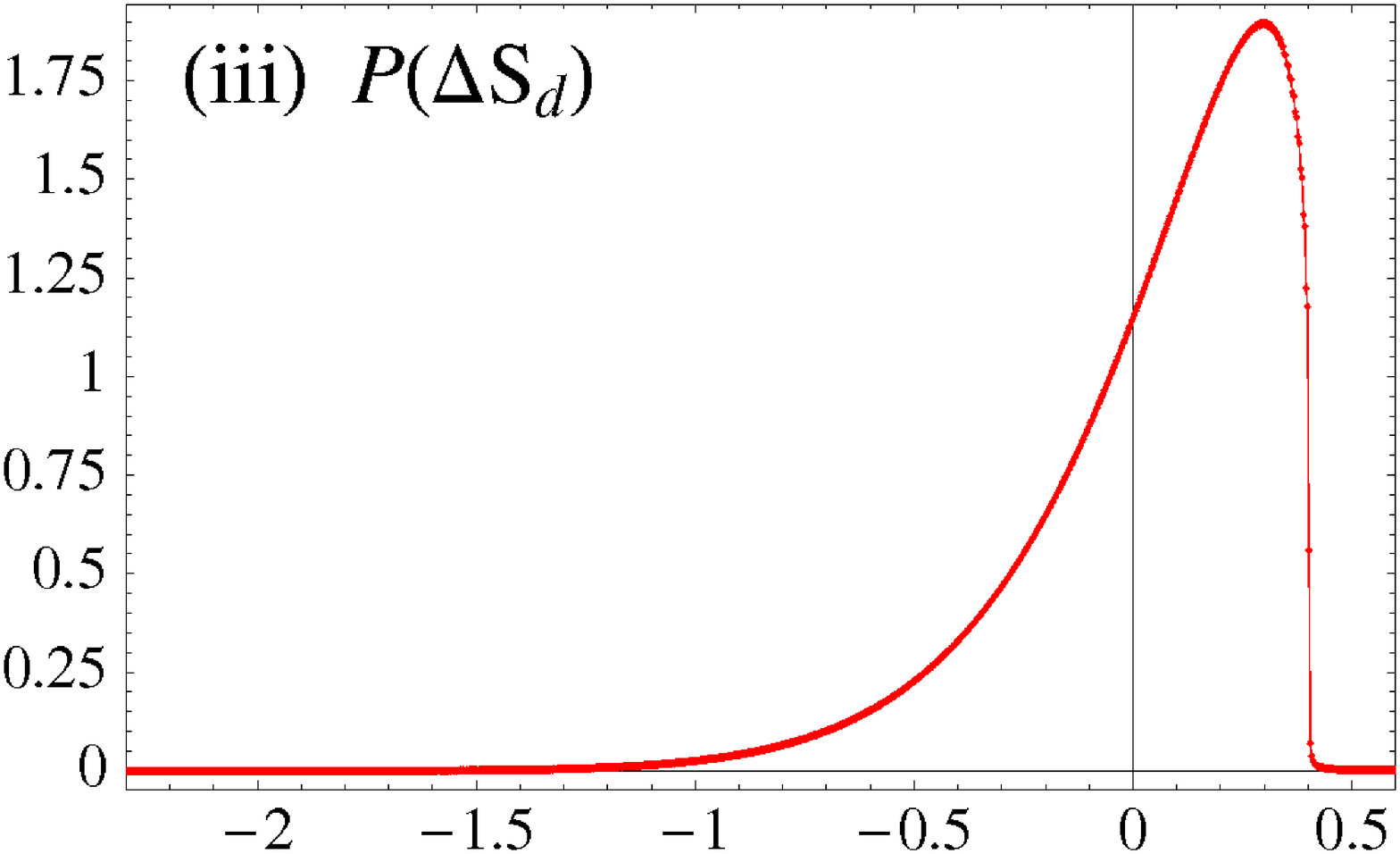}}} &
\rotatebox{0}{\scalebox{0.25}{\includegraphics{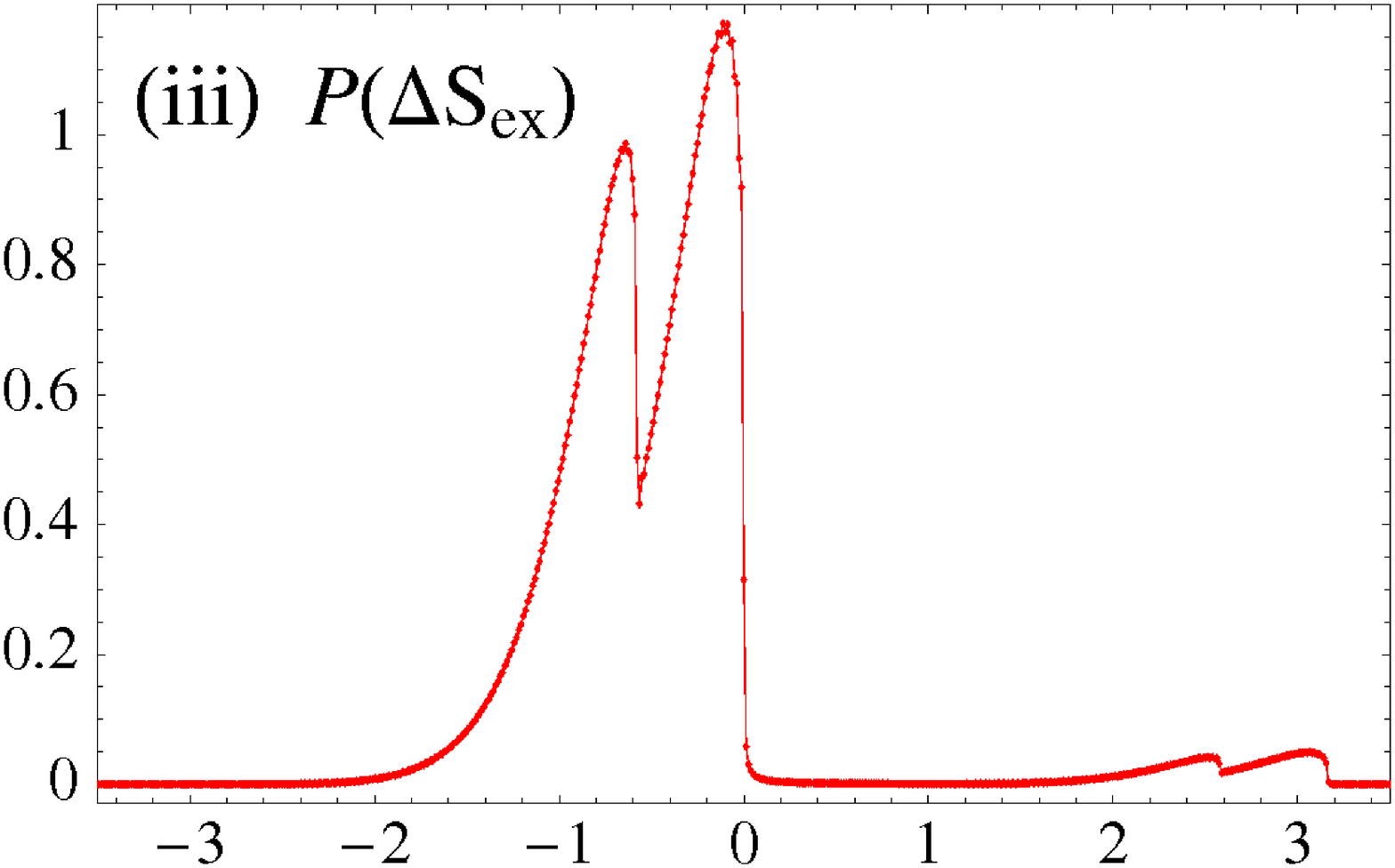}}} \\
\rotatebox{0}{\scalebox{0.25}{\includegraphics{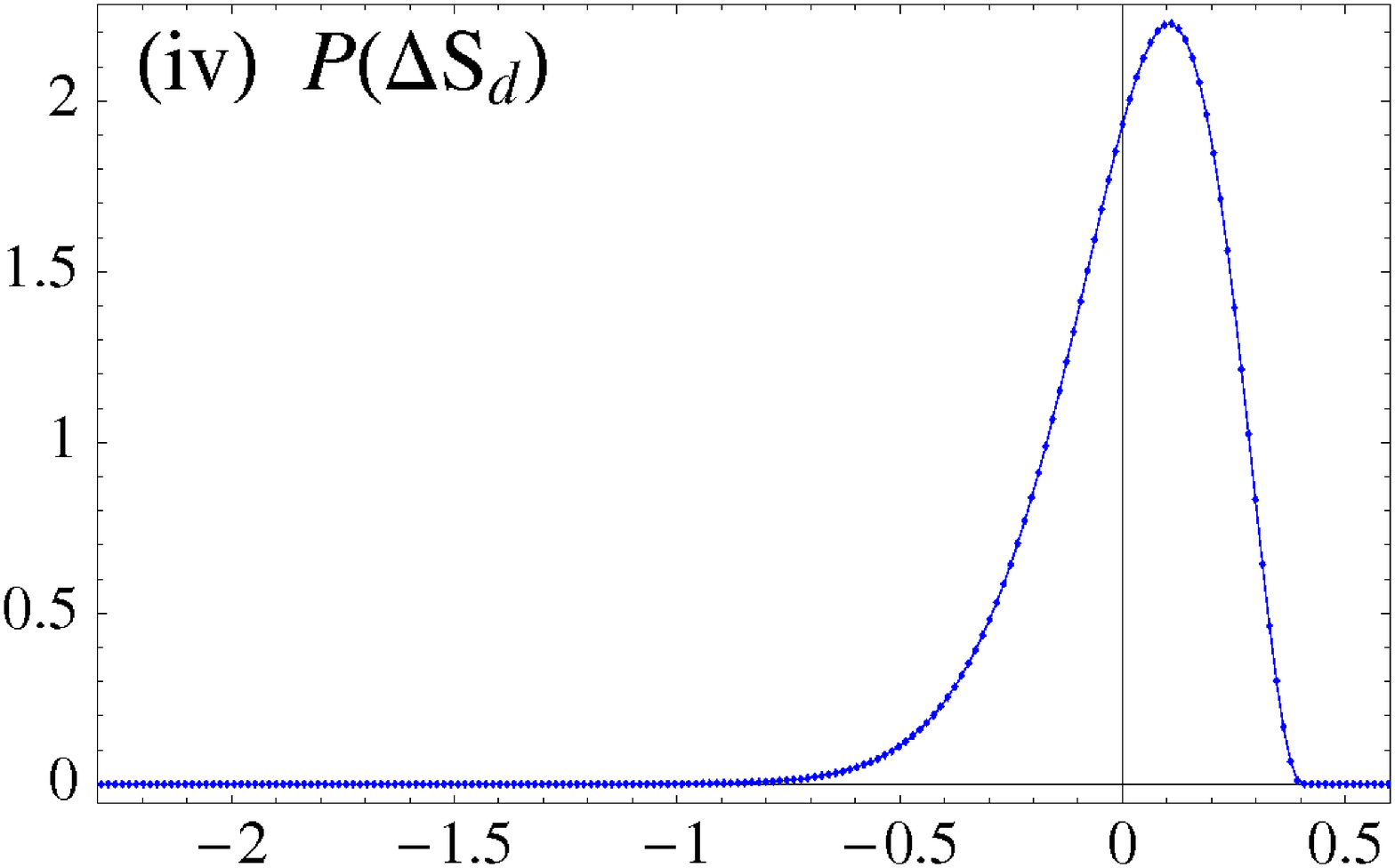}}} &
\rotatebox{0}{\scalebox{0.25}{\includegraphics{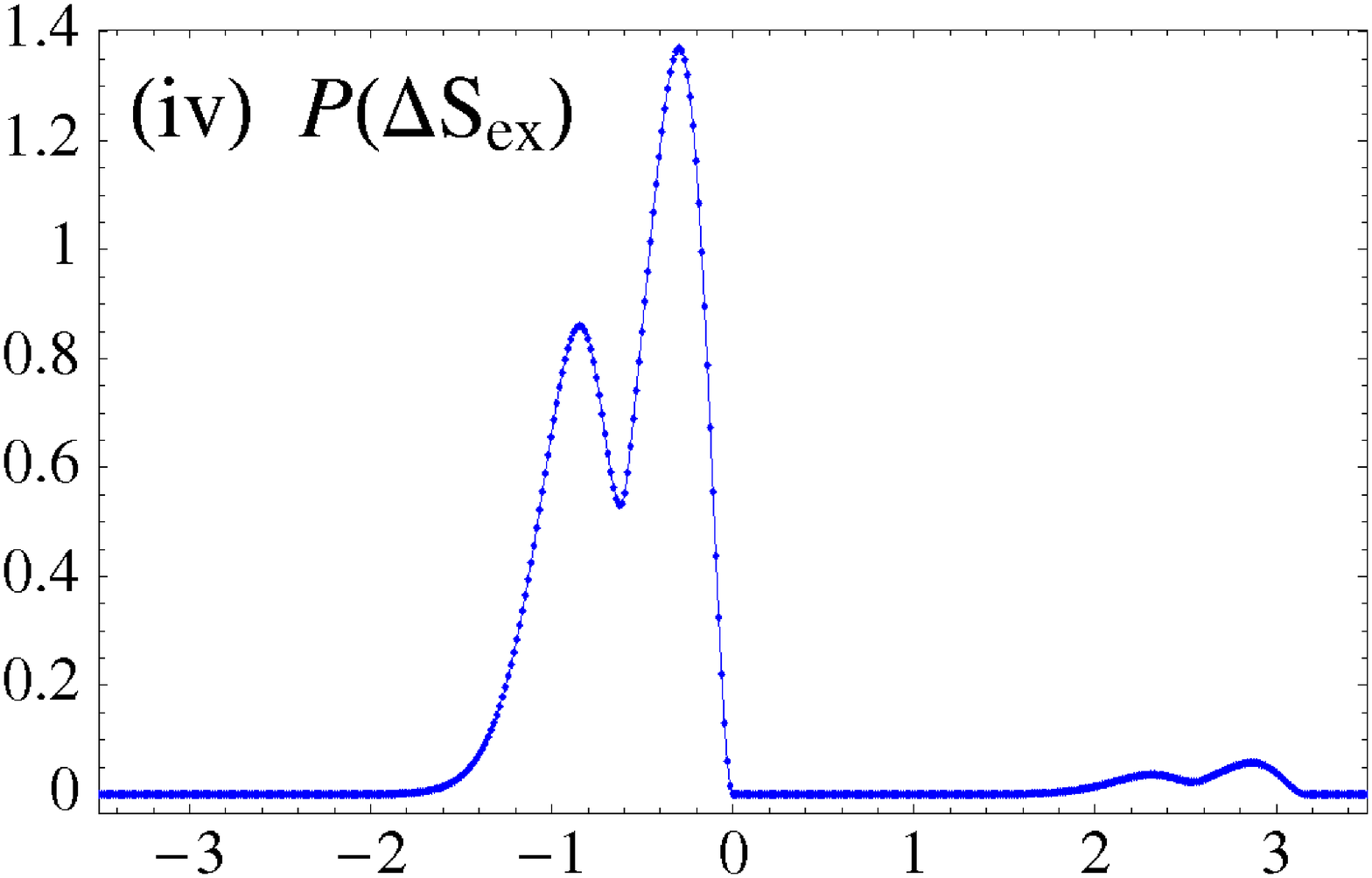}}} \\
\rotatebox{0}{\scalebox{0.25}{\includegraphics{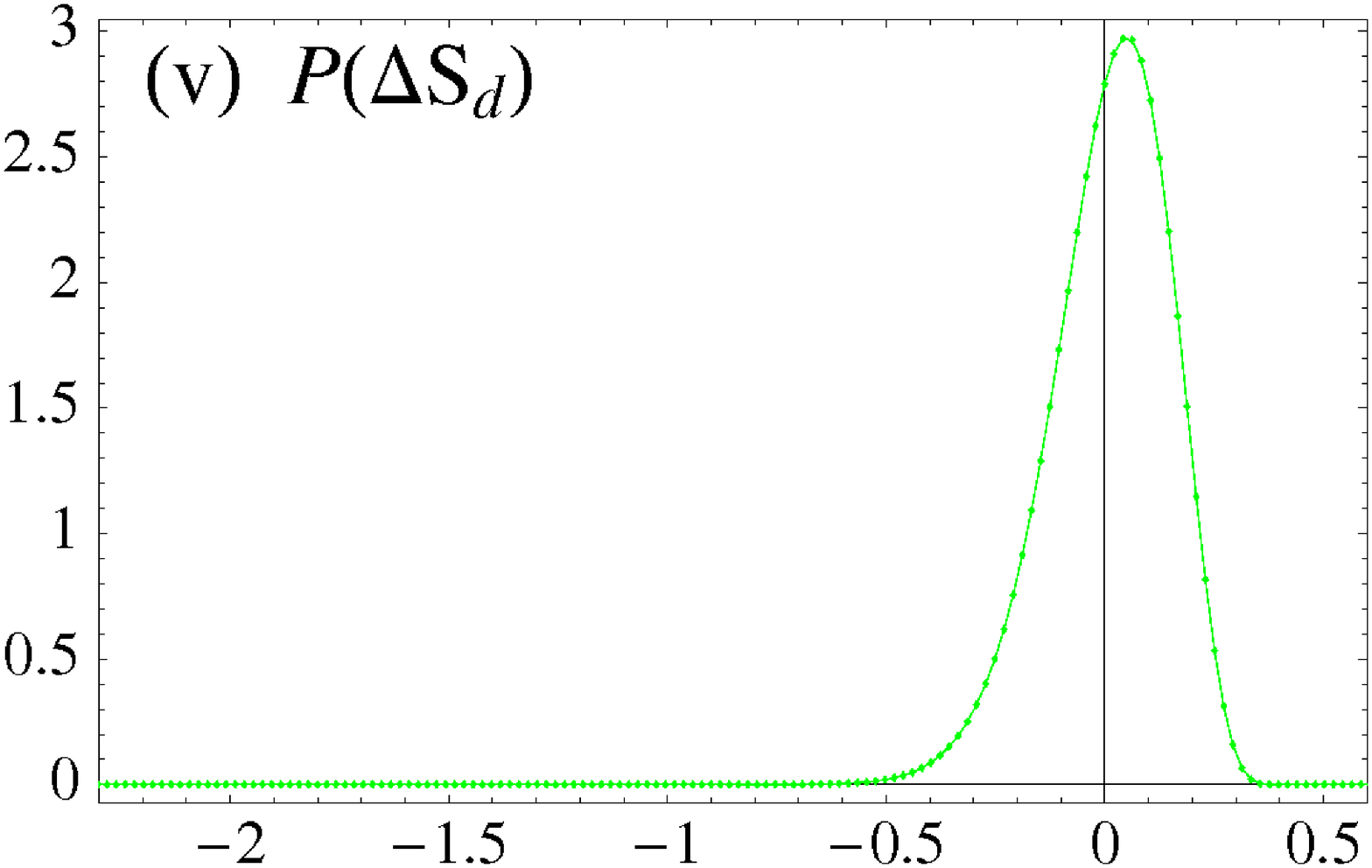}}} &
\rotatebox{0}{\scalebox{0.25}{\includegraphics{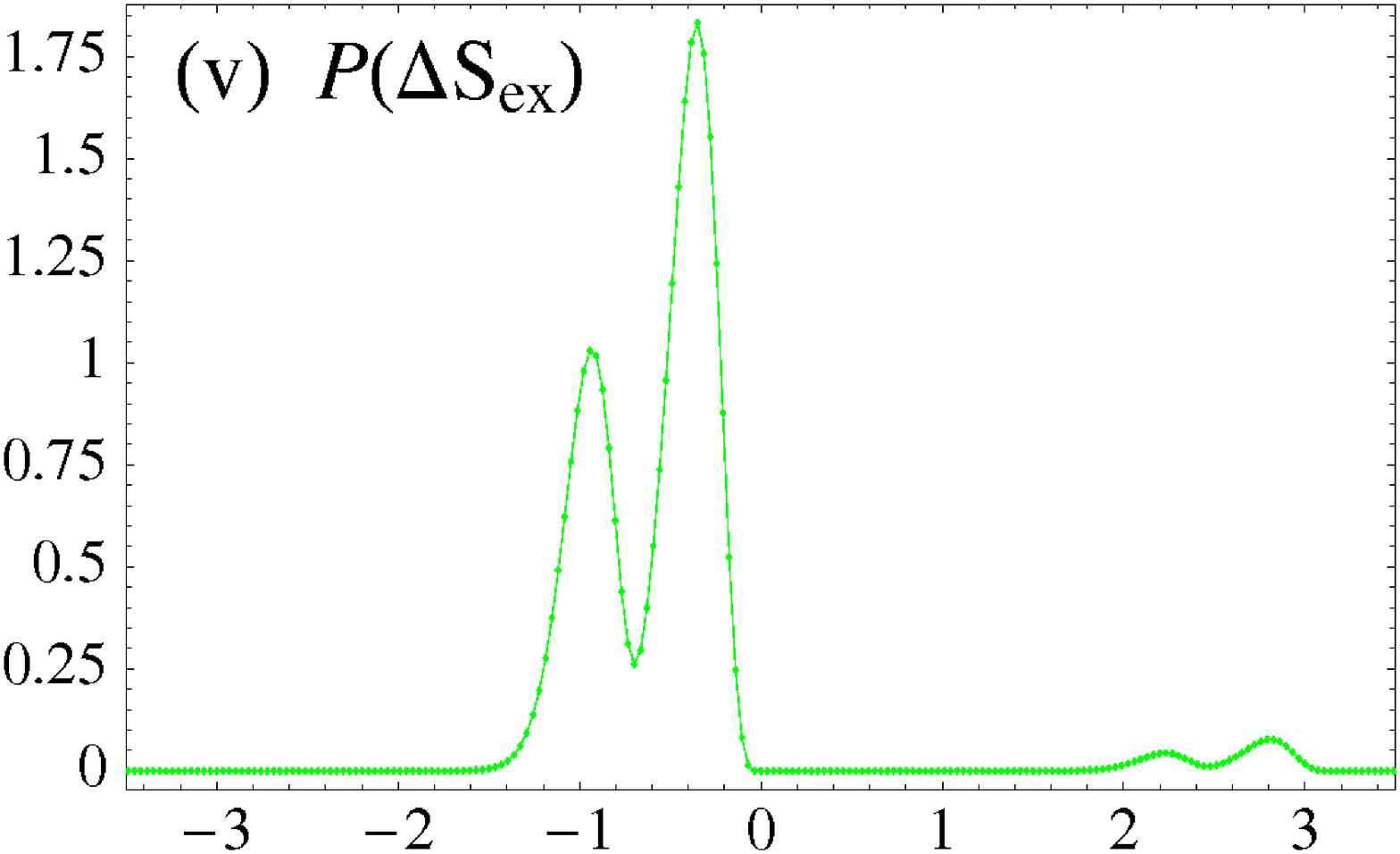}}} 
\end{tabular}
\caption{(Color online)
Probability distributions of the change in the excess
and driving TEP for the protocols in Fig. \ref{profilefig}.
All curves on the left column satisfy the FT (\ref{FTx}).}
\label{PSexPSd}
\end{figure}
%%%%%%%%%%%%%%%%%%%%%%%%%
%%%%%%%%%%%%%%%%%%%%%%%%%
\begin{figure}[p]
\centering
\begin{tabular}{c@{\hspace{0.5cm}}c@{\hspace{0.5cm}}c}
\rotatebox{0}{\scalebox{0.25}{\includegraphics{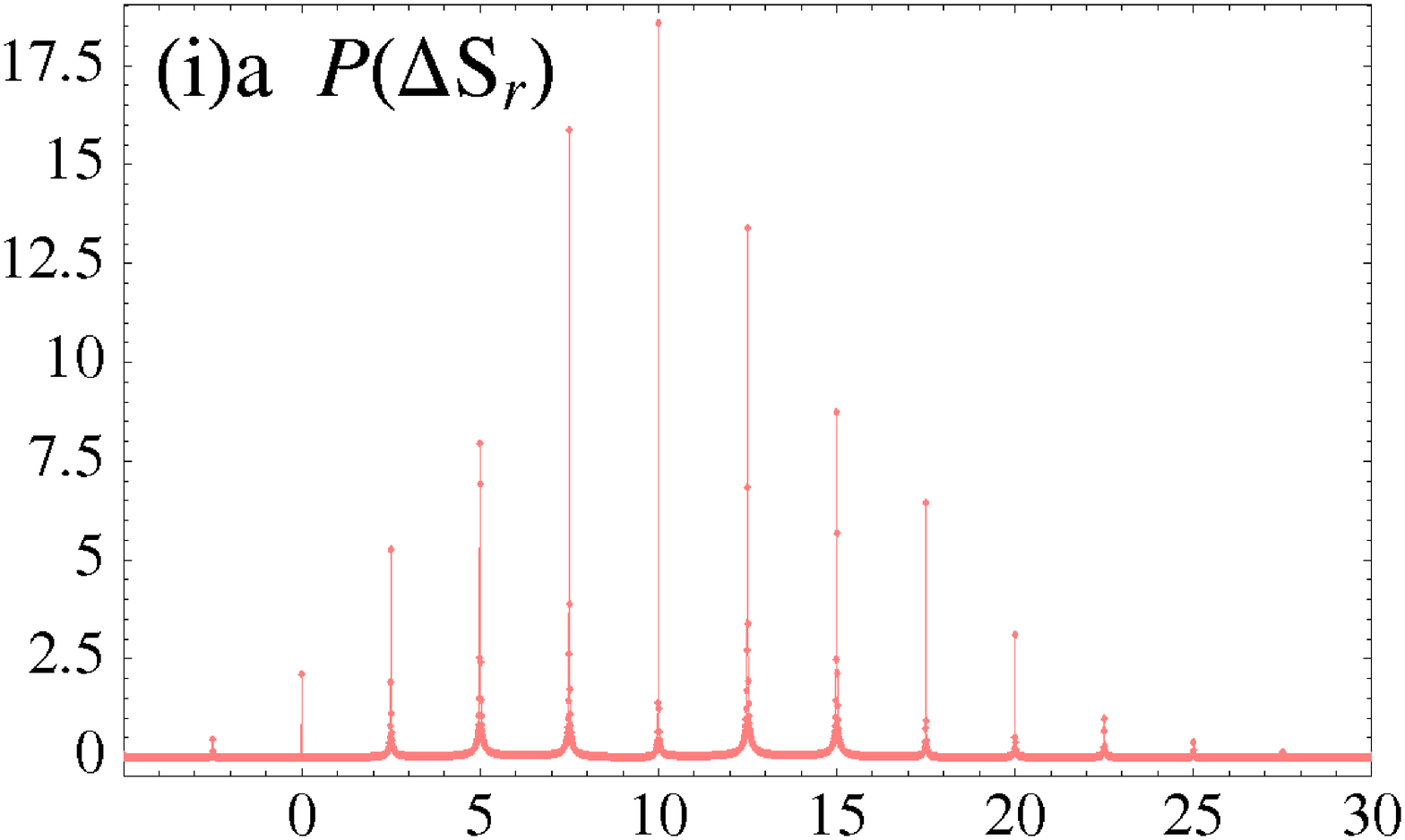}}} &
\rotatebox{0}{\scalebox{0.25}{\includegraphics{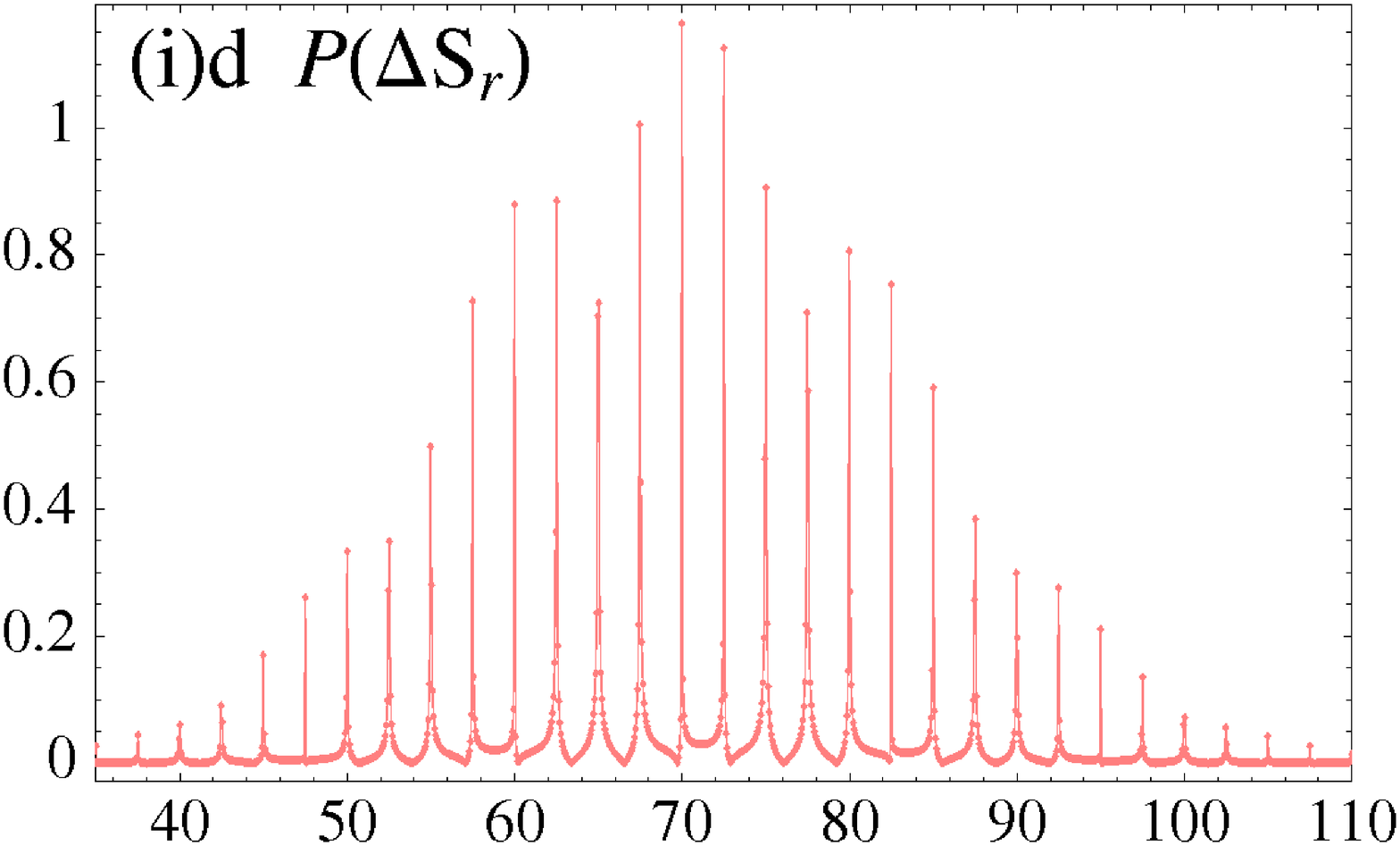}}} \\
\rotatebox{0}{\scalebox{0.25}{\includegraphics{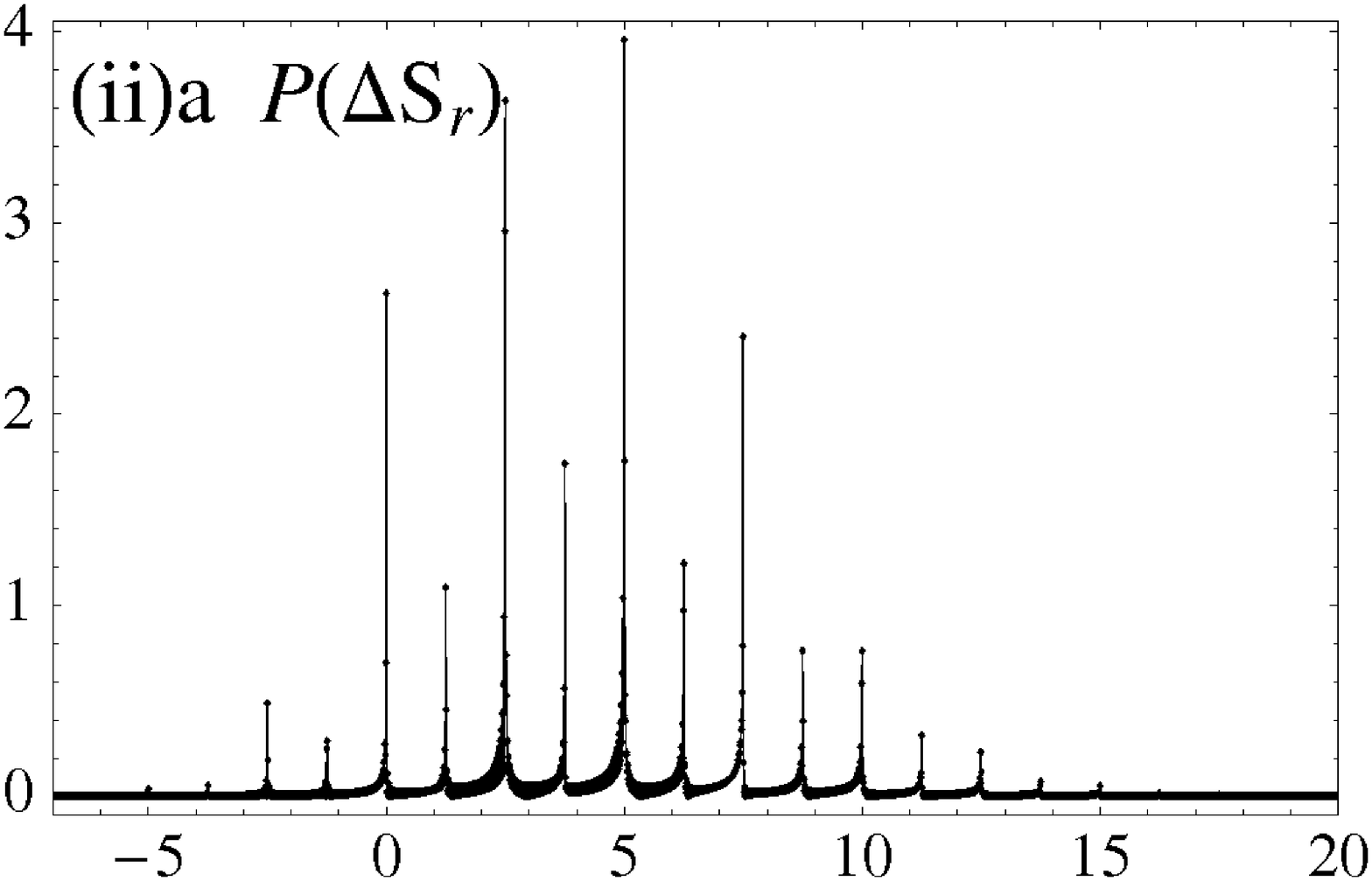}}} &
\rotatebox{0}{\scalebox{0.25}{\includegraphics{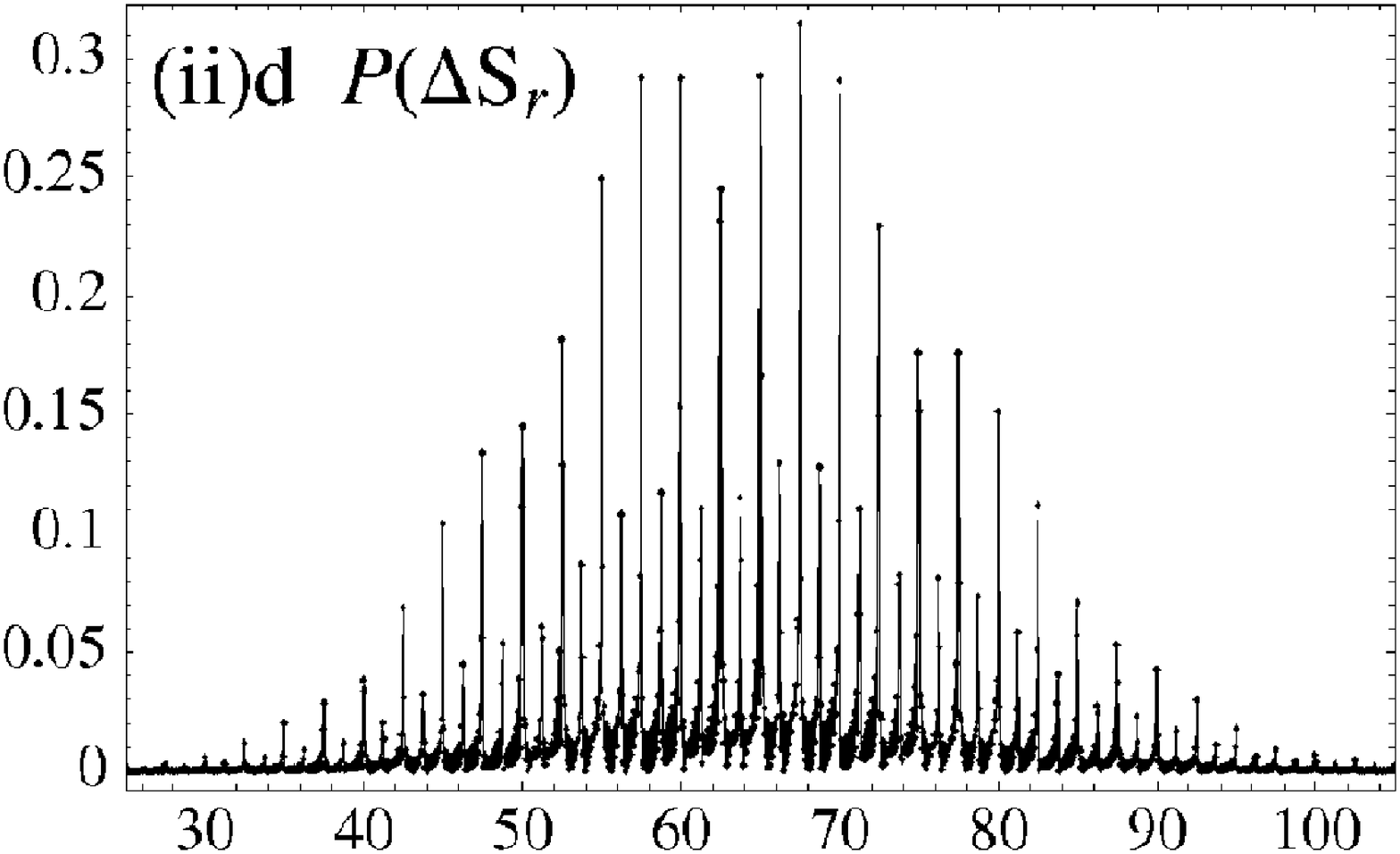}}} \\
\rotatebox{0}{\scalebox{0.25}{\includegraphics{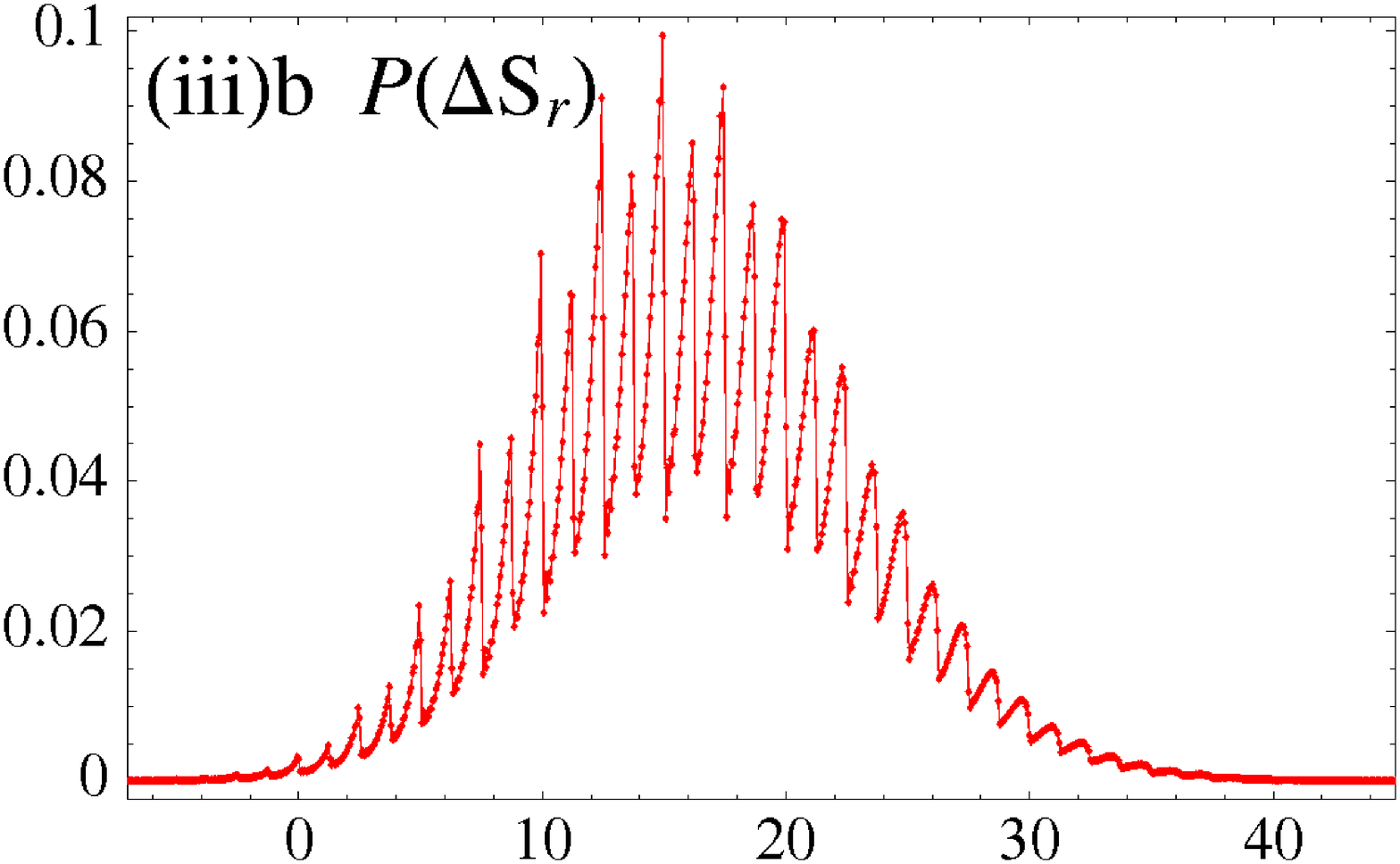}}} &
\rotatebox{0}{\scalebox{0.25}{\includegraphics{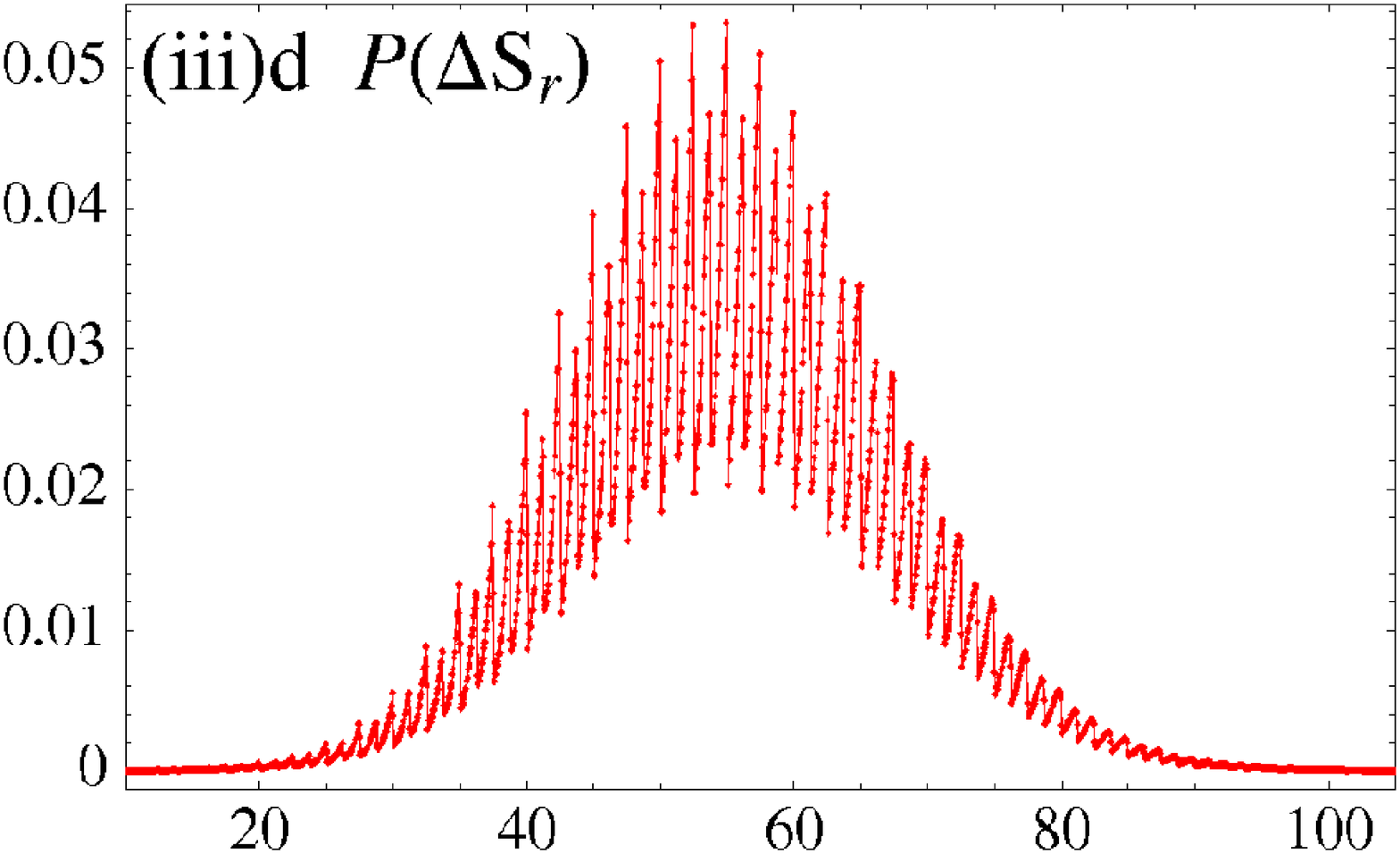}}} \\
\rotatebox{0}{\scalebox{0.25}{\includegraphics{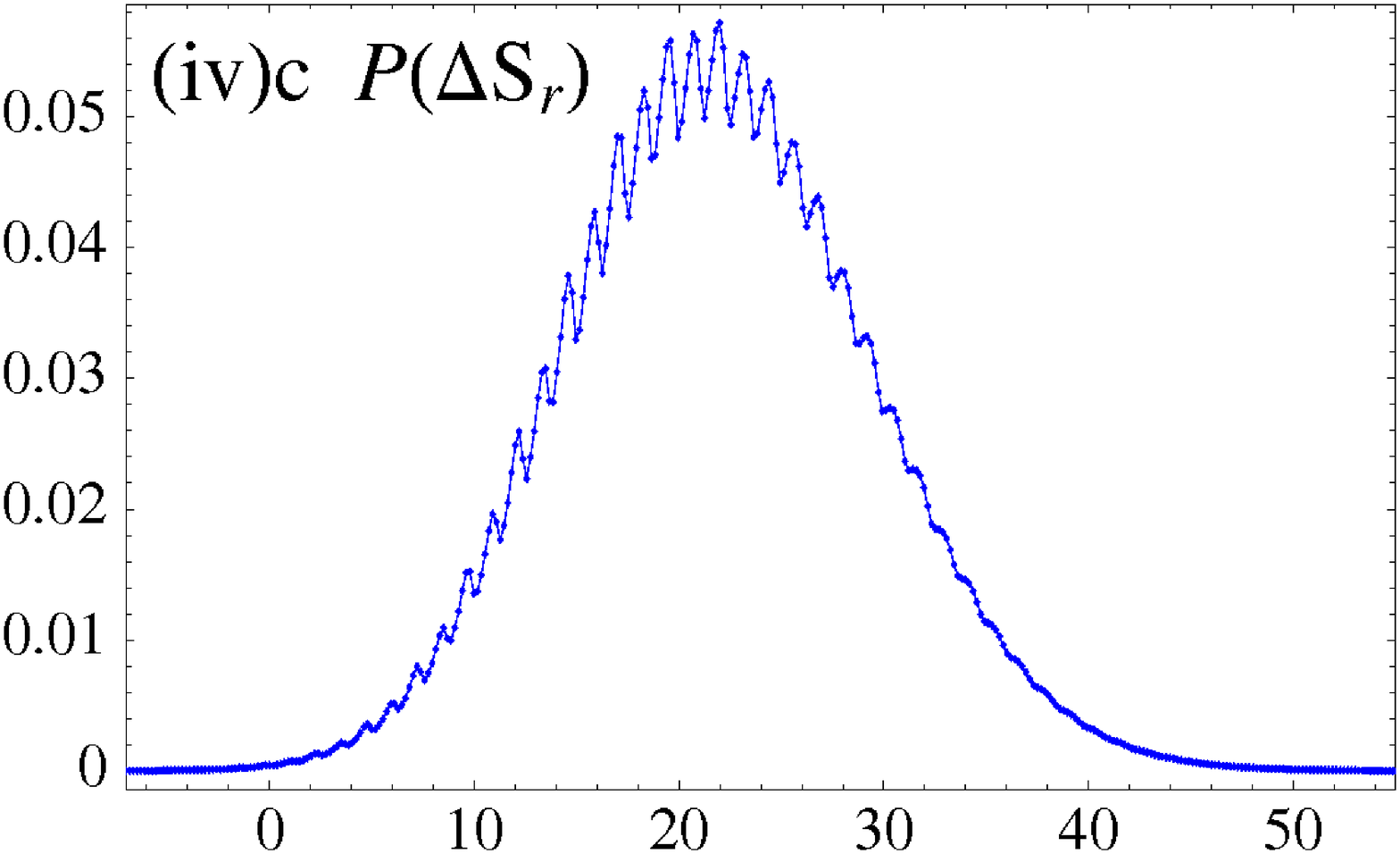}}} &
\rotatebox{0}{\scalebox{0.25}{\includegraphics{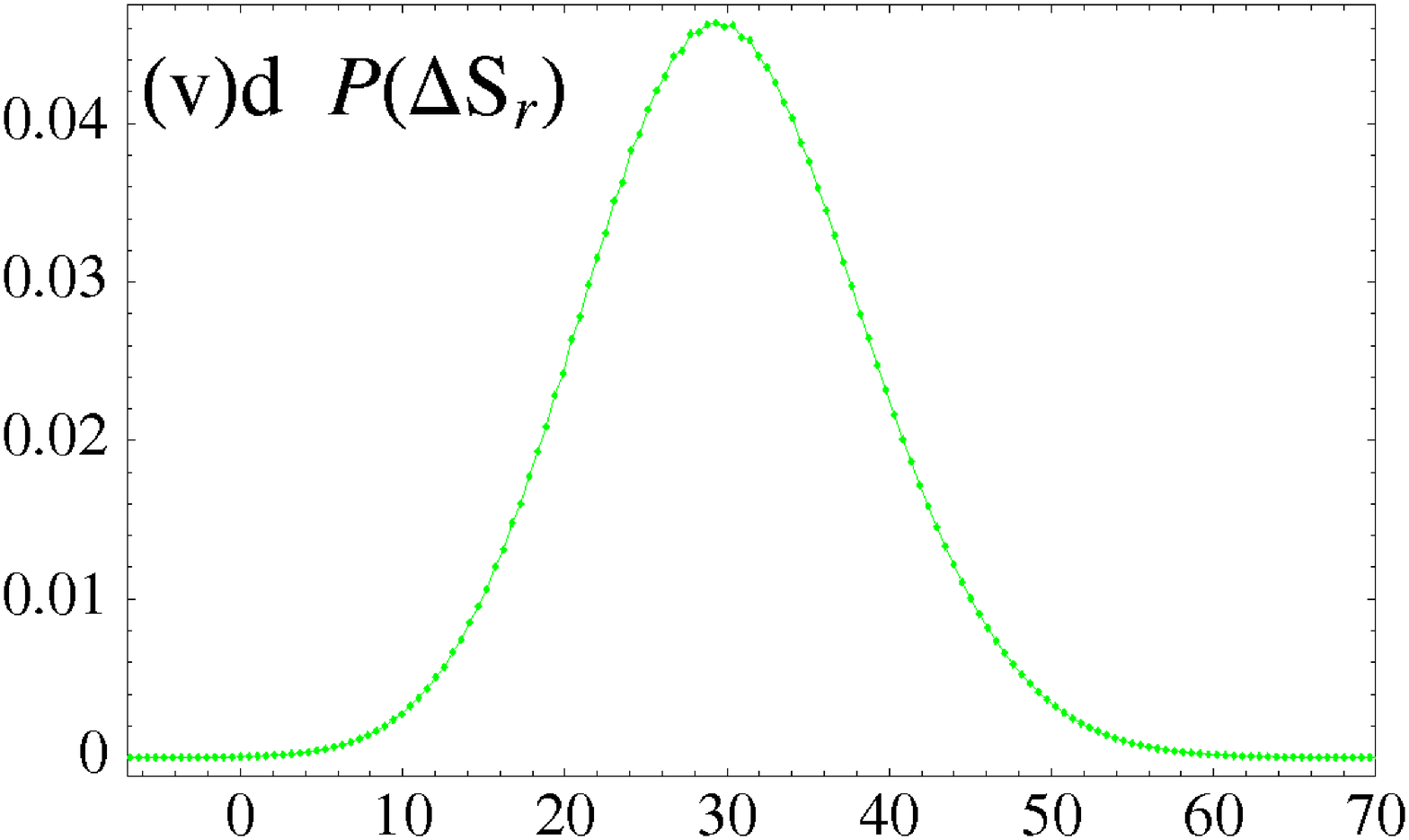}}}
\end{tabular}
\caption{(Color online)
Probability distributions of the change in the reservoir TEP 
for the driving protocols and measurement times shown in Fig. \ref{profilefig}. 
All curves [except (i)a and (id) where the driving starts at the same time 
as the measurement] satisfy the FT (\ref{SSaaaai}).}
\label{PSe}
\end{figure}
%%%%%%%%%%%%%%%%%%%%%%%%%
%%%%%%%%%%%%%%%%%%%%%%%%%
\begin{figure}[p]
\centering
\begin{tabular}{c@{\hspace{0.5cm}}c@{\hspace{0.5cm}}c}
\rotatebox{0}{\scalebox{0.25}{\includegraphics{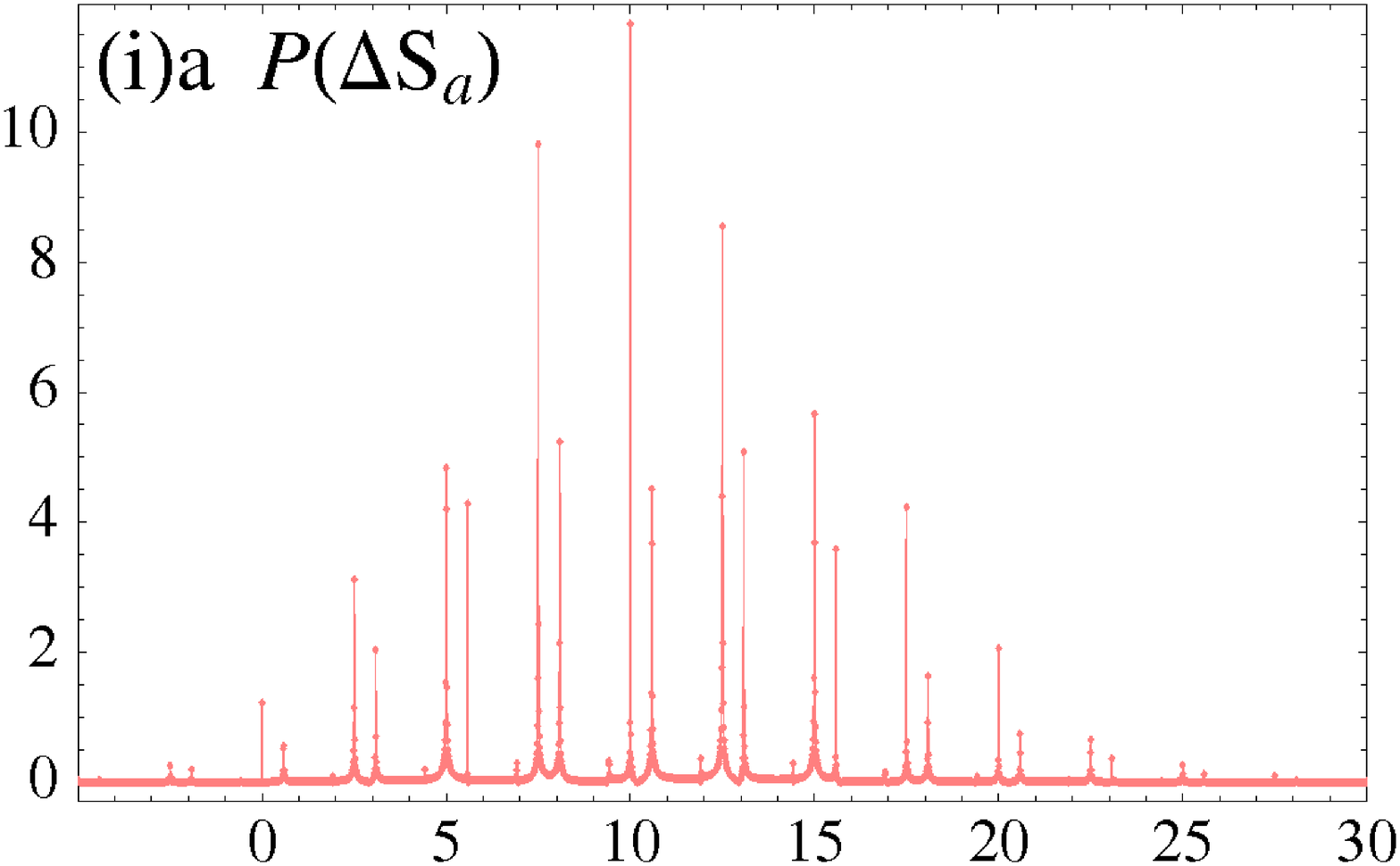}}} &
\rotatebox{0}{\scalebox{0.25}{\includegraphics{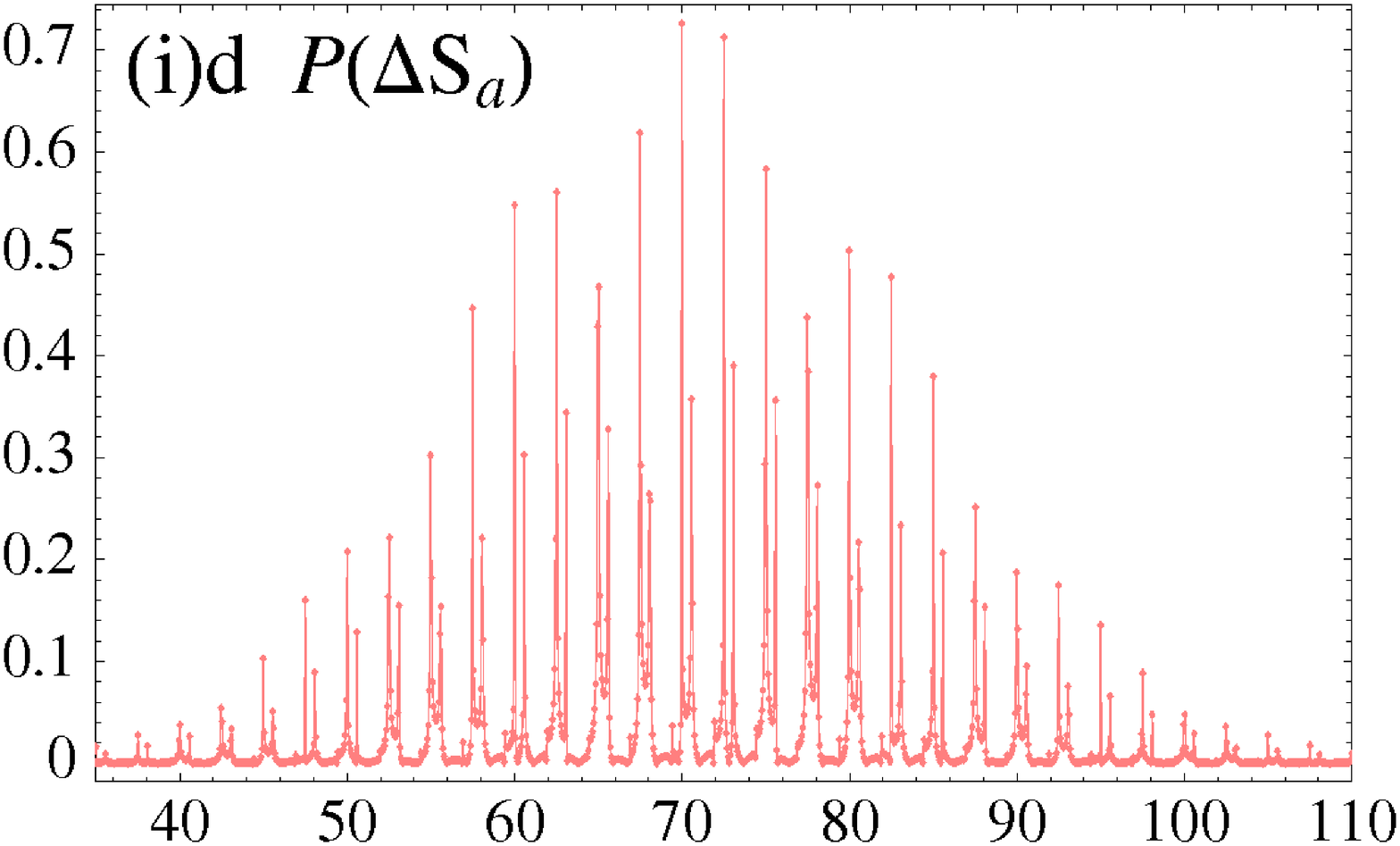}}} \\
\rotatebox{0}{\scalebox{0.25}{\includegraphics{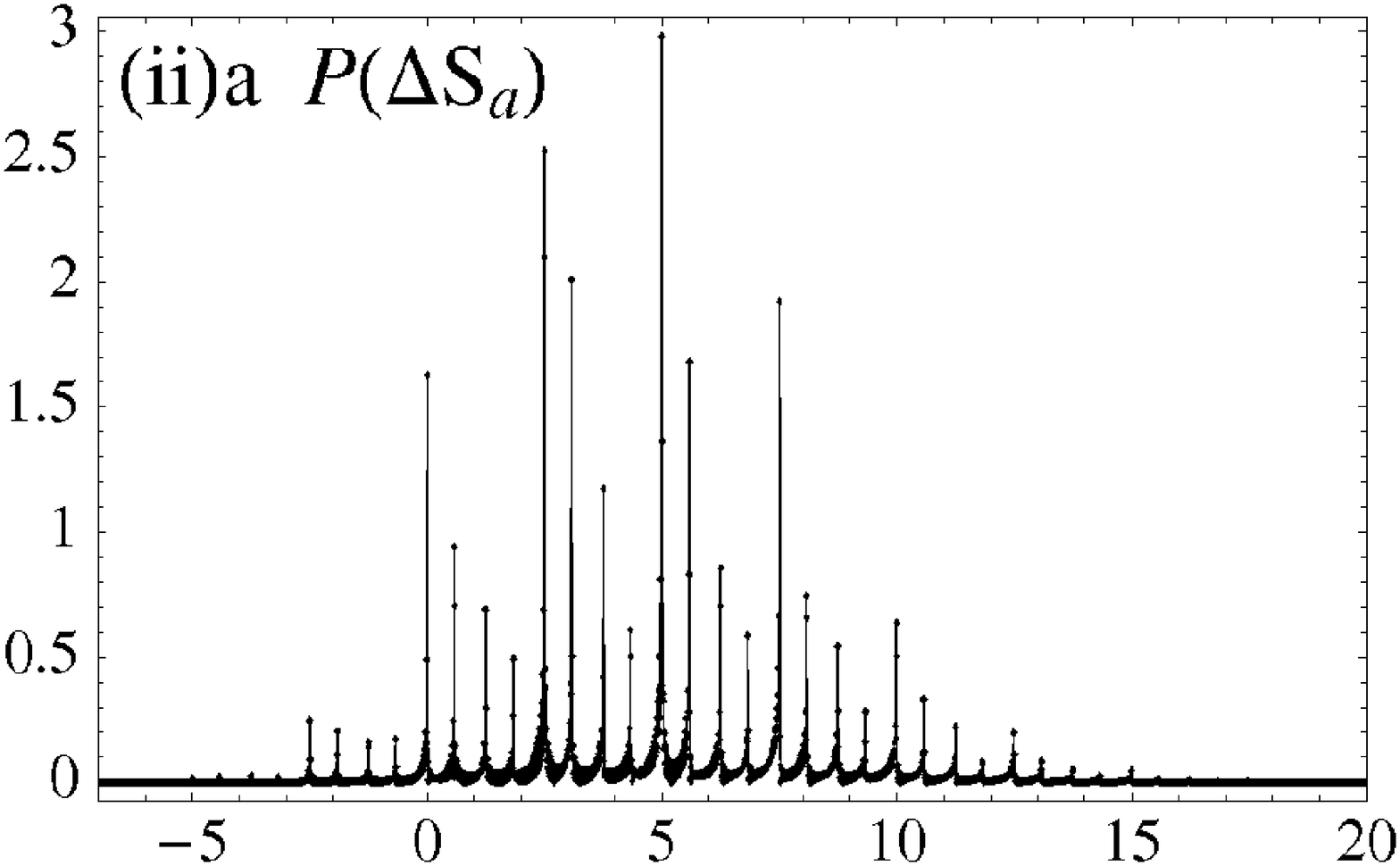}}} &
\rotatebox{0}{\scalebox{0.25}{\includegraphics{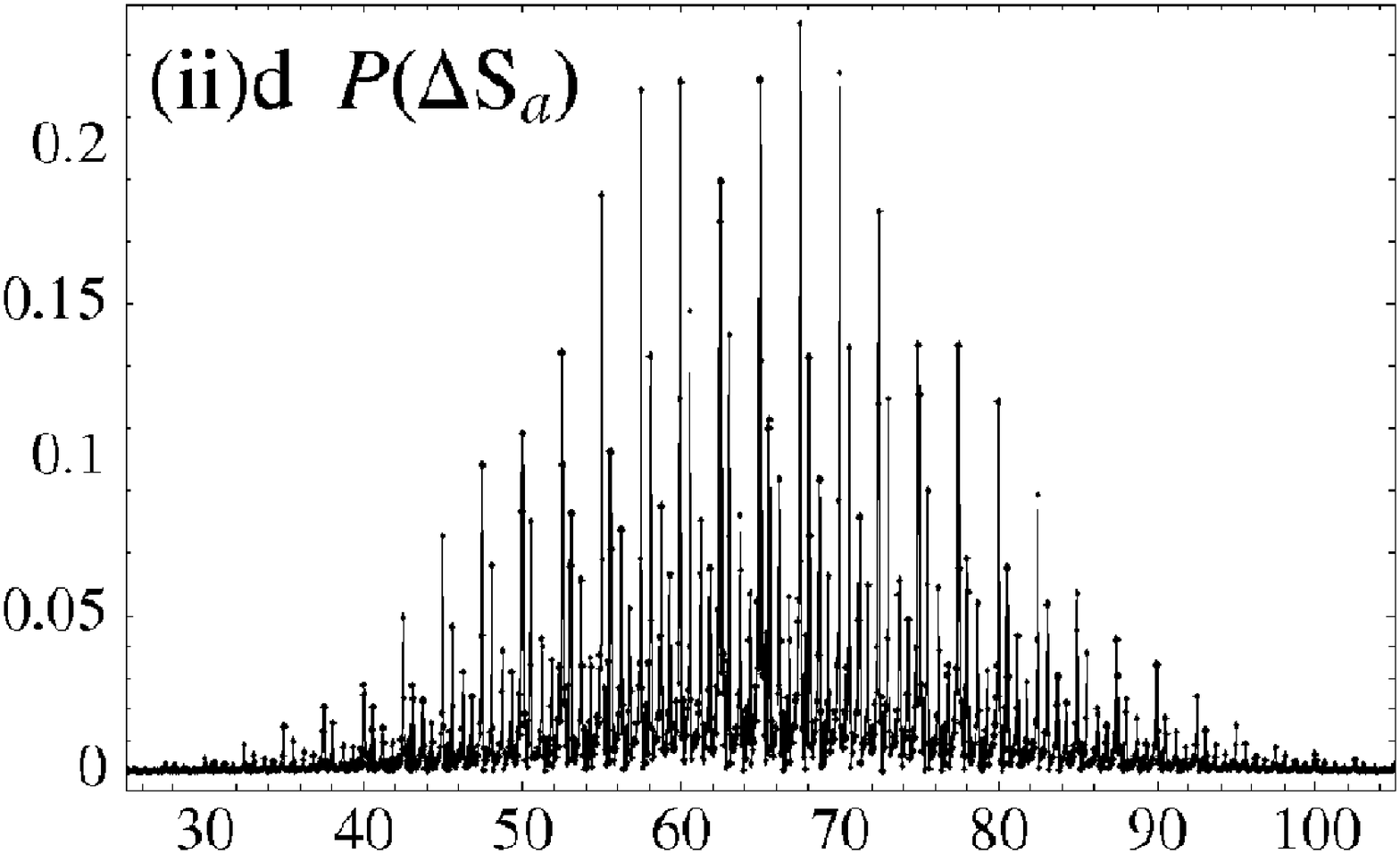}}} \\
\rotatebox{0}{\scalebox{0.25}{\includegraphics{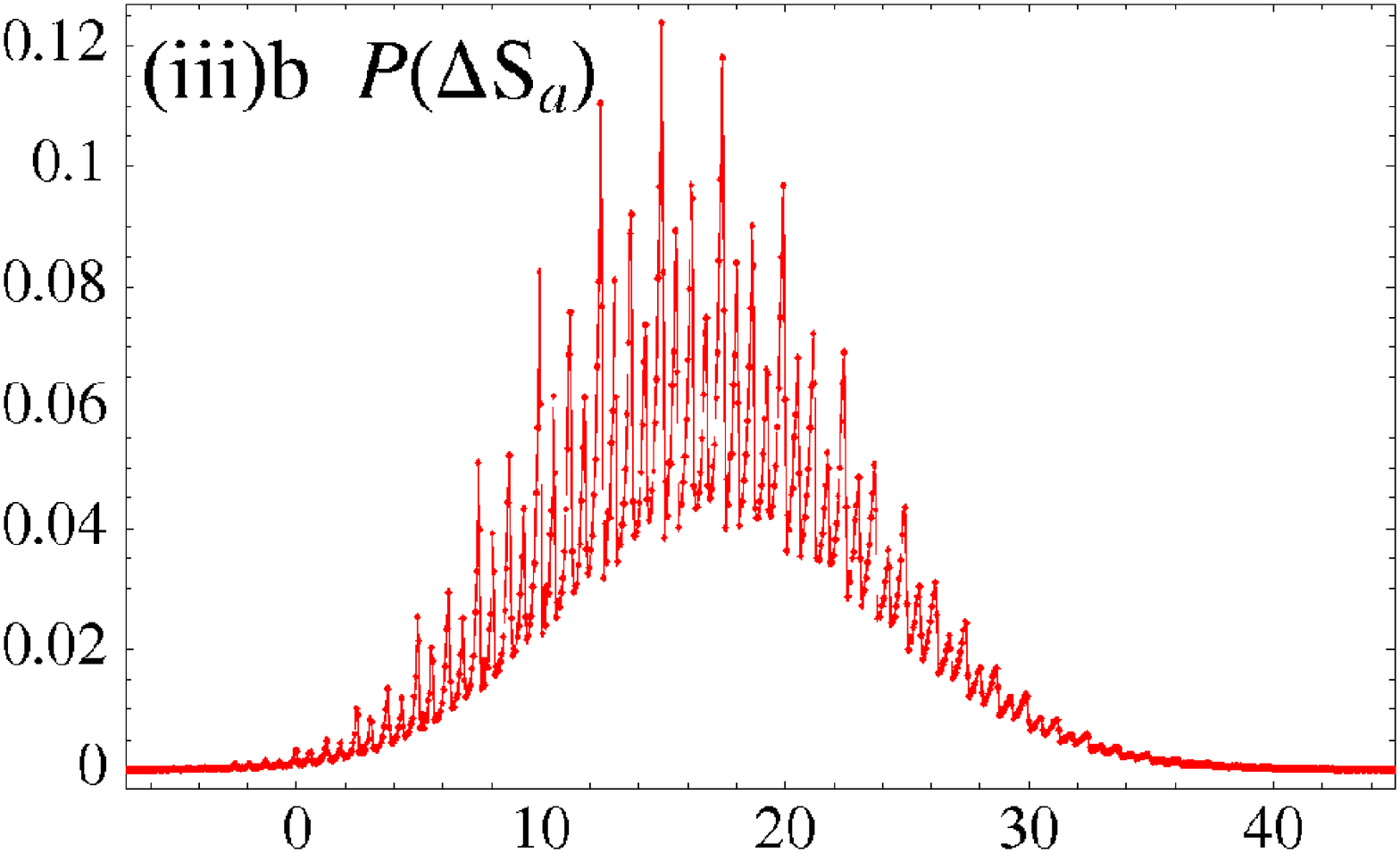}}} &
\rotatebox{0}{\scalebox{0.25}{\includegraphics{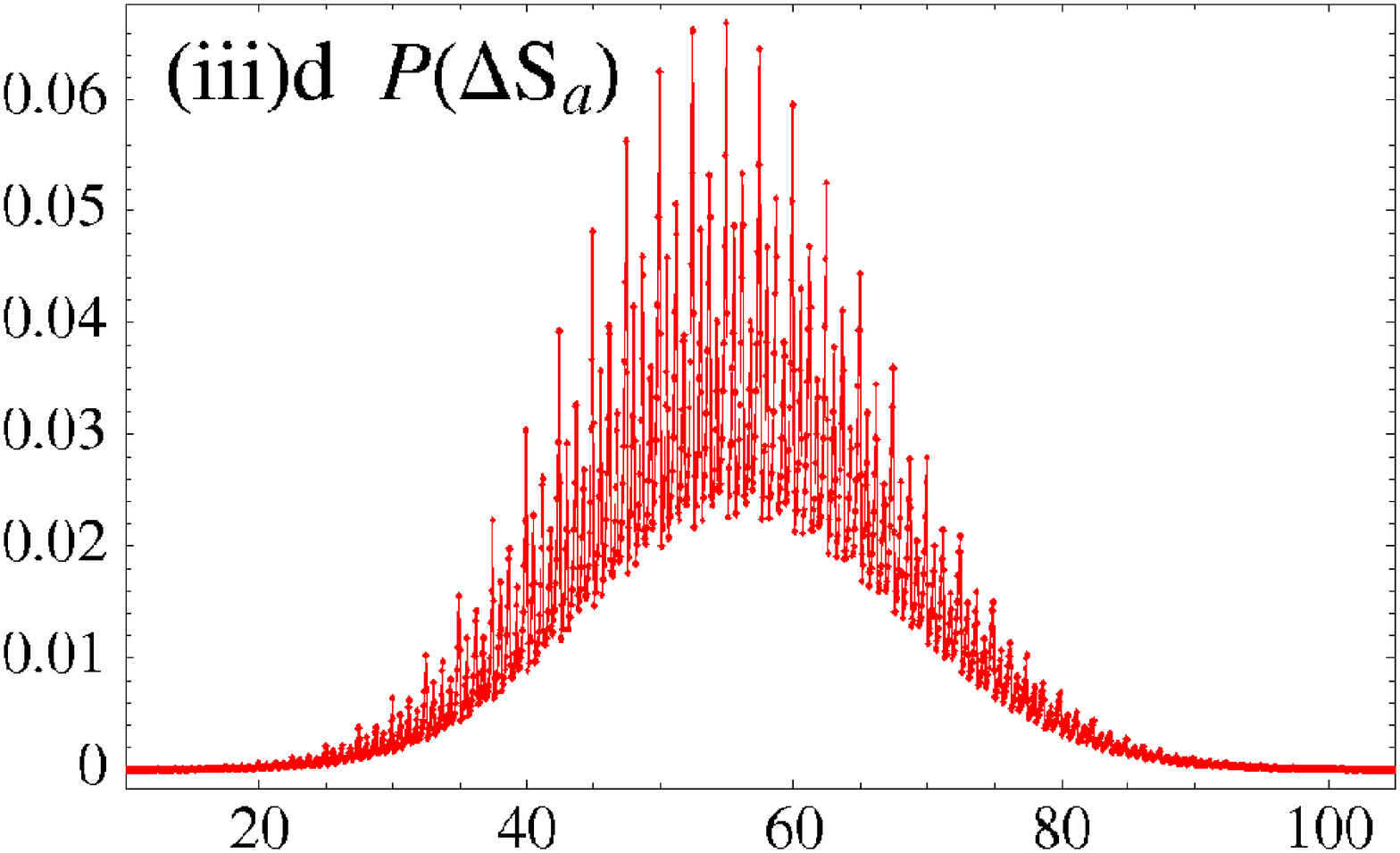}}} \\
\rotatebox{0}{\scalebox{0.25}{\includegraphics{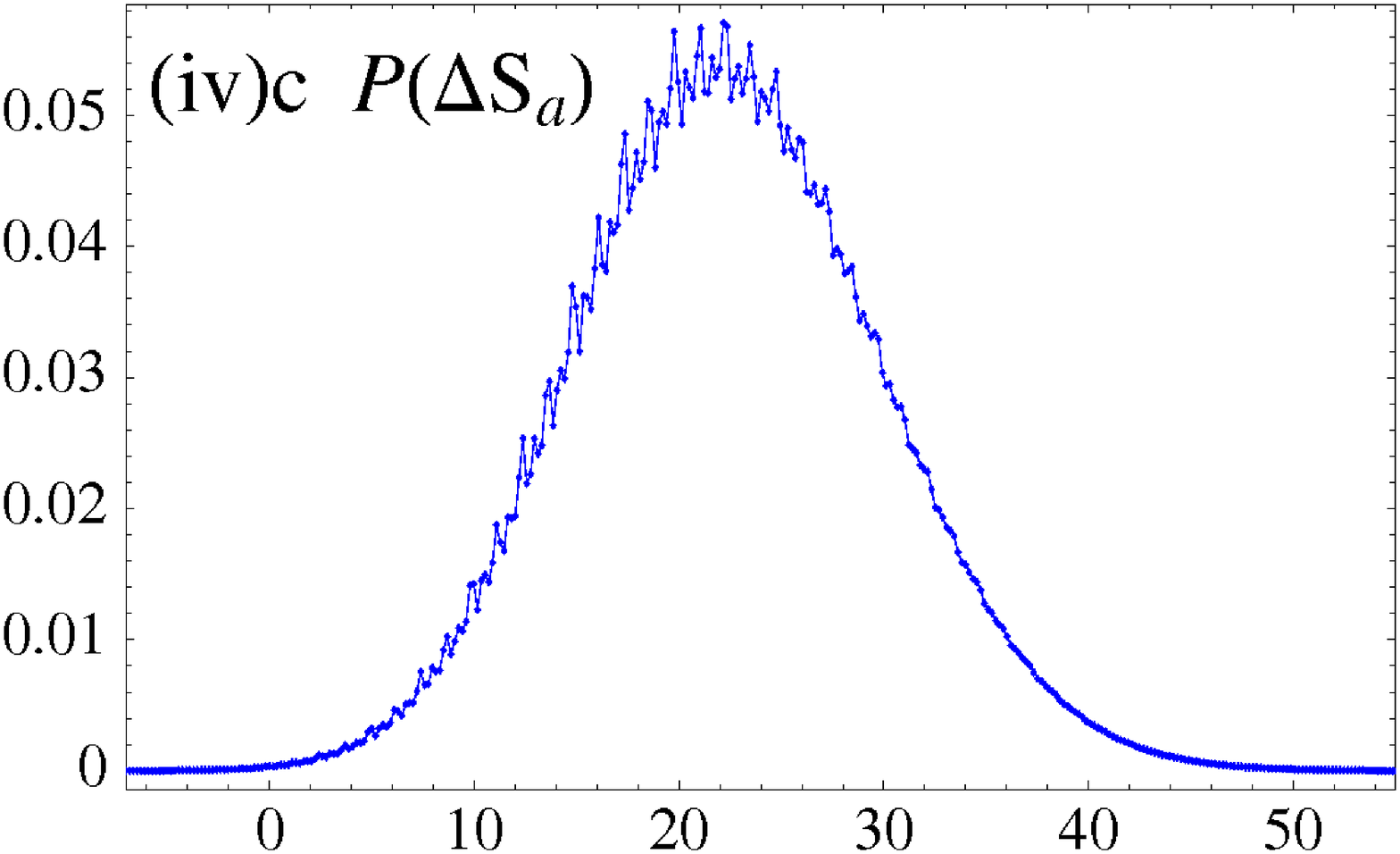}}} &
\rotatebox{0}{\scalebox{0.25}{\includegraphics{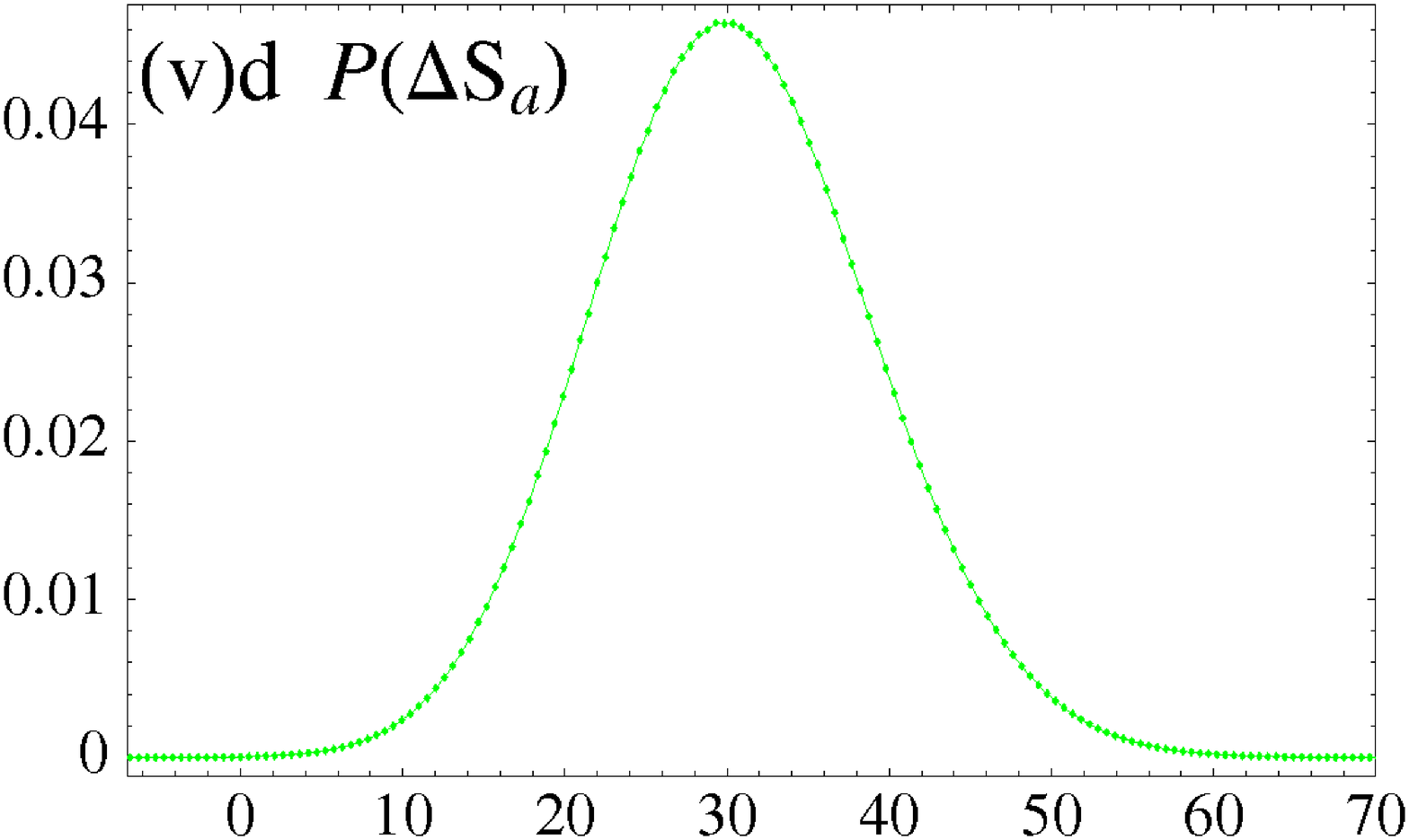}}}
\end{tabular}
\caption{(Color online)
Probability distributions of the change in the adiabatic 
TEP for the driving protocols and measurement times shown in Fig. \ref{profilefig}.
All curves satisfy the FT (\ref{FThk}).}
\label{PSa}
\end{figure}
%%%%%%%%%%%%%%%%%%%%%%%%%
%%%%%%%%%%%%%%%%%%%%%%%%%
\begin{figure}[p]
\centering
\begin{tabular}{c@{\hspace{0.5cm}}c@{\hspace{0.5cm}}c}
\rotatebox{0}{\scalebox{0.25}{\includegraphics{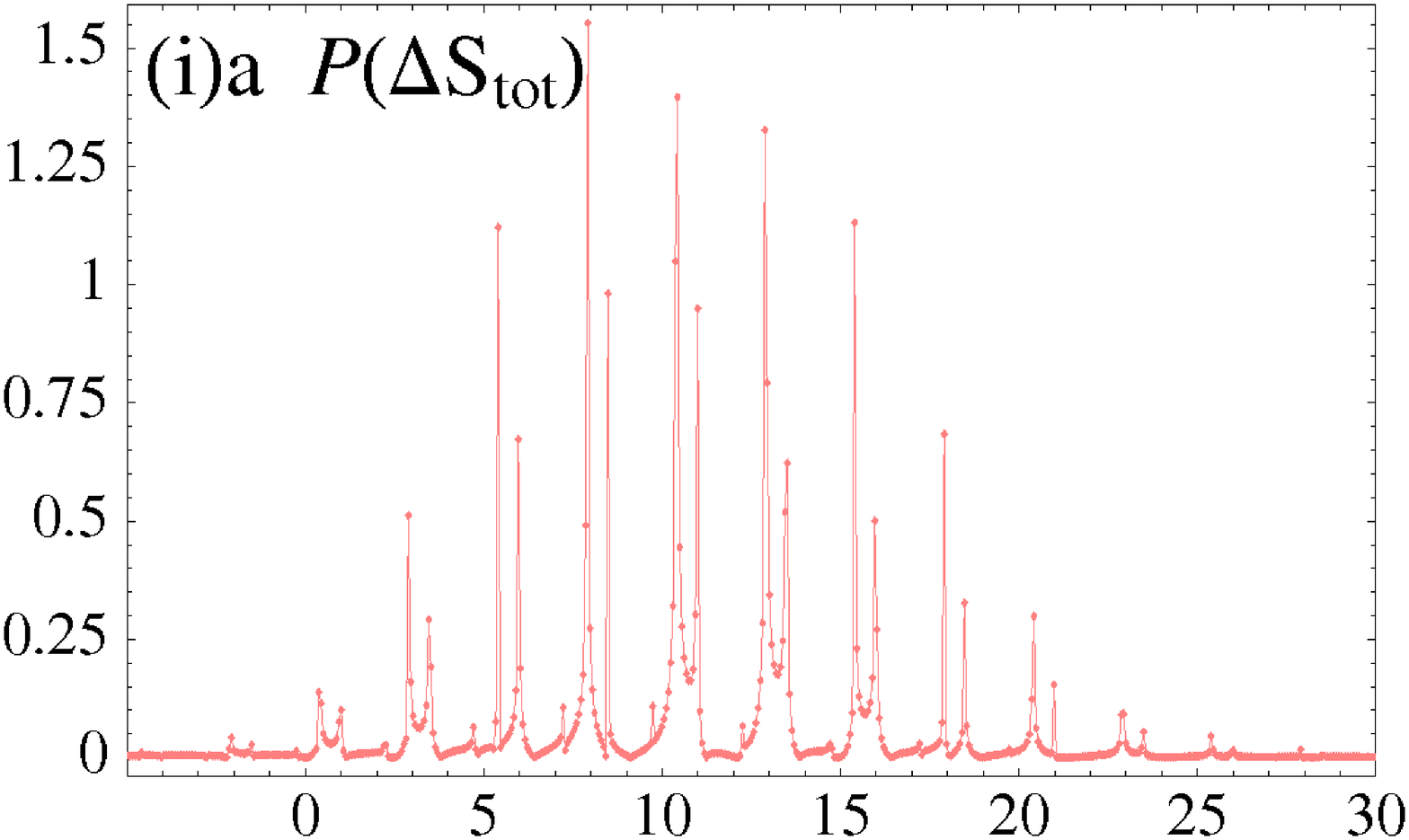}}} &
\rotatebox{0}{\scalebox{0.25}{\includegraphics{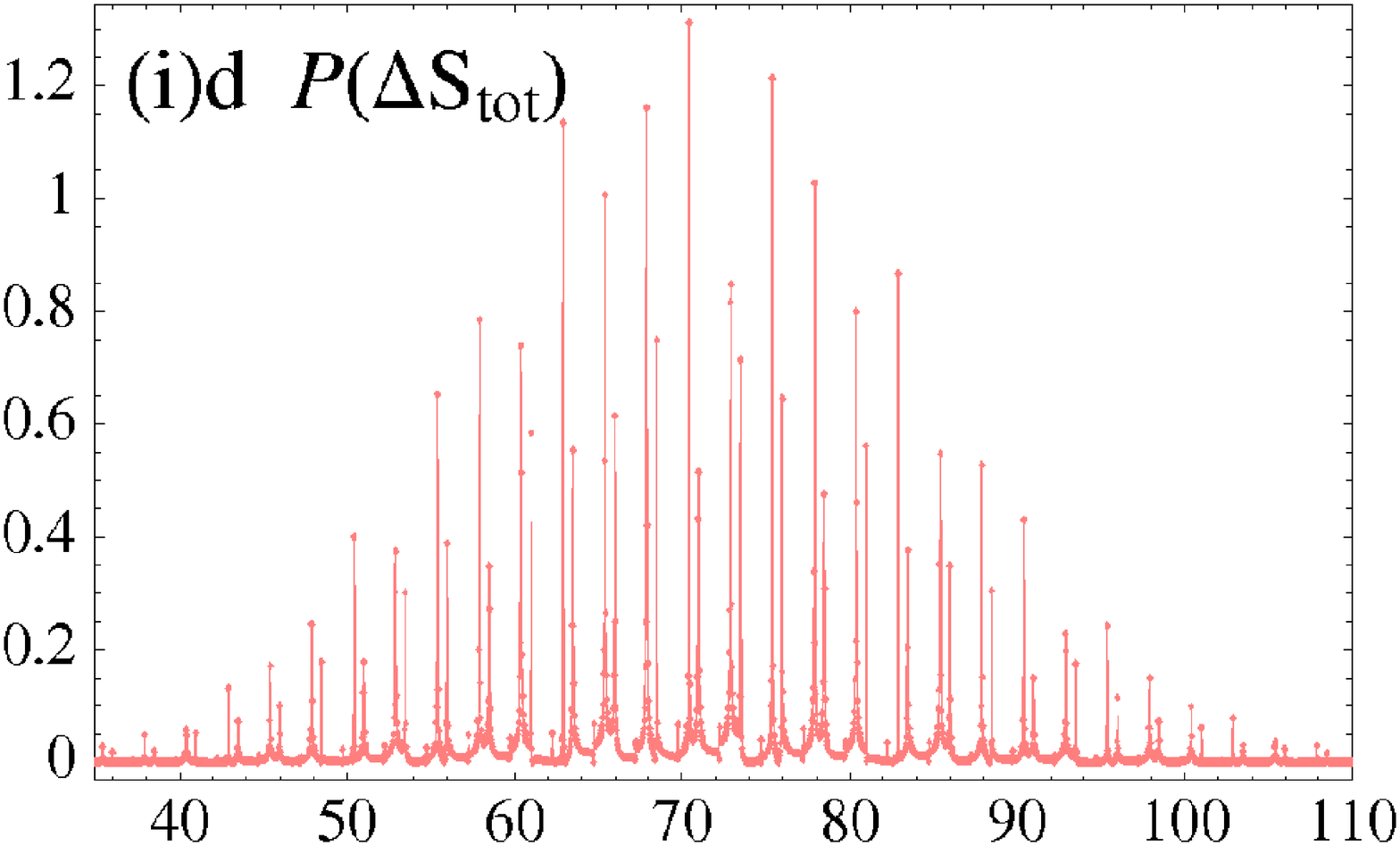}}} \\
\rotatebox{0}{\scalebox{0.25}{\includegraphics{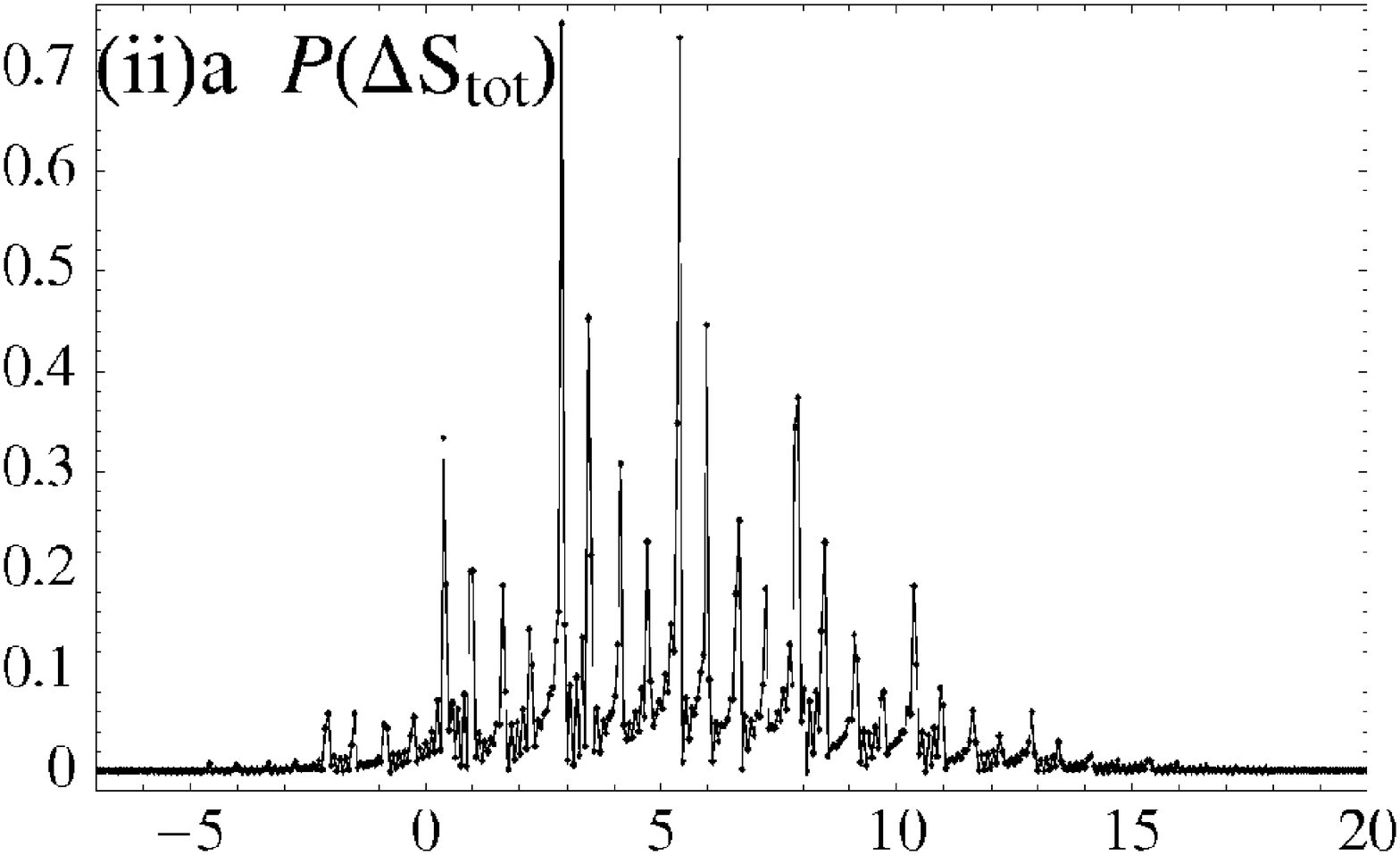}}} &
\rotatebox{0}{\scalebox{0.25}{\includegraphics{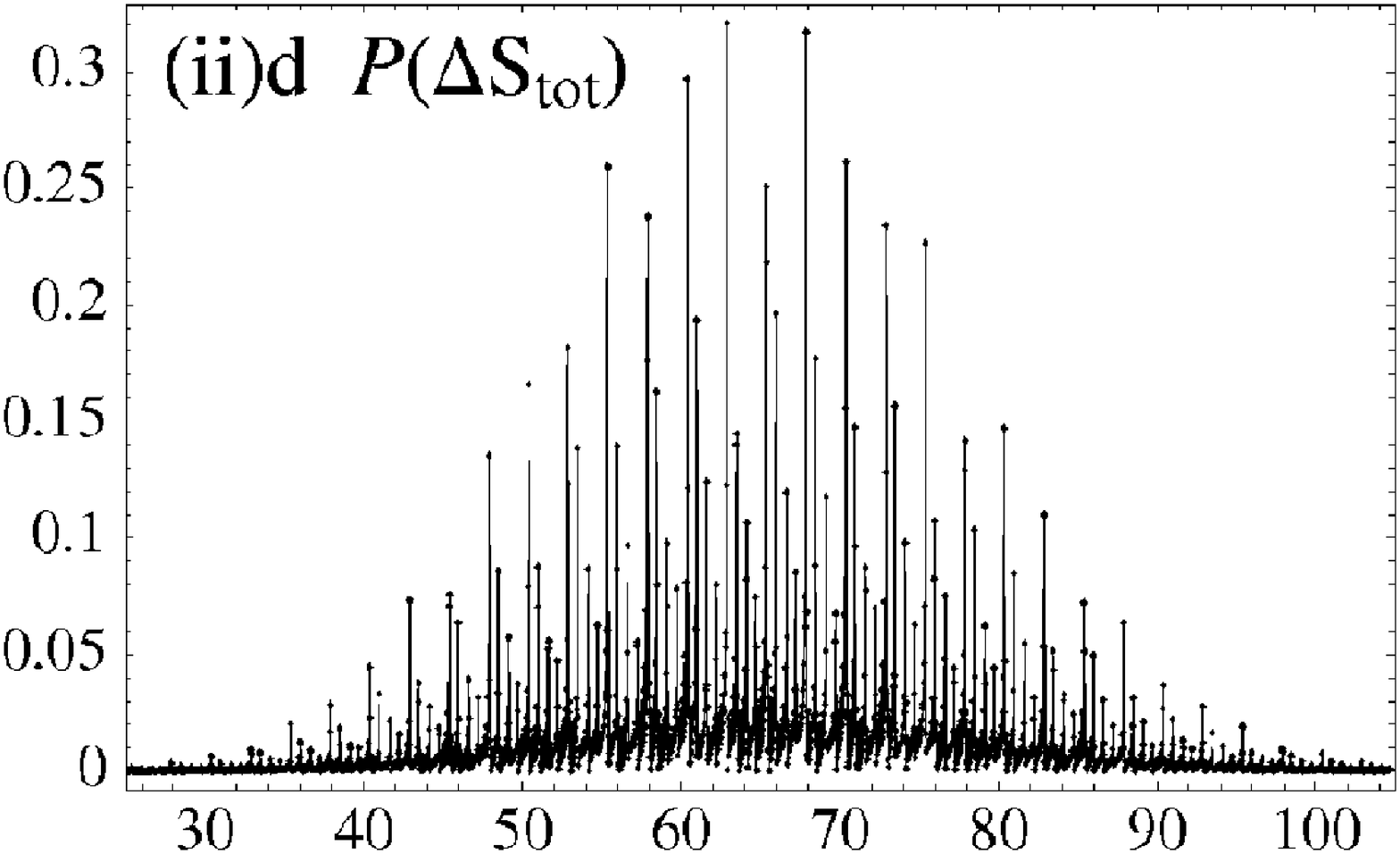}}} \\
\rotatebox{0}{\scalebox{0.25}{\includegraphics{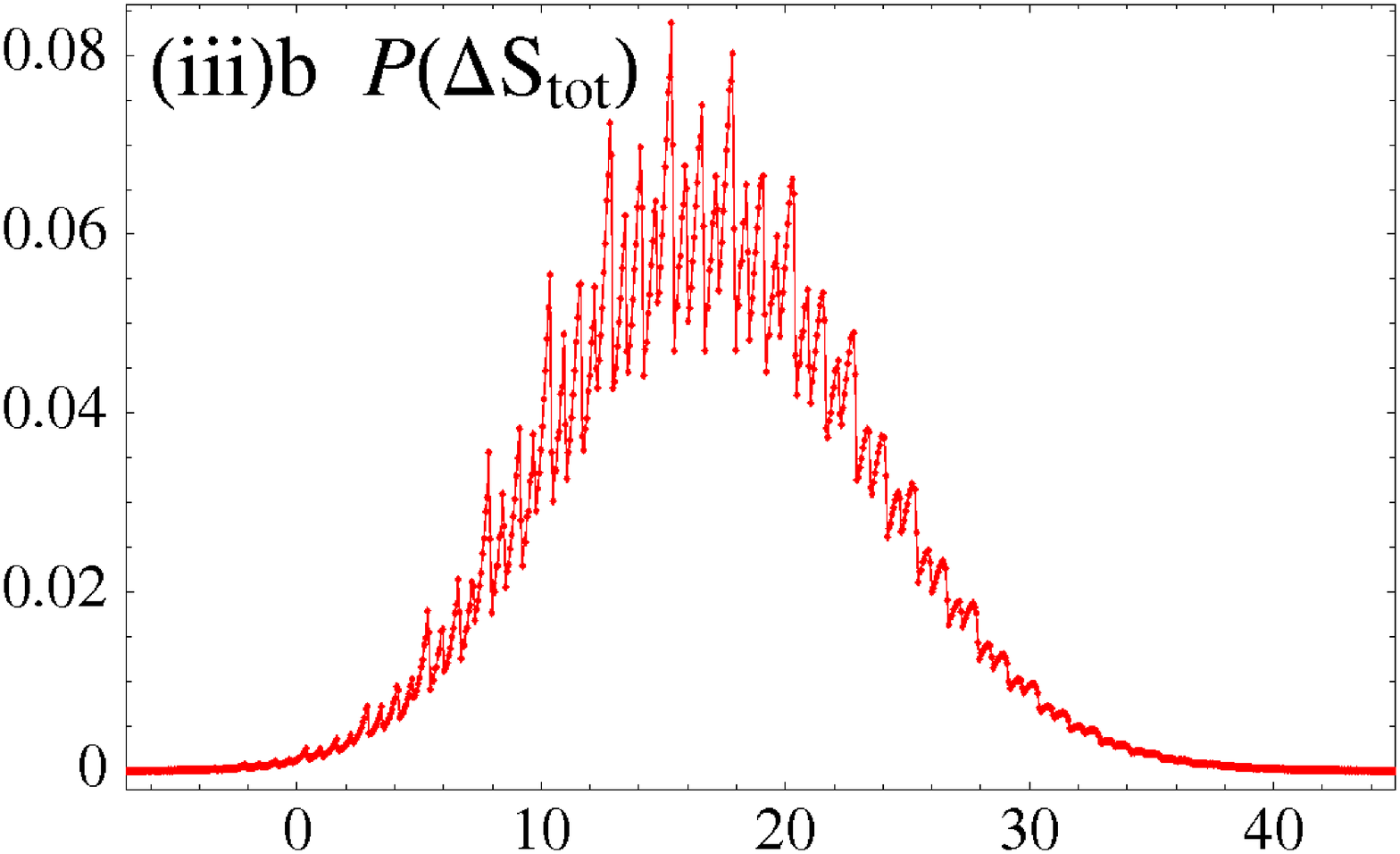}}} &
\rotatebox{0}{\scalebox{0.25}{\includegraphics{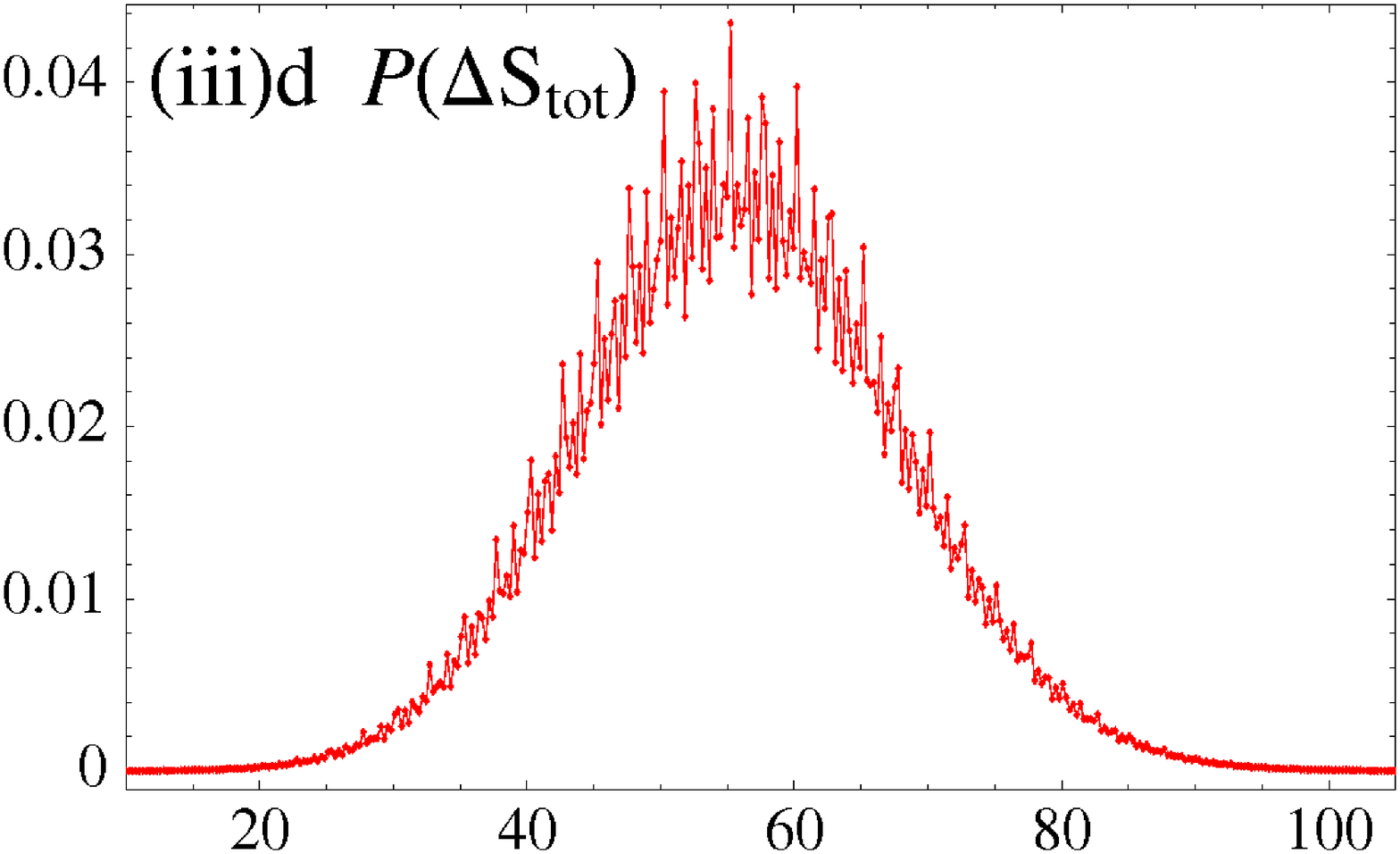}}} \\
\rotatebox{0}{\scalebox{0.25}{\includegraphics{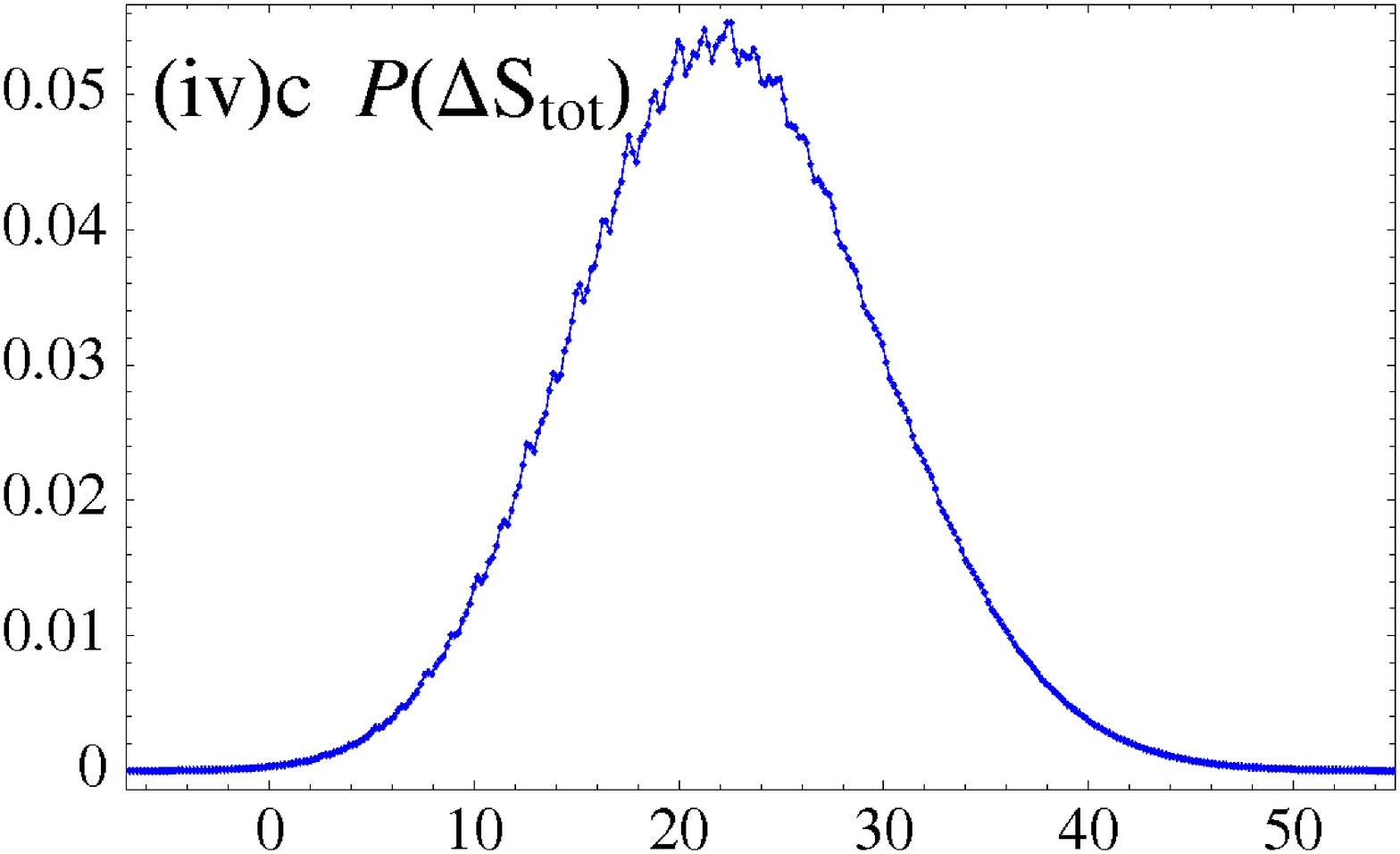}}} &
\rotatebox{0}{\scalebox{0.25}{\includegraphics{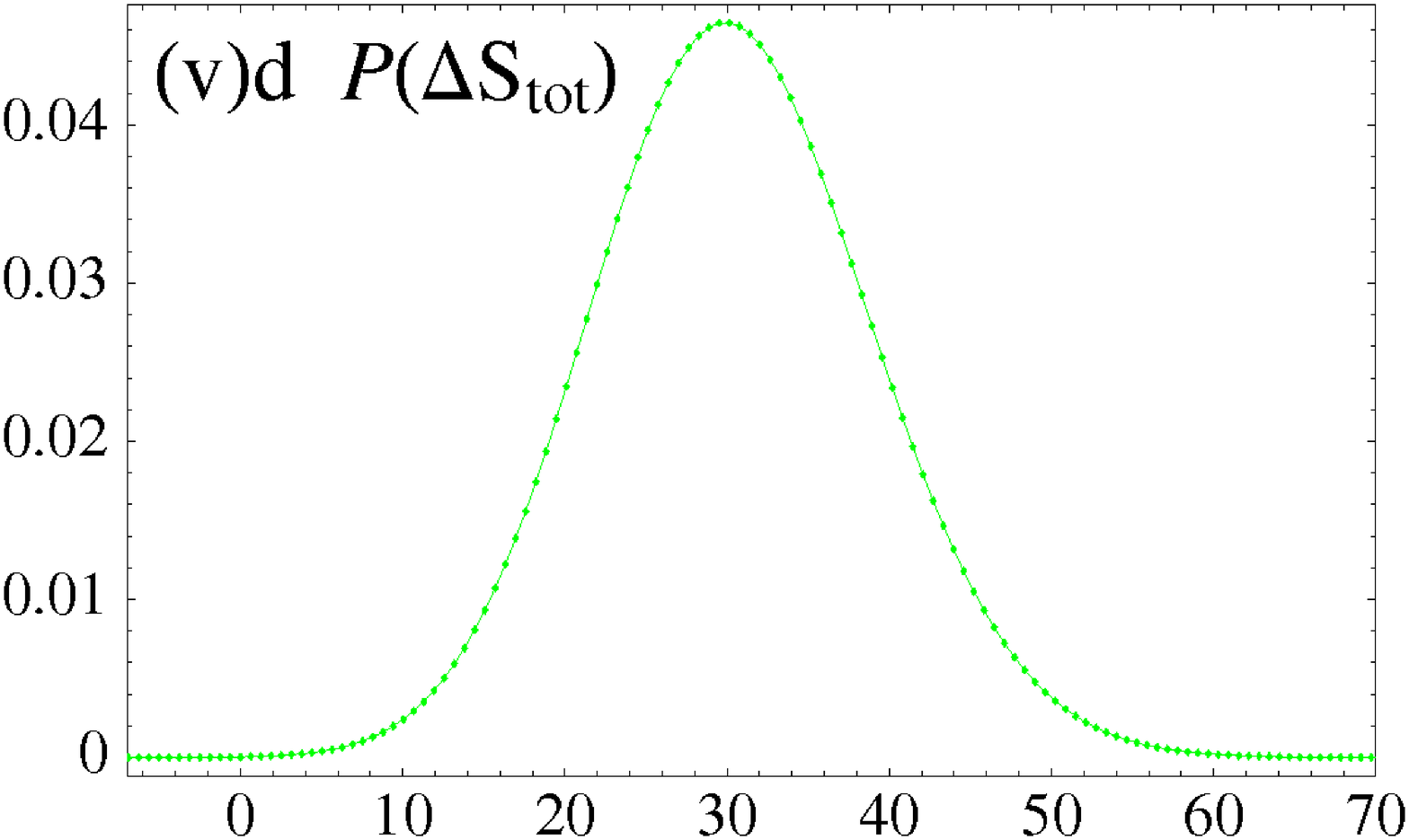}}}
\end{tabular}
\caption{(Color online)
Probability distributions of the change in the total TEP for 
the driving protocols and measurement times shown in Fig. \ref{profilefig}.
All curves satisfy the FT (\ref{FTi}).}
\label{PSi}
\end{figure}
%%%%%%%%%%%%%%%%%%%%%%%%%
%%%%%%%%%%%%%%%%%%%%%%%%%
%\begin{figure}[p]
%\centering
%\begin{tabular}{c@{\hspace{0.5cm}}c@{\hspace{0.5cm}}c}
%\rotatebox{0}{\scalebox{0.25}{\includegraphics{fign/PSi1v.eps}}} &
%\rotatebox{0}{\scalebox{0.25}{\includegraphics{fign/PSa1v.eps}}} \\
%\rotatebox{0}{\scalebox{0.25}{\includegraphics{fign/PSe1v.eps}}} &
%\rotatebox{0}{\scalebox{0.25}{\includegraphics{fign/PSex1v.eps}}} 
%\end{tabular}
%\caption{Staedy state}
%\label{}
%\end{figure}
%%%%%%%%%%%%%%%%%%%%%%%%%
%%%%%%%%%%%%%%%%%%%%%%%%%
\begin{figure}[p]
\centering
\rotatebox{0}{\scalebox{1}{\includegraphics{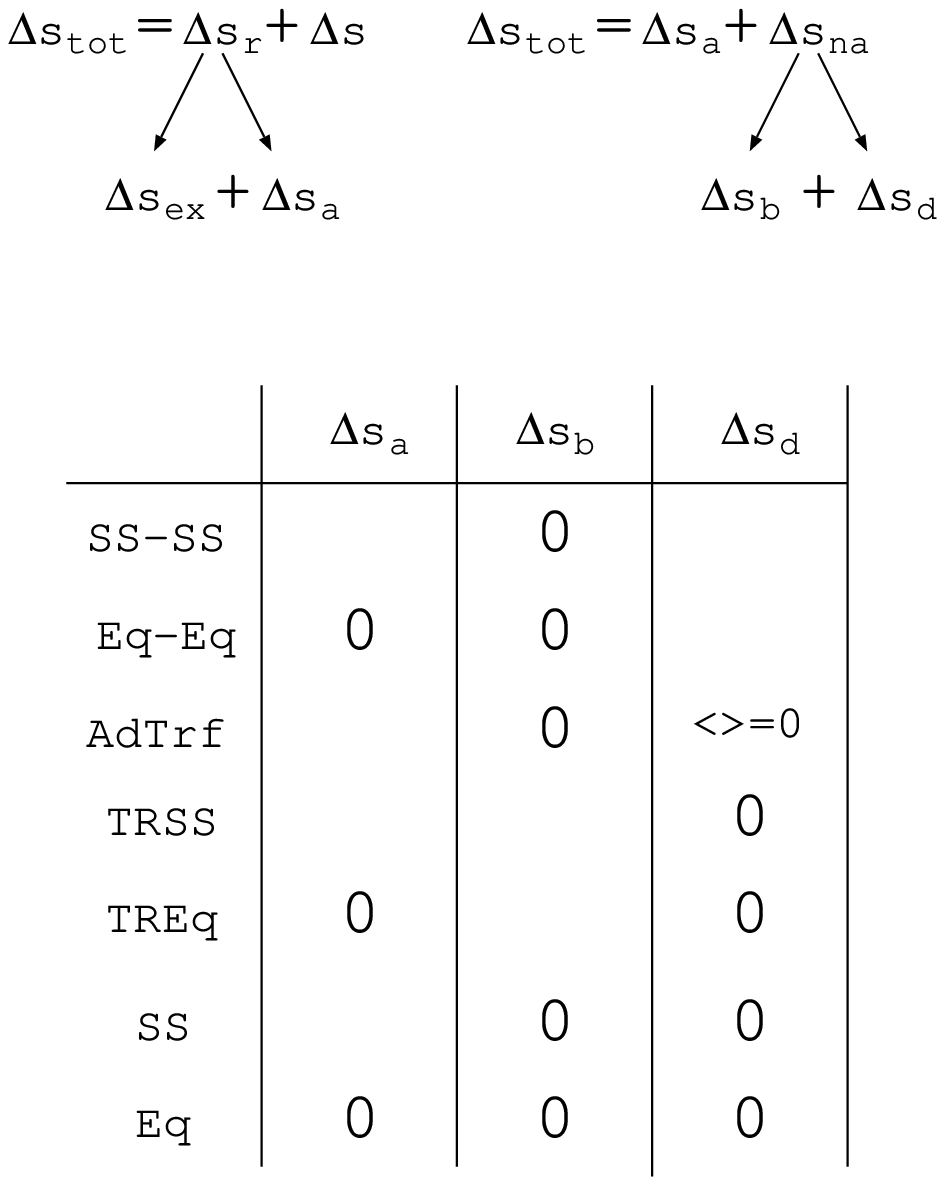}}}
\caption{Summary of the two splittings of the total TEP
in different parts and of the type of transformations during which these parts are zero.
SS-SS: transition between steady states. Eq-Eq: transition between equilibrium states.
AdTrf: adiabatic transformation. TRSS: transient relaxation to steady state.
TREq: transient relaxation to equilibrium. SS: steady state. Eq: equilibrium. 
$<>=0$ means that the ensemble average vanishes.} 
\label{shema}
\end{figure}
%%%%%%%%%%%%%%%%%%%%%%%%%
%%%%%%%%%%%%%%%%%%%%%%%%%%%%%%%%%%%%%%%%%%%%%%%%%%%%%%%%%%%%%%%%%%%%%%%%%%%%%%%%%%%%%%%%%%%%%%%%%%%%%%%%%%
\end{document}